\documentclass[aps,prd,showkeys,superscriptaddress,singlecolumn,nofootinbib,floatfix]{revtex4-2}

\usepackage{subfiles}
\usepackage{amsmath}
\usepackage{amssymb}
\usepackage{amsthm}
\usepackage{mathrsfs}
\usepackage{graphicx}
\usepackage{epstopdf}
\usepackage{fancyhdr}
\usepackage{array}
\usepackage{physics}
\usepackage[all]{xy}
\usepackage{eufrak}
\usepackage{euscript}
\usepackage{enumerate}
\usepackage{slashed}
\usepackage{hyperref}
\usepackage{subfigure}
\usepackage{graphicx}
\usepackage{caption}
\usepackage{soul}
\usepackage{xcolor} 
\usepackage{epstopdf} 

\graphicspath{{./figures/}} 
\hypersetup{pdftex,colorlinks=true,linkcolor=blue,citecolor=blue,menucolor=black,urlcolor=blue,filecolor=blue}

\begin{document}

\title{New purely damped pairs of quasinormal modes in a hot and dense strongly-coupled plasma}

\author{Gustavo de Oliveira}
\email{gustav.o.liveira@discente.ufg.br}
\affiliation{Instituto de F\'{i}sica, Universidade Federal de Goi\'{a}s, Av. Esperan\c{c}a - Campus Samambaia, CEP 74690-900, Goi\^{a}nia, Goi\'{a}s, Brazil}

\author{Romulo Rougemont}
\email{rougemont@ufg.br}
\affiliation{Instituto de F\'{i}sica, Universidade Federal de Goi\'{a}s, Av. Esperan\c{c}a - Campus Samambaia, CEP 74690-900, Goi\^{a}nia, Goi\'{a}s, Brazil}

\begin{abstract}
Perturbed black holes exhibit damped oscillations whose eigenfrequencies define their quasinormal modes (QNMs). In the case of asymptotically Anti-de Sitter (AdS) black holes, the spectra of QNMs are related to the near-equilibrium behavior of specific strongly interacting quantum field theories via the holographic gauge-gravity duality. In the present work, we numerically obtain the spectra of homogeneous non-hydrodynamic QNMs of a top-down holographic construction called the 2 R-Charge Black Hole (2RCBH) model, which describes a hot and dense strongly-coupled plasma. The main result is the discovery of a new structure of pairs of purely imaginary QNMs. Those new purely damped QNMs dominate the late time equilibration of the strongly-coupled plasma at large values of the chemical potential, while at lower values the fundamental QNMs are instead ordinary poles with imaginary and real parts describing oscillatory decaying perturbations. We also observe a new phenomenon of asymptotic pole fusion for different pairs of purely imaginary QNMs at asymptotically large values of the chemical potential. This phenomenon corresponds to the asymptotic merging of the two poles within each pair of purely imaginary QNMs, with the different pairs of merged poles being evenly spaced by a constant value of $4\pi$ in all the different perturbation channels associated to different irreducible representations of the spatial $SO(3)$ rotation symmetry of the medium. In particular, this indicates that characteristic equilibration times for the plasma develop upper bounds that cannot be surpassed by further doping the medium with increasing values of the chemical potential.
\end{abstract}

\maketitle
\tableofcontents


\section{Introduction}
\label{sec:intro}


The holographic gauge-gravity duality \cite{Maldacena:1997re,Gubser:1998bc,Witten:1998qj,Witten:1998zw} (see also e.g. \cite{Aharony:1999ti,Nastase:2007kj,Ramallo:2013bua,Natsuume:2014sfa} for reviews) comprises a detailed mathematical dictionary that maps the evaluation of physical processes and observables of some strongly-coupled quantum field theories (QFTs) into calculations involving semi-classical gravity defined on asymptotically AdS spacetimes with at least one extra dimension (corresponding to the so-called holographic radial coordinate). The target QFTs live at the conformally flat boundary of the curved bulk spacetime.

A given holographic model is defined by a specific bulk gravity action with negative cosmological constant, possibly coupled to some bulk matter fields. Depending on the ansatze and the boundary conditions considered for the bulk fields, one may define the dual QFT under different circumstances at the boundary. For instance, if the ansatze for the bulk fields only depend on the holographic radial coordinate and the bulk metric features no blackening function, the dual QFT is generally defined in vacuum. On the other hand, if the ansatz for the bulk metric comprises a blackening function such that black hole (or, more generally, black brane) solutions are possible, one may define the dual QFT at the boundary in a given thermal state in equilibrium. Different black hole geometries in the bulk are obtained as solutions of the bulk field equations by evolving the bulk fields in the holographic radial coordinate from the black hole horizon up to the boundary, for a given choice of horizon data for those fields; different choices for the horizon data give rise to different black hole geometries, each of them corresponding to some definite thermal state of the dual QFT at the boundary of the considered bulk geometry. More generally, if one considers ansatze for the bulk fields with nontrivial dependence also on coordinates parallel to the boundary, then one can study the dual QFT at the boundary undergoing different out-of-equilibrium dynamics.

A particular case of great physical interest corresponds to considering plane-wave ansatze in the coordinates parallel to the boundary for linearized perturbations of the bulk fields around the black hole backgrounds in thermodynamic equilibrium, which is associated with taking the dual QFT slightly out-of-equilibrium. The corresponding black hole perturbations typically exhibit exponentially damped oscillations, known as quasinormal modes (QNMs), which describe the process of equilibration of the perturbed background geometries \cite{Vishveshwara:1970zz,Davis:1971gg,Nollert:1999ji,Kokkotas:1999bd,Berti:2009kk,Konoplya:2011qq}.

QNMs are of great relevance in cosmology and astrophysics because they describe the late time ringdown of the remnants of mergers of black holes and binary stars, which were of fundamental importance for the first direct detections of gravitational waves emitted in such processes \cite{LIGOScientific:2016aoc,LIGOScientific:2016sjg}. Within the context of the gauge-gravity correspondence, the QNMs of a given gravity-matter model correspond to the poles of the retarded thermal correlators of gauge-invariant operators of the dual QFT \cite{Starinets:2002br,Kovtun:2005ev}, encoding a large amount of information about the near-equilibrium properties of the boundary theory. In particular, the hydrodynamic QNMs\footnote{A hydrodynamic QNM corresponds to a mode satisfying a dispersion relation with vanishing frequency in the limit of zero wavenumber, $\omega(\mathbf{k}\to\mathbf{0})= 0$.} of the system can be employed to derive some hydrodynamic transport coefficients of the medium, alternatively to the more direct use of Kubo's formulas \cite{Policastro:2002se,Policastro:2002tn,Heller:2013fn,Janik:2016btb}. On the other hand, the non-hydrodynamic QNMs\footnote{A non-hydrodynamic QNM corresponds to a mode satisfying a dispersion relation with non-vanishing frequency in the limit of zero wavenumber, $\omega(\mathbf{k}\to\mathbf{0})\neq 0$.} of the system provide essential information on upper bounds for characteristic equilibration times of the medium near thermodynamic equilibrium \cite{Horowitz:1999jd}, being also of fundamental importance in the analysis of the domain of applicability of the asymptotic hydrodynamic gradient series \cite{Heller:2013fn,Heller:2015dha,Buchel:2016cbj}. Quasinormal modes in several different holographic models have been analyzed, e.g. in Refs.~\cite{Janik:2015waa,Rougemont:2015wca,Janiszewski:2015ura,Janik:2015iry,Attems:2016ugt,Gursoy:2016ggq,Demircik:2016nhr,Finazzo:2016psx,Betzios:2017dol,Critelli:2017euk,Rougemont:2018ivt}, which is by no means an exhaustive list.

In the present work, we numerically derive the spectra of homogeneous non-hydrodynamic QNMs of the so-called 2 R-Charge Black Hole (2RCBH) model, which is a top-down holographic construction describing a strongly coupled quantum fluid at finite temperature and density \cite{DeWolfe:2011ts,DeWolfe:2012uv}. As shall be discussed in detail, we discover new features associated with the emergence of pairs of purely imaginary QNMs at nonzero R-charge chemical potential, which describe purely damped perturbations with no oscillatory contribution, besides the usual holographic structure of pairs of ordinary quasinormal modes with mirrored nonzero real parts.

The manuscript is organized as follows. In section \ref{sec:2} we briefly review the two top-down holographic models to be comparatively analyzed in the present work, their background black hole solutions in equilibrium and the associated thermodynamic observables, while in section \ref{sec:3} we briefly review the classification of the diffeomorphism and gauge-invariant combinations of homogeneous linearized fluctuations of the bulk fields, which are organized under different irreducible representations of the $SO(3)$ rotation symmetry group of the isotropic black hole backgrounds. Our main results are presented in sections \ref{sec:4}, \ref{sec:5}, and \ref{sec:6}, where we numerically derive the spectra of homogeneous non-hydrodynamic QNMs for the three different $SO(3)$ channels of the 2RCBH model, also comparing them with previous results for the QNM spectra of the so-called 1 R-Charge Black Hole (1RCBH) model \cite{Finazzo:2016psx,Critelli:2017euk}. Finally, in section \ref{sec:conc} we conclude with the main observations disclosed in the present work, and also point out future perspectives for developments based on the new findings reported here.

In the present work we use natural units with $\hbar = c = k_B = 1$ and a mostly plus metric signature.

\section{2RCBH and 1RCBH models}
\label{sec:2}

Since both holographic models to be discussed in the present work are described by five-dimensional bulk actions of the Einstein-Maxwell-Dilaton (EMD) class, we begin by writing down a generic EMD action in five dimensions,\footnote{More generally, one could also consider the addition of a topological five-dimensional Chern-Simons term of the form $A_1\wedge F_2\wedge F_2$, where $F_2=dA_1$ is the Maxwell strength 2-form. However, as discussed in \cite{DeWolfe:2010he}, this term vanishes for the field configurations considered here for both the thermodynamic backgrounds and their fluctuations.}
\begin{equation}
\label{EMDaction}
    S=\frac{1}{2\kappa_5^2}\int_{\mathcal{M}_5} \dd^5x~\sqrt{-g}\left[R-\frac{f(\phi)}{4}F_{\mu\nu}F^{\mu\nu}-\frac{1}{2}(\partial_\mu\phi)^2-V(\phi)\right],
\end{equation}
where $g_{\mu\nu}$ is the bulk spacetime metric with associated Ricci scalar curvature $R$, $A_{\mu}$ is a Maxwell field with associated strength tensor $F_{\mu\nu}=\partial_\mu A_{\nu}-\partial_\nu A_\mu$, $\phi$ is a real scalar field called the dilaton, with an associated potential $V(\phi)$, $f(\phi)$ is the Maxwell-dilaton coupling function, and $\kappa_5^2\equiv 8\pi G_5$, with $G_5$ being the five-dimensional Newton's constant. The bulk action \eqref{EMDaction} is accompanied by two boundary terms: the traditional Gibbons-Hawking-York action \cite{York:1972sj,Gibbons:1976ue} required for the well-posedness of the Dirichlet boundary condition problem in spacetimes with boundaries \cite{Poisson:2009pwt} (which is the case of asymptotically AdS geometries used in holography), and a counterterm action \cite{Critelli:2017euk} constructed via holographic renormalization \cite{Bianchi:2001kw,Skenderis:2002wp,deHaro:2000vlm,Papadimitriou:2011qb,Lindgren:2015lia,Elvang:2016tzz} with the purpose of systematically and consistently removing the divergences of the full on-shell boundary action. Since these two boundary terms will not be needed in the calculations to be discussed in the present work, we do not bother to explicitly writing them down here.

The corresponding field equations obtained by extremizing the action \eqref{EMDaction} with respect to the bulk EMD fields are \cite{Critelli:2017euk},
\begin{subequations}
\label{EqsofMotion}
    \begin{align}
\label{eq:Einstein} R_{\mu\nu}-\frac{g_{\mu\nu}}{3}\left[V(\phi)-\frac{f(\phi)}{4}F_{\alpha\beta}F^{\alpha\beta}\right]-\frac{1}{2}\partial_\mu \phi \partial_\nu \phi-\frac{f(\phi)}{2} F_{\mu\rho}F_\nu\,^\rho &=0,\\
   \label{eq:Maxwell} \partial_\mu\left(\sqrt{-g}f(\phi) F^{\mu\nu}\right)&=0,\\
   \label{eq:Dilaton} \frac{1}{\sqrt{-g}} \partial_\mu\left(\sqrt{-g} g^{\mu\nu} \partial_\nu\phi\right) - \frac{\partial_\phi f(\phi)}{4}F_{\mu\nu}F^{\mu\nu}-\partial_\phi V(\phi)&=0.
\end{align}
\end{subequations}

In order to analyze equilibrium properties of spatially homogeneous and isotropic charged EMD black hole backgrounds in thermodynamic equilibrium, we take the following ansatze for the bulk EMD fields \cite{DeWolfe:2010he},
\begin{align}
\label{AnsatzEqs}
\dd s^2=e^{2 A(r)}\left[-h(r)\dd t^2+\dd \mathbf{x}^2\right]+\frac{e^{2B(r)}}{h(r)}\dd r^2,\qquad A_\mu = \Phi(r)\delta_\mu^0,\qquad \phi=\phi(r),
\end{align}
where all the bulk fields only depend on the holographic radial coordinate, $r$, the metric is isotropic in spatial directions parallel to the boundary, and the Maxwell field has only a nontrivial time component (which will be identified with the chemical potential of the dual strongly-coupled quantum field theory at the boundary located at $r\to\infty$). Plugging the ansatze \eqref{AnsatzEqs} into the field equations \eqref{EqsofMotion}, one obtains the following set of coupled ordinary differential equations of motion \cite{DeWolfe:2010he,Rougemont:2015wca},
\begin{subequations}
\label{backODEs}
    \begin{align}
    A''-A'B'+\frac{1}{6}\phi'^2&=0,\\
    h''+(4A'-B')h'-e^{-2A}f(\phi)\Phi'^2&=0,\\
   \label{eq:MaxwellAnsatz} \Phi''+\left[2A'-B'+\frac{\partial_\phi f(\phi)}{f(\phi)}\phi'\right]\Phi'&=0,\\
    \phi''+\left[4 A'-B'+ \frac{h'}{h}\right]\phi'-\frac{e^{2B}}{h} \left[\partial_\phi V(\phi)-\frac{e^{-2(A+B)}}{2}\partial_\phi f(\phi) \Phi'^2\right]&=0,\\
    h\left(24A'^2-\phi'^2\right)+6A'h'+2e^{2B}V(\phi)+e^{-2A}f(\phi)\Phi'^2 &=0,
\end{align}
\end{subequations}
where the prime denotes a derivative with respect to the holographic radial coordinate, $r$. Any isotropic and translationally invariant charged EMD black hole background in thermodynamic equilibrium must satisfy the set of equations of motion \eqref{backODEs}.

Of particular interest for us here, there are the 2RCBH and 1RCBH models, which are both described by different EMD actions. These are two top-down holographic models obtained as different particular cases of a more general holographic construction called the STU model \cite{Behrndt:1998jd,Cvetic:1999ne}.\footnote{The bosonic part of five-dimensional $\mathcal{N}=2$ gauged supergravity is given by Eq. (1) of \cite{Behrndt:1998jd}, with the potential of the scalar fields of the theory being given by Eq. (2) of that same reference, satisfying the general constraint in Eq. (3). When that constraint is given the particular realization in Eq. (38) of \cite{Behrndt:1998jd}, one defines the STU model, where S, T, U are three scalar fields of which only 2 are independent, due to the constraint Eq. (38), $\textrm{STU} = 1$. Although it originally appears as a particular case of five-dimensional $\mathcal{N}=2$ gauged supergravity with the aforementioned constraint, the STU model can also be embedded into five-dimensional maximally supersymmetric $\mathcal{N}=8$ gauged supergravity \cite{Behrndt:1998jd,Cvetic:1999ne}.} The STU model describes five-dimensional black brane solutions\footnote{Although the term ``black hole'' is being used here and in the related holographic literature, we are actually dealing with black brane backgrounds.} which are charged only under the $U(1)\times U(1)\times U(1)$ Cartan subgroup of the global $SU(4)$ R-symmetry group of the holographic dual strongly-coupled $\mathcal{N}=4$ Supersymmetric Yang-Mills (SYM) theory in flat four spacetime dimensions. In its general form, the STU model comprises three different conserved Abelian R-charges associated with three Maxwell gauge fields, besides two scalar fields and the metric field. By setting two of the three R-charges of the STU model to zero one obtains the 1RCBH model, while by setting one of the three R-charges to zero and by further identifying the remaining two R-charges one gets the 2RCBH model \cite{DeWolfe:2011ts,DeWolfe:2012uv}. Both models reduce to the purely thermal SYM plasma when the nontrivial R-charge left in each theory is taken to zero.

Although being both descendants of the STU model, the 1RCBH and 2RCBH models describe rather different realizations of the strongly-coupled SYM plasma at finite temperature and R-charge density. In fact, as discussed in \cite{DeWolfe:2011ts,DeWolfe:2012uv,Finazzo:2016psx}, while the 1RCBH model can only probe a limited range of values of R-charge chemical potential over temperature, $\mu/T\in[0,\pi/\sqrt{2}]$, with $\mu/T=\pi/\sqrt{2}$ being a critical point of its phase diagram, the 2RCBH model probes values of $\mu/T\in[0,\infty)$ and has no critical point in its phase diagram. Moreover, for each value of $\mu/T$ in the 1RCBH model there are two different black hole solutions, one thermodynamically stable and another one unstable, while in the 2RCBH model there is only one branch of black hole solutions for each value of $\mu/T$.

Due to the critical point featured in its phase diagram, and also due to its simplicity and the fact that it is a rigorous top-down holographic construction defined at finite temperature and density, the 1RCBH model has attracted considerable attention in the literature in recent years. Indeed, the thermodynamics of the model was analyzed in \cite{DeWolfe:2011ts,DeWolfe:2012uv,Finazzo:2016psx}, some hydrodynamic transport coefficients were derived in \cite{DeWolfe:2011ts,Asadi:2021hds}, the spectra of QNMs were obtained in \cite{Critelli:2017euk,Finazzo:2016psx}, several observables related to the theory of quantum information were evaluated in \cite{Ebrahim:2018uky,Ebrahim:2020qif,Amrahi:2021lgh}, chaotic properties and the pole-skipping phenomenon were addressed in \cite{Amrahi:2023xso,Karan:2023hfk}, while the holographic renormalization of the model and far-from-equilibrium numerical simulations of homogeneous isotropization dynamics and the inhomogeneous Bjorken flow were discussed in \cite{Critelli:2017euk,Critelli:2018osu,Rougemont:2022piu,Rougemont:2024hpf}.

On the other hand, the physics of the 2RCBH model remained up to now largely unexplored in the literature. In fact, besides its thermodynamics \cite{DeWolfe:2011ts,DeWolfe:2012uv} and a few hydrodynamic transport coefficients \cite{DeWolfe:2011ts}, little is known about the behavior of different physical observables in the 2RCBH model. In the present work, besides briefly reviewing and plotting the thermodynamic observables of the 2RCBH model, comparing them to the corresponding results of the 1RCBH construction, we have as our main purpose the investigation of the spectra of QNMs of the 2RCBH model, which shall reveal new and very interesting features, as we are going to discuss in detail. Moreover, the systematic comparison between two holographic models defined at finite temperature and density, both descending from the same top-down construction, but with rather different properties, will be also important to obtain insights on the behavior of different physical observables in strongly-coupled quantum gauge theories in the absence or in the presence of a critical point in the phase diagram of the model.

Denoting by the $k=1,2$ index, respectively, the 1RCBH and 2RCBH models, the corresponding bulk actions are both of the EMD class given by Eq.~\eqref{EMDaction}, with the following particular forms for the dilaton potential $V_k(\phi)$ and the Maxwell-dilaton coupling function $f_k(\phi)$ \cite{DeWolfe:2011ts,DeWolfe:2012uv},
\begin{align}
V_1(\phi) =V_2(\phi)= -\frac{1}{L^2} \left(8 e^{\phi/\sqrt{6}} + 4 e^{-\sqrt{2/3}\,\phi} \right), \qquad f_1(\phi) = e^{- 2\sqrt{2/3}\,\phi}, \qquad f_2(\phi) = e^{\sqrt{2/3}\,\phi},
\label{eq:Vandf}
\end{align}
where $L$ is the radius of the asymptotically AdS$_5$ background geometry. Concerning the 1RCBH and 2RCBH models viewed as particular cases of the STU model \cite{DeWolfe:2012uv}, as aforementioned, the 1RCBH model is obtained by setting two of the three Abelian R-charges of the STU model to zero, $Q_2=Q_3=0$, while the 2RCBH model is defined by setting $Q_1=0$ and by further identifying $Q_3=Q_2$. Notice also from the ultraviolet, near-boundary expansion of the dilaton potential~\eqref{eq:Vandf}, $V_k(\phi)=[-12-2\phi^2+\mathcal{O}(\phi^4)]/L^2$, that the mass of the dilaton field is tachyonic, $m_\phi^2=\partial_\phi^2 V_k(\phi=0)=-4/L^2$, and satisfies the Breitenlohner-Freedman (BF) bound for massive scalar fields on stable asymptotically AdS$_{d+1}$ spacetimes, $m_\phi^2\ge -d^2/4L^2$ \cite{Breitenlohner:1982jf,Breitenlohner:1982bm} (see also \cite{Ramallo:2013bua}). According to the holographic dictionary, the scaling dimension $\Delta_\phi$ of a quantum field theory operator at the boundary, dual to a bulk scalar field with mass $m_\phi$, satisfies the relation $m_\phi^2 L^2 = \Delta_\phi(\Delta_\phi-4)$, which implies that the boundary scalar field operator dual to the bulk dilaton field has scaling dimension $\Delta_\phi=2$ for both 1RCBH and 2RCBH models.

\subsection{Background Fields}

By plugging $V_k(\phi)$ and $f_k(\phi)$ from Eq.~\eqref{eq:Vandf} into the field equations~\eqref{backODEs}, one obtains the specific equations of motion for the 1RCBH and 2RCBH models in thermodynamic equilibrium, whose solutions are obtained as analytical functions of the charge $Q_k$ and the mass $M$ of the black hole backgrounds as follows \cite{DeWolfe:2011ts,DeWolfe:2012uv,Finazzo:2016psx},\footnote{There is a typo corresponding to a wrong factor of $1/2$ (which should be $\sqrt{2}$) for $\Phi_2(r)$ in Eq.~(14) of \cite{DeWolfe:2012uv}, and also a typo corresponding to a wrong factor of $1/2$ (which should be 1) for $\Phi_1(r)$ in Eq.~(17) of \cite{DeWolfe:2012uv}. On the other hand, the signs of $\Phi_1(r)$ and $\Phi_2(r)$ in \cite{DeWolfe:2011ts} are inverted relatively to the ones used here and in \cite{DeWolfe:2012uv,Finazzo:2016psx}: those are still solutions of the field equations, but would imply instead in negative R-charge chemical potentials. Notice also the different notation used in \cite{DeWolfe:2011ts}, where the black hole mass $M$ and the dual gauge theory chemical potential $\mu_k$ are instead denoted by $\sqrt{\mu}$ and $\Omega_k$, respectively.}
\begin{subequations}
\label{eq:AnsatzAll}
    \begin{align}
   &A_1(r)=\ln \frac{r}{L} +\frac{1}{6}\ln\left(1+\frac{Q_1^2}{r^2}\right),\qquad & &A_2(r)=\ln \frac{r}{L} +\frac{1}{3}\ln\left(1+\frac{Q_2^2}{r^2}\right),\\ 
   &B_1(r)=-\ln \frac{r}{L} -\frac{1}{3}\ln\left(1+\frac{Q_1^2}{r^2}\right),\qquad & &B_2(r)=-\ln \frac{r}{L} -\frac{2}{3}\ln\left(1+\frac{Q_2^2}{r^2}\right),\\
   &h_1(r)=1-\frac{M^2L^2}{r^2(r^2+Q_1^2)} \qquad & &h_2(r)=1-\frac{M^2L^2}{(r^2+Q_2^2)^2}\nonumber\\
&\qquad\,\,=1-\frac{r_{H,1}^2(r_{H,1}^2+Q_1^2)}{r^2(r^2+Q_1^2)},&
  &\qquad\,\,=1-\frac{(r_{H,2}^2+Q_2^2)^2}{(r^2+Q_2^2)^2},\\
   &\phi_1(r)=-\sqrt{\frac{2}{3}}\ln\left(1+\frac{Q_1^2}{r^2}\right),& &\phi_2(r)=\sqrt{\frac{2}{3}}\ln\left(1+\frac{Q_2^2}{r^2}\right),\\
   &\Phi_1(r) = -\frac{MQ_1}{r^2+Q_1^2}+\frac{MQ_1}{r_{H,1}^2+Q_1^2} &&\Phi_2(r) = -\frac{\sqrt{2}MQ_2}{r^2+Q_2^2}+\frac{\sqrt{2}MQ_2}{r_{H,2}^2+Q_2^2}\nonumber\\
&\qquad\,\,=\frac{r_{H,1}Q_1}{L\sqrt{r^2_{H,1}+Q_1^2}} \left(1-\frac{r^2_{H,1}+Q_1^2}{r^2+Q_1^2}\right),&
  &\qquad\,\,=\frac{\sqrt{2}Q_2}{L} \left(1-\frac{r^2_{H,2}+Q_2^2}{r^2+Q_2^2}\right),
\end{align}
\end{subequations}
where the radial position of the black brane horizon is given by the largest real zero of the blackening function $h_k(r)$,
\begin{equation}
    r_{H,1}=\sqrt{\frac{1}{2}\left(\sqrt{Q_1^4+4M^2L^2}-Q_1^2\right)}\qquad \qquad\text{and}\qquad \qquad r_{H,2}=\sqrt{LM-Q_2^2}.
\end{equation}
Consequently, both 1RCBH and 2RCBH models backgrounds are described in terms of two non-negative parameters, namely $(Q,M)$ or, equivalently, $(Q,r_H)$. The Hawking temperature of the black brane and the $U(1)$ R-charge chemical potential take the form,
\begin{subequations}
\begin{align}
&T_1=\frac{\sqrt{-(g_{tt})'(g^{rr})'}}{4\pi}\Bigg|_{r=r_{H,1}}=\frac{Q_1^2+2r_{H,1}^2}{2\pi L^2\sqrt{Q_1^2+r_{H,1}^2}}, &&T_2=\frac{r_{H,2}}{\pi L^2},\\
&\mu_1=\lim_{r\to\infty}\frac{\Phi(r)}{L}=\frac{r_{H,1}Q_1}{L^2\sqrt{Q_1^2+r_{H,1}^2}}, &&\mu_2=\frac{\sqrt{2}Q_2}{L^2}.
\end{align}
\end{subequations}
From now on, for simplicity, we set $L=1$.

\subsection{Thermodynamics}
\label{sec:thermo}

As it is well known, a strongly-coupled SYM plasma with a large-rank gauge group $SU(N_c)$ satisfies the following relation between the number of colors $N_c$ in the quantum gauge theory and the gravitational constant $\kappa_5^2=8\pi G_5$ of its gravity dual \cite{Gubser:1996de,Natsuume:2014sfa},
\begin{equation}
\label{eq:k5Nc}
    \frac{1}{\kappa_5^2}=\frac{N_c^2}{4\pi^2}.
\end{equation}
By using Eq.~\eqref{eq:k5Nc} into the Bekenstein-Hawking's relation between the entropy of a black hole in thermodynamic equilibrium and the area of its event horizon \cite{Bekenstein:1973ur,Hawking:1975vcx},
\begin{align}
S=\frac{A_H}{4G_5}= \frac{N_c^2}{2\pi}\int_{\mathcal{M}_H}d^3x \sqrt{|\gamma_H|} = \frac{N_c^2}{2\pi} g_{xx}^{3/2}(r_H)\int_{\mathcal{M}_H}d^3x = \frac{N_c^2}{2\pi} g_{xx}^{3/2}(r_H) V_H,
\label{eq:sBH}
\end{align}
one obtains the following expressions for the entropy density, $s\equiv S/V_H$, of the 1RCBH and 2RCBH plasmas, respectively,
\begin{equation}
    \frac{s_1}{N_c^2 T^3}=\frac{\pi^2}{16}\left[3\pm \sqrt{1-\left(\frac{\mu/T}{\pi/\sqrt{2}}\right)^2}\right]^2\left[1\mp \sqrt{1-\left(\frac{\mu/T}{\pi/\sqrt{2}}\right)^2}\right], \qquad \frac{s_2}{N_c^2 T^3}=\frac{\pi^2}{2}\left[1+\frac{(\mu/T)^2}{2\pi^2}\right],
\label{eq:s12}
\end{equation}
where the upper/lower signs designates the thermodynamically unstable/stable branches of black hole solutions of the 1RCBH model, while the 2RCBH model has only one branch of black hole solutions. In fact, as discussed in details in \cite{DeWolfe:2011ts,Finazzo:2016psx}, the 1RCBH model can only probe a limited range of values of R-charge chemical potential over temperature, $\mu/T\in[0,\pi/\sqrt{2}]$, where for each value of $\mu/T\in(0,\pi/\sqrt{2})$ there are two branches of black hole solutions,\footnote{For $\mu/T=0$, there are also two solutions: the uncharged AdS$_5$-Schwarzschild black hole dual to the purely thermal SYM plasma, which is in the stable branch of solutions of the 1RCBH model, and a charged solution with no horizon, which is not a black hole and corresponds instead to a supersymmetric BPS solution dubbed ``superstar'' \cite{Myers:2001aq}, which lies in the unstable branch of solutions \cite{DeWolfe:2011ts,Finazzo:2016psx}.} one of them thermodynamically stable (corresponding to solutions with $Q_1/r_{H,1}<\sqrt{2}$), and another one which is unstable (corresponding to solutions with $Q_1/r_{H,1}>\sqrt{2}$), with $\mu/T=\pi/\sqrt{2}$ being a critical point of the phase diagram of the 1RCBH model where both branches merge and where second and higher order derivatives of the pressure diverge. On the other hand, the 2RCBH model probes values of $\mu/T\in[0,\infty)$, having only a single branch of black hole solutions and no critical point in its phase diagram.\footnote{For $\mu/T=0$, the 2RCBH model in thermodynamic equilibrium reduces to the purely thermal SYM plasma.}

According to the holographic dictionary, the charge density may be evaluated as, $\rho=\lim_{r\to \infty} \delta S/\delta \Phi'$, therefore,
\begin{equation}
    \frac{\rho_1}{N_c^2 T^3}=\frac{\mu/T}{16}\left[3\pm \sqrt{1-\left(\frac{\mu/T}{\pi/\sqrt{2}}\right)^2}\right]^2, \qquad \frac{\rho_2}{N_c^2 T^3}=\frac{\mu/T}{4}\left[1+\frac{(\mu/T)^2}{2\pi^2}\right].
\label{eq:rho12}
\end{equation}

Employing the Gibbs-Duhem thermodynamic relation, $\dd p=s\dd T+\rho \dd \mu$, one may integrate Eqs. \eqref{eq:s12} and \eqref{eq:rho12} to obtain the plasma pressure,
\begin{equation}
    \frac{p_1}{N_c^2 T^4}=\frac{\pi^2}{128}\left[3\pm \sqrt{1-\left(\frac{\mu/T}{\pi/\sqrt{2}}\right)^2}\right]^3\left[1\mp \sqrt{1-\left(\frac{\mu/T}{\pi/\sqrt{2}}\right)^2}\right],\qquad \frac{p_2}{N_c^2 T^4}=\frac{\pi^2}{8}\left[1+\frac{(\mu/T)^2}{2\pi^2}\right]^2.
\label{eq:p12}
\end{equation}

\begin{figure}
\centering  
\subfigure[Entropy Density]{\includegraphics[width=0.45\linewidth]{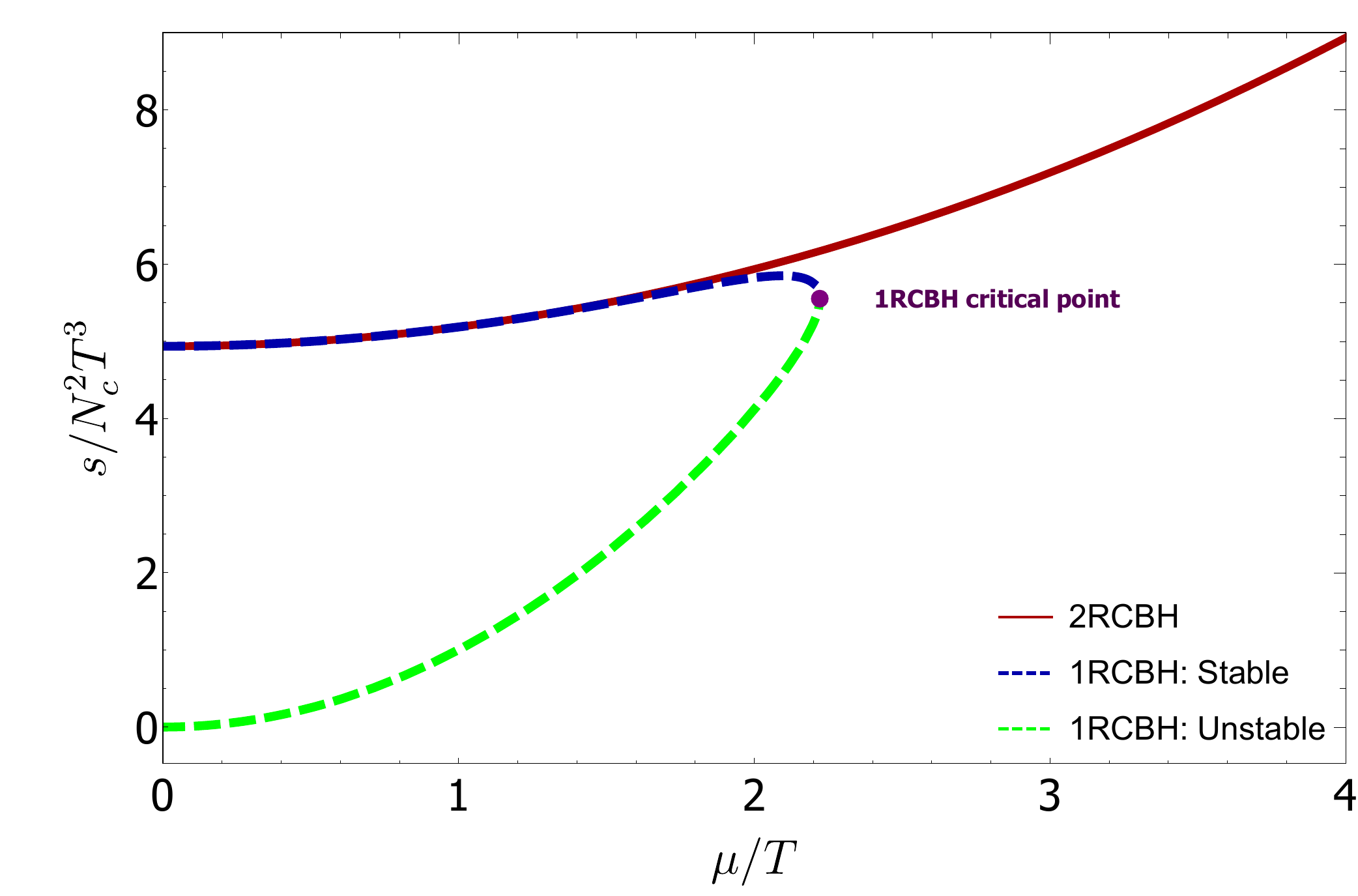}}
\subfigure[Charge Density]{\includegraphics[width=0.45\linewidth]{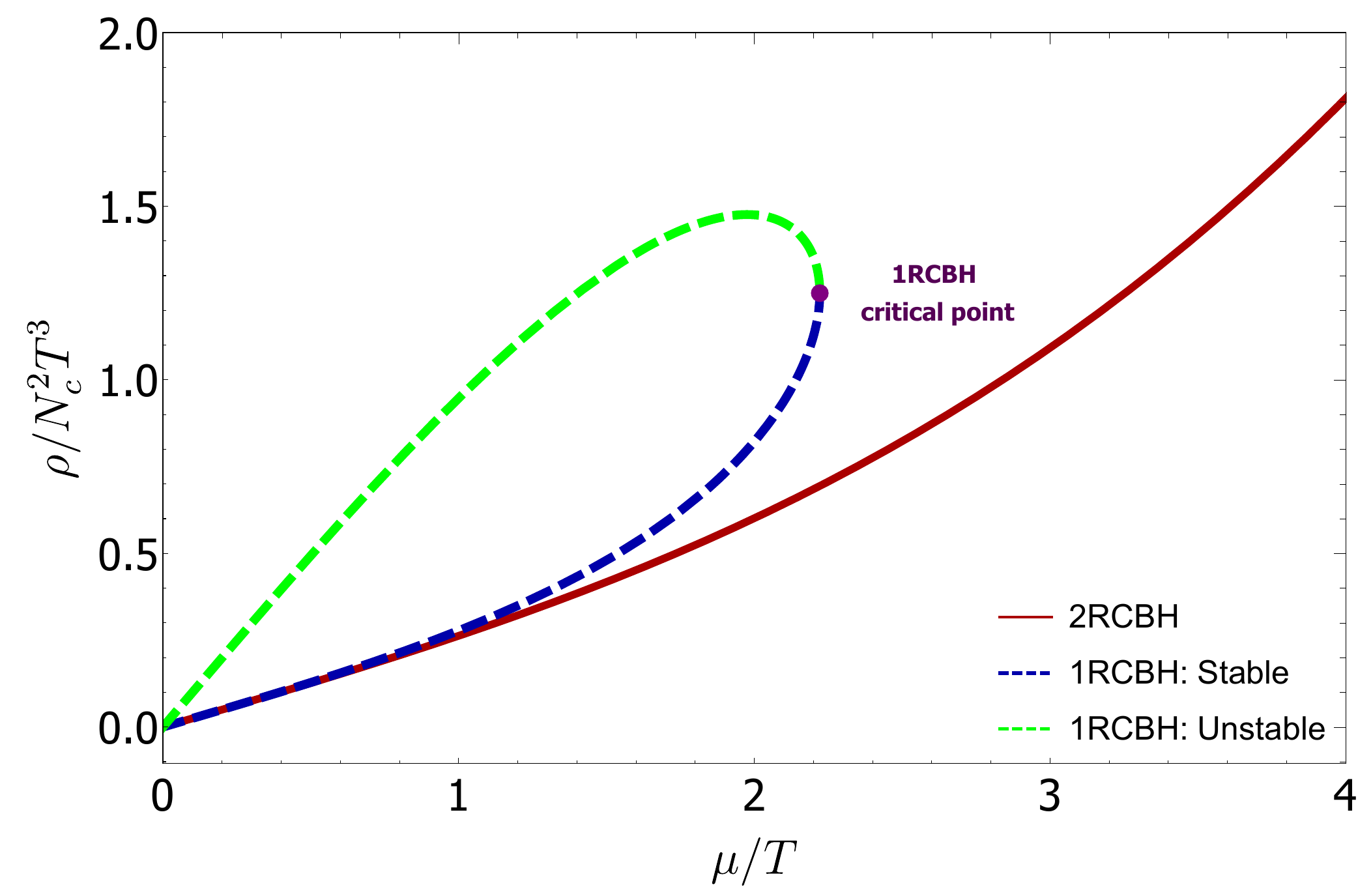}}
\subfigure[Pressure]{\includegraphics[width=0.45\linewidth]{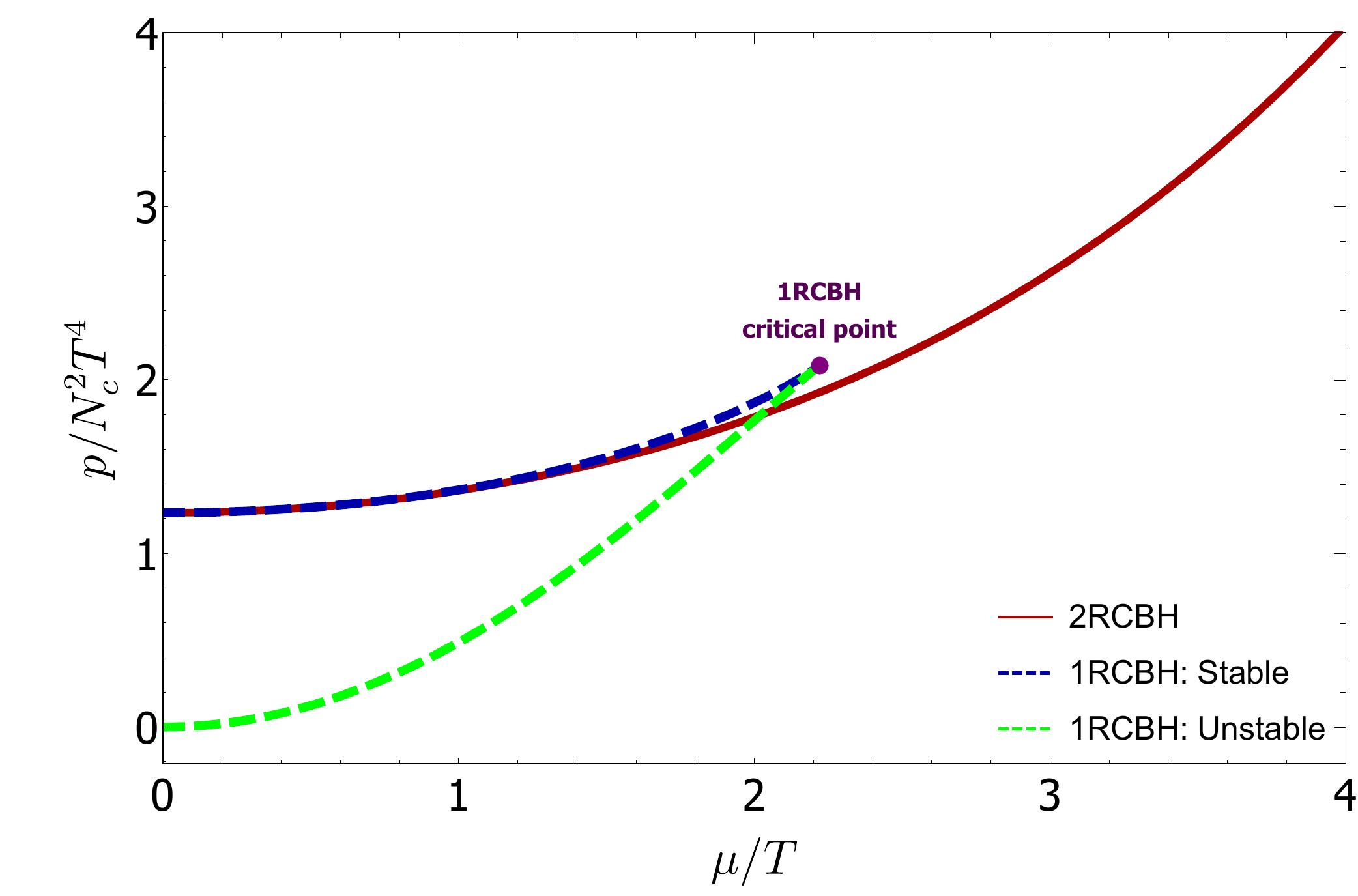}}
\subfigure[Specific Heat]{\includegraphics[width=0.45\linewidth]{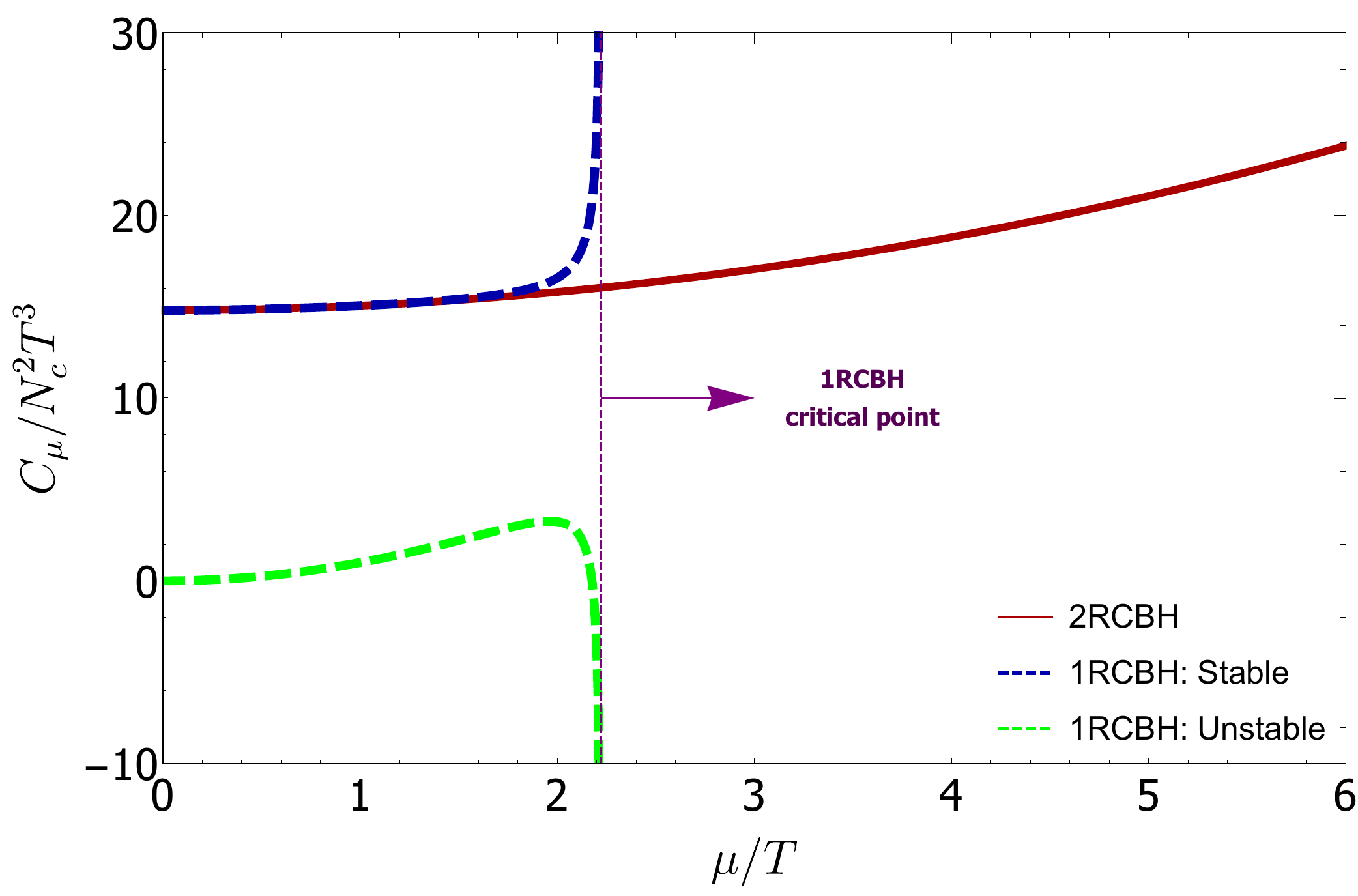}}
\caption{Equation of state and specific heat for the 1RCBH and 2RCBH models.}
\label{fig:thermo1}
\end{figure}

\begin{figure}
\centering  
\subfigure[2nd order susceptibility]{\includegraphics[width=0.45\linewidth]{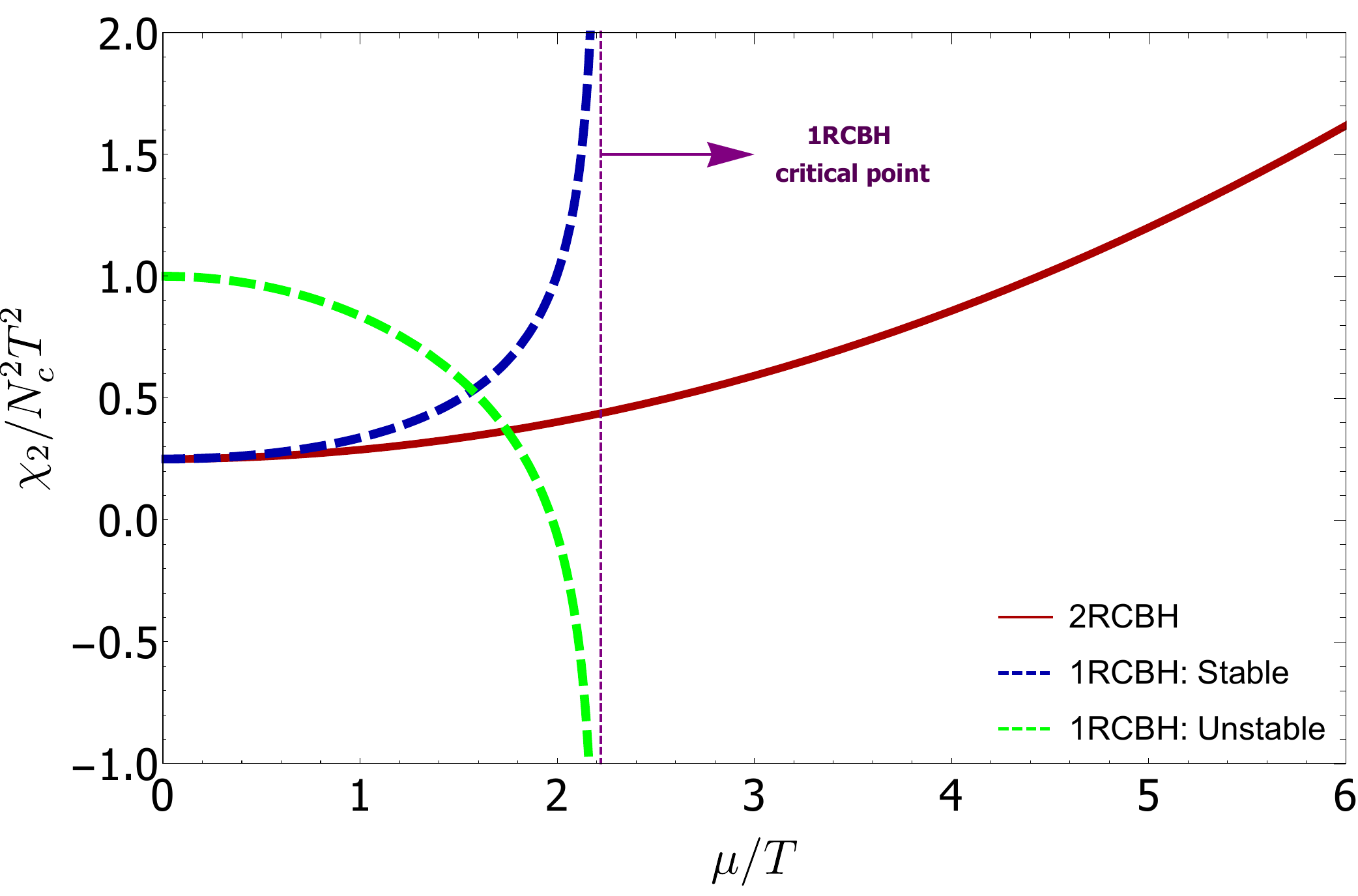}}
\subfigure[3rd order susceptibility]{\includegraphics[width=0.45\linewidth]{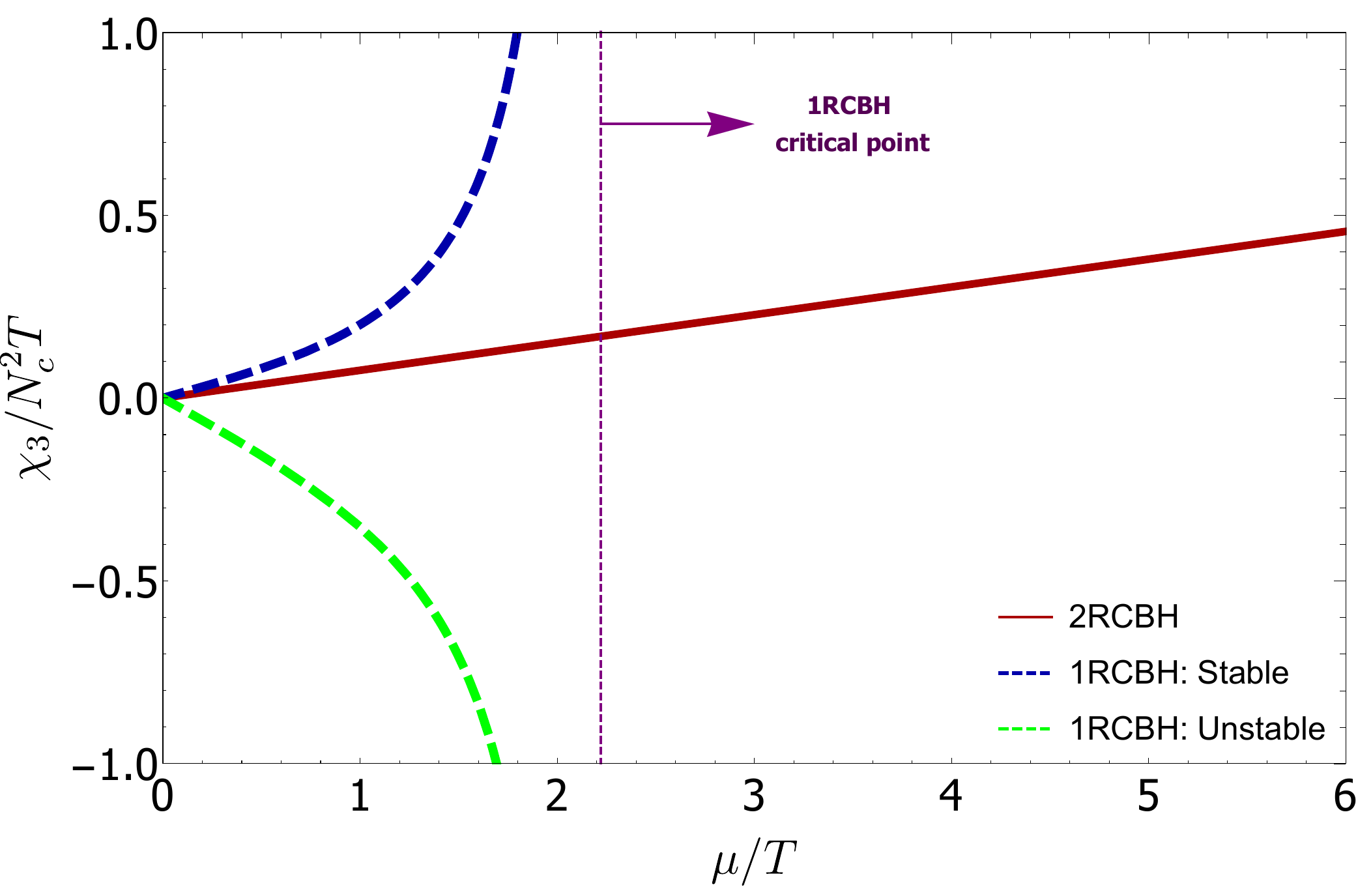}}
\subfigure[4th order susceptibility]{\includegraphics[width=0.45\linewidth]{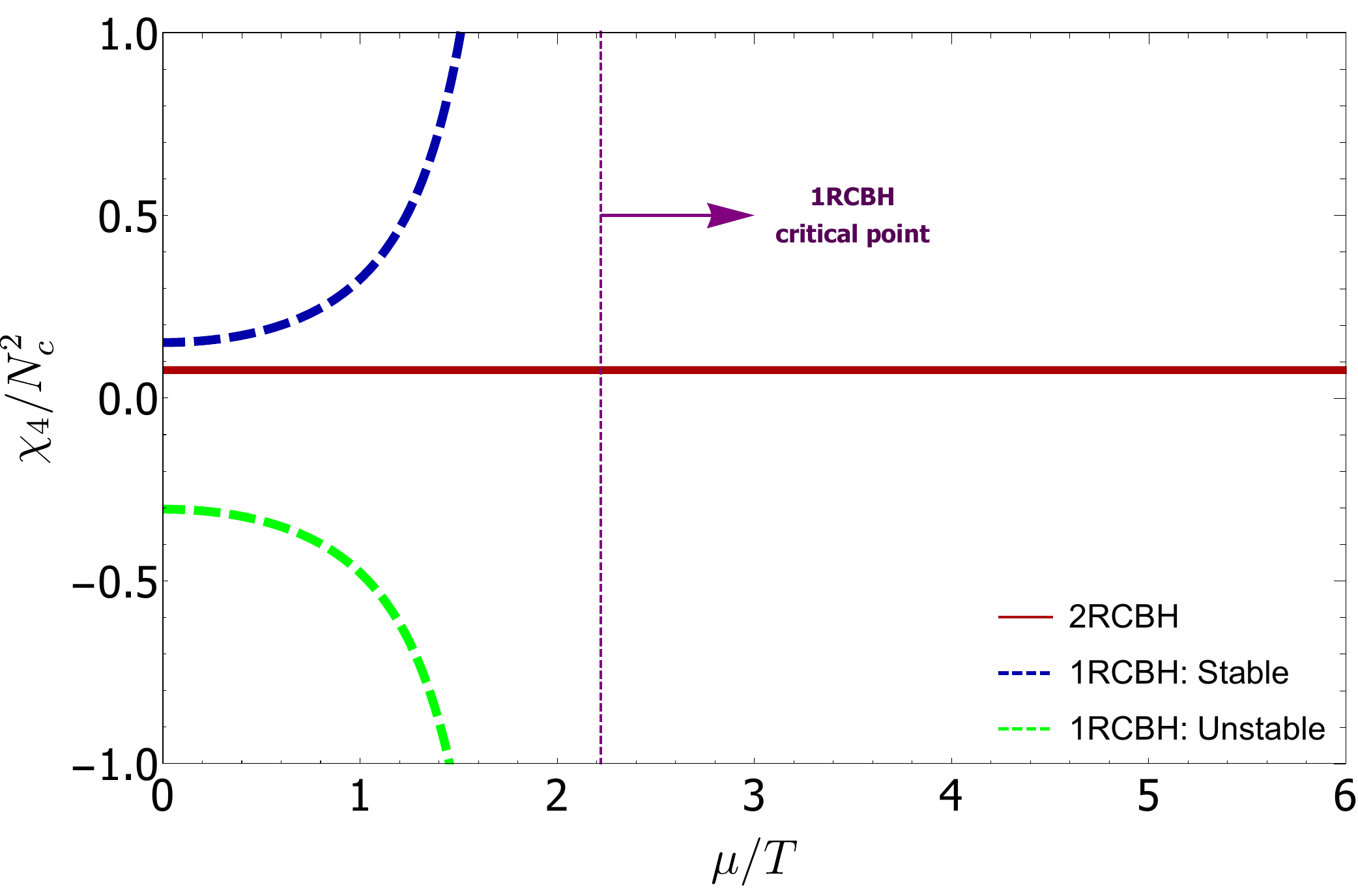}}
\subfigure[5th order susceptibility]{\includegraphics[width=0.45\linewidth]{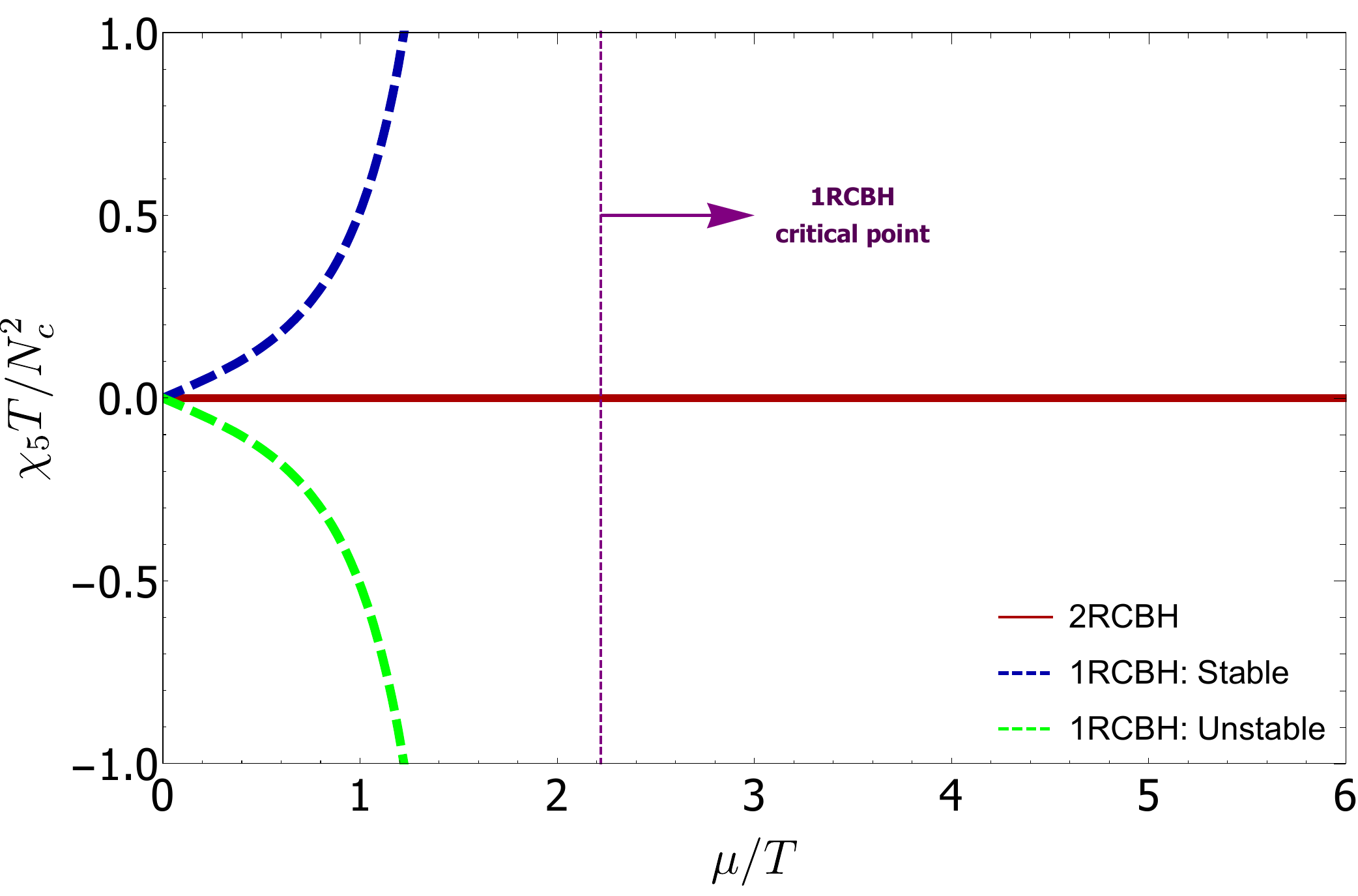}}
\subfigure[6th order susceptibility]{\includegraphics[width=0.45\linewidth]{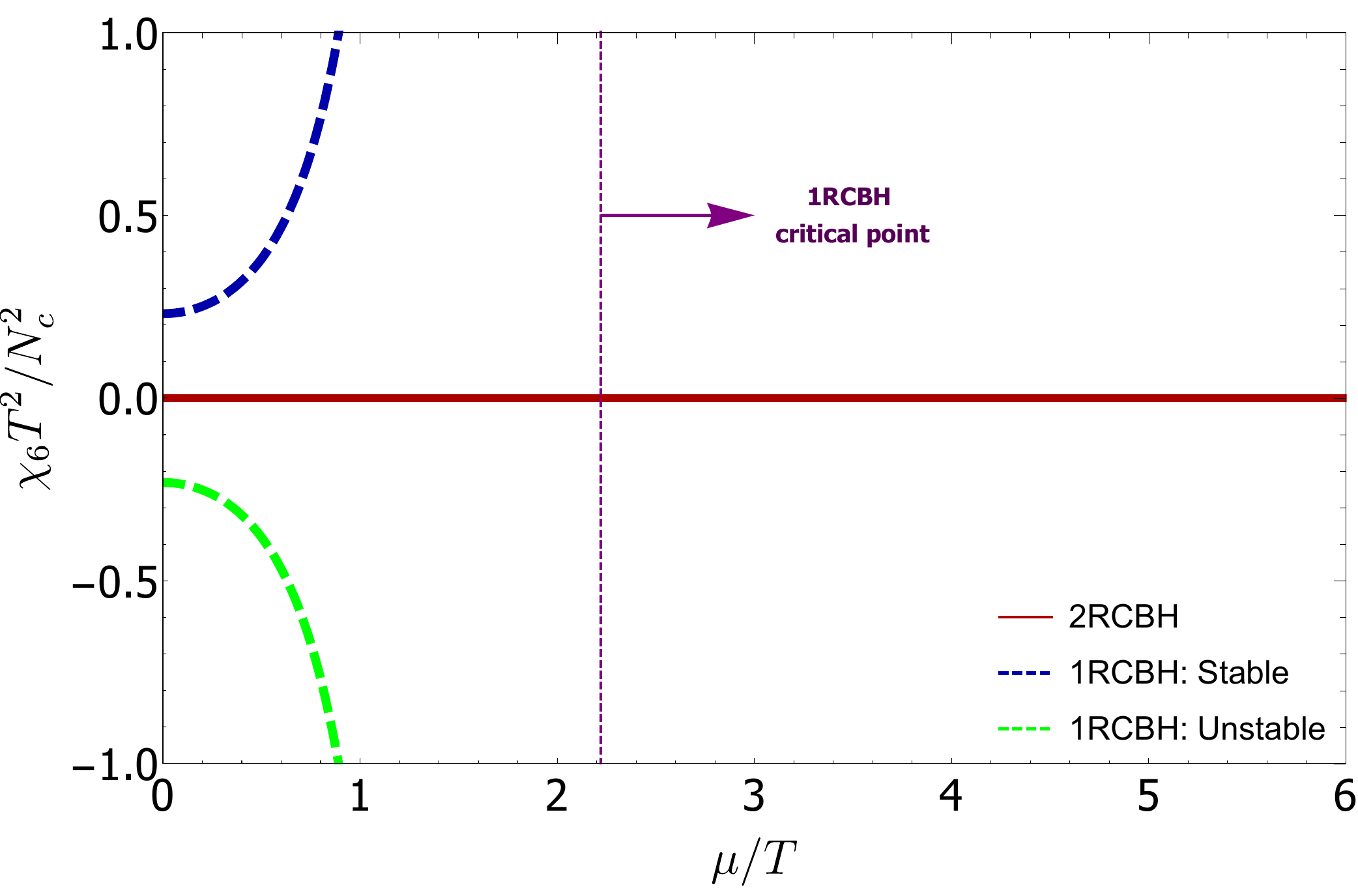}}
\subfigure[7th order susceptibility]{\includegraphics[width=0.45\linewidth]{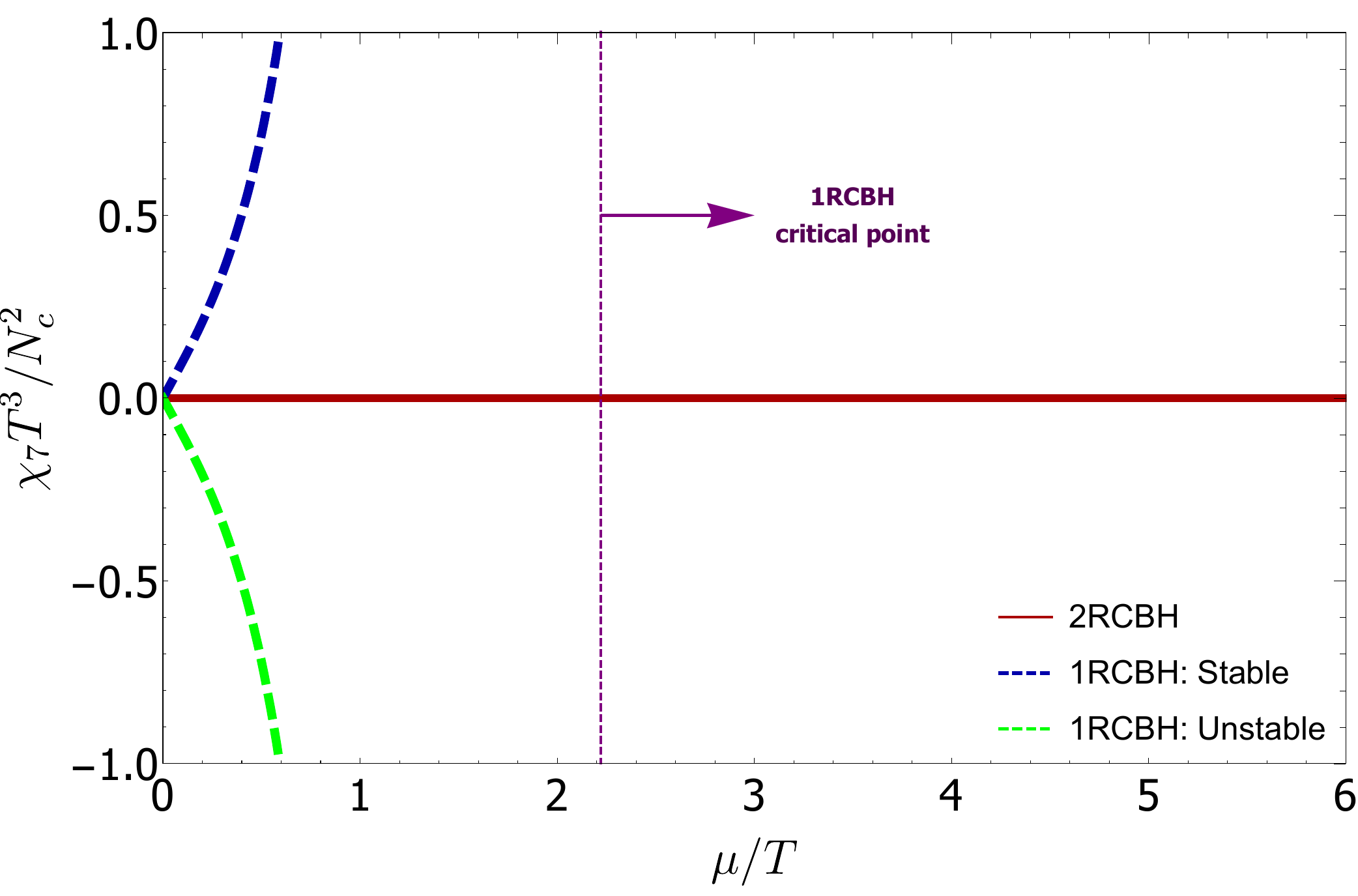}}
\caption{$n$-th order R-charge susceptibility for the 1RCBH and 2RCBH models (from the fifth order onward, the susceptibilities for the 2RCBH model vanish). }
\label{fig:thermo2}
\end{figure}

From Eq.~\eqref{eq:p12}, one may calculate the specific heat at fixed chemical potential \cite{DeWolfe:2010he}, $C_\mu=T(\partial^2p/\partial T^2)_\mu=T(\partial s/\partial T)_\mu$, and the $n$-th order R-charge susceptibility, $\chi_n=(\partial^np/\partial \mu^n)_T=(\partial^{n-1}\rho/\partial \mu^{n-1})_T$. In Figs. \ref{fig:thermo1} and \ref{fig:thermo2} we plot the results for several thermodynamic observables of the 1RCBH and 2RCBH models. For the 1RCBH model there are two branches of solutions and global thermodynamic stability implies that the preferred phase is the one that minimizes the free energy density $f$ or, correspondingly, maximizes the pressure $p=-f$ \cite{DeWolfe:2010he}: see Fig. \ref{fig:thermo1}(c). Moreover, local thermodynamic stability under small thermal fluctuations requires that the Jacobian of the matrix of susceptibilities is positive, which in turn implies that $C_\mu$ and $\chi_2$ must be positive for stable solutions: see Figs. \ref{fig:thermo1} and \ref{fig:thermo2}. Finally, for second order phase transitions (as in the case of a critical point), second and higher order derivatives of the pressure diverge, which is exactly what happens for the 1RCBH model at $\mu/T=\pi/\sqrt{2}$: see Figs. \ref{fig:thermo1} and \ref{fig:thermo2}. In the same figures one notes that the 2RCBH model has only a single branch of solutions and no phase transitions, as aforementioned.

\section{Classification of the Gauge-Invariant Fluctuating Fields}
\label{sec:3}

Insightful knowledge about the near-equilibrium behavior of a strongly-coupled quantum gauge theory may be uncovered by investigating the linear response of its dual gravitational theory under the effect of small fluctuations in the bulk background fields. The relevant information is encoded in the spectrum of damped eigenfrequencies corresponding to the quasinormal modes of the theory, which are related to the relaxation process of perturbed black hole solutions toward equilibrium.

In order to obtain the spectra of homogeneous QNMs of the 2RCBH model (and compare it with the corresponding results for the 1RCBH model \cite{Finazzo:2016psx,Critelli:2017euk}), we consider small fluctuations of the bulk EMD fields around the background black hole solutions in thermodynamic equilibrium \eqref{AnsatzEqs}, taking the following homogeneous plane-wave ansatze preserving the $SO(3)$ rotation symmetry of the background \cite{DeWolfe:2011ts},
\begin{align}
\label{FluctuationEqs}
\delta\phi=\Re{\varphi(r) e^{-i\omega t}},\qquad
\delta A_\mu=\Re{a_\mu(r)e^{-i\omega t}},\qquad
\delta g_{\mu\nu}=\Re{e^{2A(r)}h_{\mu\nu}(r)e^{-i\omega t}}.
\end{align}

The problem of finding the QNMs is much more conveniently handled if we organize the relevant linearized perturbations in terms of diffeormorphism and gauge-invariant variables. In terms of infinitesimal diffeomorphisms, $x^\mu \to x^\mu +\epsilon^\mu$, and the $U(1)$ gauge transformations, $A_\mu\to A_\mu +\partial_\mu \Lambda$, the metric, the vector and the scalar field fluctuations transform as \cite{DeWolfe:2011ts},\footnote{In order to derive the transformations of the scalar field, the vector field and the metric field in terms of the infinitesimal diffeomorphism, $x^\mu\to x^\mu+\epsilon^\mu(x)$, one transforms the fields at first order in $\epsilon^\mu$ such that,
 \begin{align*}
     \phi(x)&\to \phi(x+\epsilon)=\phi(x)+\epsilon^\rho\partial_\rho \phi(x)+\mathcal{O}(\epsilon^2);\\
     A_\mu(x) \dd x^{\mu}&\to  A_\mu(x+\epsilon)\left(\dd x^\mu+\dd \epsilon^\mu\right)=\left[A_\mu(x)+\epsilon^\rho \partial_\rho A_\mu+\mathcal{O}(\epsilon^2)\right]\left(\dd x^\mu+\partial_\rho \epsilon^\mu\dd x^\rho\right)\\
     &=\left[A_\mu+A_\rho\partial_\mu\epsilon^\rho+\epsilon^\rho\partial_\rho A_\mu+\mathcal{O}(\epsilon^2)\right]\dd x^\mu;\\
     g_{\mu\nu}(x)\dd x^\mu\dd x^\nu &\to  g_{\mu\nu}(x+\epsilon)\left(\dd x^\mu+\dd \epsilon^\mu \right)\left(\dd x^\nu+\dd \epsilon^\nu\right)\\
     &=\left[g_{\mu\nu}(x)+\epsilon^\rho \partial_\rho g_{\mu\nu}(x)+\mathcal{O}(\epsilon^2)\right]\left(\dd x^\mu+\partial_\rho \epsilon^\mu \dd x^\rho \right)\left(\dd x^\nu+\partial_\sigma \epsilon^\nu \dd x^\sigma\right)\\
     &=\left[g_{\mu\nu}+\epsilon^\rho \partial_\rho g_{\mu\nu}+g_{\rho\nu}\partial_\mu \epsilon^\rho+g_{\mu\sigma}\partial_\nu \epsilon^\sigma+\mathcal{O}(\epsilon^2)\right]\dd x^\mu\dd x^\nu.
 \end{align*}}
\begin{subequations}
\label{GaugeTransf}
    \begin{align}
    \phi &\to   \phi + \epsilon^\mu \partial_\mu  \phi,\\
     A_\mu &\to A_\mu + \partial_\mu \Lambda +\epsilon^\nu \partial_\nu A_\mu+\left(\partial_\mu \epsilon^\nu\right)A_\nu,\\
    g_{\mu\nu}&\to  g_{\mu\nu}+\epsilon^\rho \partial_\rho g_{\mu\nu}+\left(\partial_\mu \epsilon^\rho\right)g_{\rho\nu}+\left(\partial_\nu \epsilon^\rho\right)g_{\mu\rho}, 
\end{align}
\end{subequations}
where the transformation variables also take the plane-wave form, $\epsilon^\mu =e^{-i\omega t}\xi^\mu(r)$ and $\Lambda =e^{-i\omega t}\lambda(r)$.

Ref.~\cite{DeWolfe:2011ts} identified diffeormorphism and gauge-invariant linear combinations of the EMD field fluctuations which are classified into three different irreducible representations of the $SO(3)$ rotation group: the quintuplet (spin 2), the triplet (spin 1), and the singlet (spin 0) channels. The $SO(3)$ quintuplet channel is represented by any of the five independent traceless spatial components of the metric perturbation (the graviton), $\chi\equiv h_{ij}$. In the case of the $SO(3)$ triplet channel, the invariant pertubation is the variable $a\equiv a_i$, representing any of the spatial components of the vector field fluctuation. Finally, the invariant perturbation of the $SO(3)$ singlet channel is given by the following linear combination of the metric and the dilaton fluctuations: $\mathcal{S}\equiv \varphi -\left(\phi'/2A'\right)\left[\left(h_{xx}+h_{yy}+h_{zz}\right)/3\right]$.

In order to compute the QNM spectra for the above perturbations, we will proceed in the next sections by deriving linearized equations of motion for each class of field fluctuations, taking into account the plane-wave profiles in Eq.~\eqref{FluctuationEqs}. The ultraviolet, near-boundary expansions of the perturbed bulk fields typically involve a leading order non-normalizable mode and a subleading normalizable mode for each field perturbation. According to the holographic dictionary \cite{Maldacena:1997re,Gubser:1998bc,Witten:1998qj,Witten:1998zw,Aharony:1999ti}, the leading modes serve as sources for local gauge-invariant operators of the dual quantum field theory at the boundary, such as the energy-momentum tensor, conserved vector currents, and scalar field operators. On the other hand, the subleading modes correspond to the expectation values of those QFT operators.

By setting the subleading modes to zero at the boundary and by imposing the causal infalling wave condition at the black brane horizon, the corresponding solutions of the linearized field equations for the bulk field fluctuations can be employed to evaluate the (renormalized) on-shell boundary action and derive retarded thermal field correlators of the dual strongly-coupled QFT, which are related via Kubo's formulas to transport coefficients of the strongly-coupled medium described by the dual QFT \cite{Son:2002sd}.

On the other hand, and akin to our purposes in the present work, since the retarded thermal field correlators of the dual strongly-coupled QFT are given by minus the ratio between the subleading and the leading modes of the bulk field fluctuations \cite{Son:2002sd}, by setting instead the leading modes to zero at the boundary and by imposing the causal infalling wave condition at the black brane horizon, we get the poles of those retarded field correlators. As the frequency eigenvalue problem for quasinormal modes in asymptotically AdS backgrounds is defined by the Dirichlet boundary condition that sets to zero at the boundary the leading modes of the bulk fields fluctuations \cite{Kovtun:2005ev}, one concludes that the QNMs describing the exponentially modulated decay of linear perturbations of asymptotically AdS black holes correspond to the poles of the retarded thermal field correlators of the strongly-coupled dual QFT. Moreover, since the spectra of QNMs of the theory is obtained by solving the linearized equations of motion for the bulk perturbations subjected to the Dirichlet boundary condition that such perturbations vanish at the boundary of the bulk spacetime, the corresponding field equations for the bulk fluctuations will generally only admit solutions for a discrete set of complex eigenfrequencies $\{\omega\}$, which define the QNMs of the theory as solutions of eigenfrequency problems with Dirichlet boundary conditions.

\section{Quintuplet Channel}
\label{sec:4}

\subsection{Field Equation for the Perturbation}

Since, to first order in the perturbation parameter, the metric fluctuation $\delta g_{\mu\nu}$ does not mix with either the Maxwell or the dilaton fluctuation fields, the equation of motion for the perturbation $\chi$ of the $SO(3)$ quintuplet channel can be obtained by substituting any of the spatial components of the metric fluctuation $h_{ij}$ from Eq.~\eqref{FluctuationEqs} into the Einstein's field equations \eqref{eq:Einstein}. This results in the following equation of motion~\cite{DeWolfe:2011ts},
\begin{equation}
\chi''+\left(4A'-B'+\frac{h'}{h}\right)\chi'+\frac{e^{-2(A-B)}}{h^2}\omega^2 \chi=0.
\label{eq:chidiag}
\end{equation}
Defining the Eddington-Finkelstein (EF) time coordinate,
\begin{equation}
\label{eq:EEcoordinate}
    \dd v = \dd t +\sqrt{-\frac{g_{rr}}{g_{tt}}}\dd r = \dd t +\frac{e^{B-A}}{h}\dd r,
\end{equation}
one can rewrite the metric \eqref{AnsatzEqs} in EF coordinates as follows, 
\begin{equation}
    \dd s^2 = e^{2A}\left[-h\dd v^2+\dd \mathbf{x}^2\right]+2 e^{A+B}\dd v \dd r.
\label{eq:EFmetric}
\end{equation}
The purpose of going from the diagonal domain-wall coordinates~\eqref{AnsatzEqs} to the non-diagonal EF coordinates~\eqref{eq:EFmetric} is that, in terms of the latter, the imposition of the aforementioned causal infalling wave condition at the black brane horizon simply translates to the requirement of regularity of the bulk field fluctuations at the horizon.

\begin{figure}
\centering  
\subfigure[1st mode]{\includegraphics[width=0.45\linewidth]{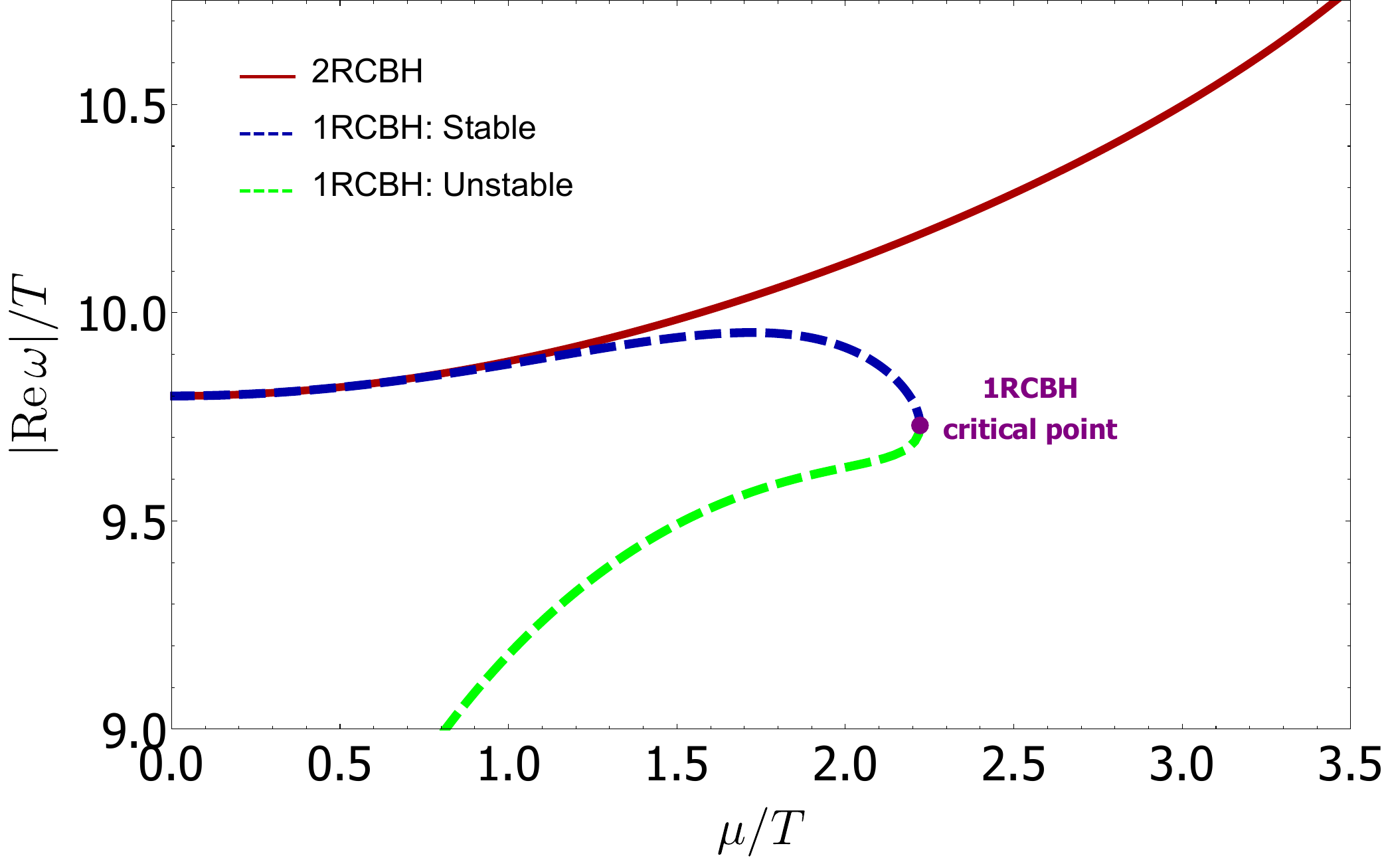}}
\subfigure[2nd mode]{\includegraphics[width=0.45\linewidth]{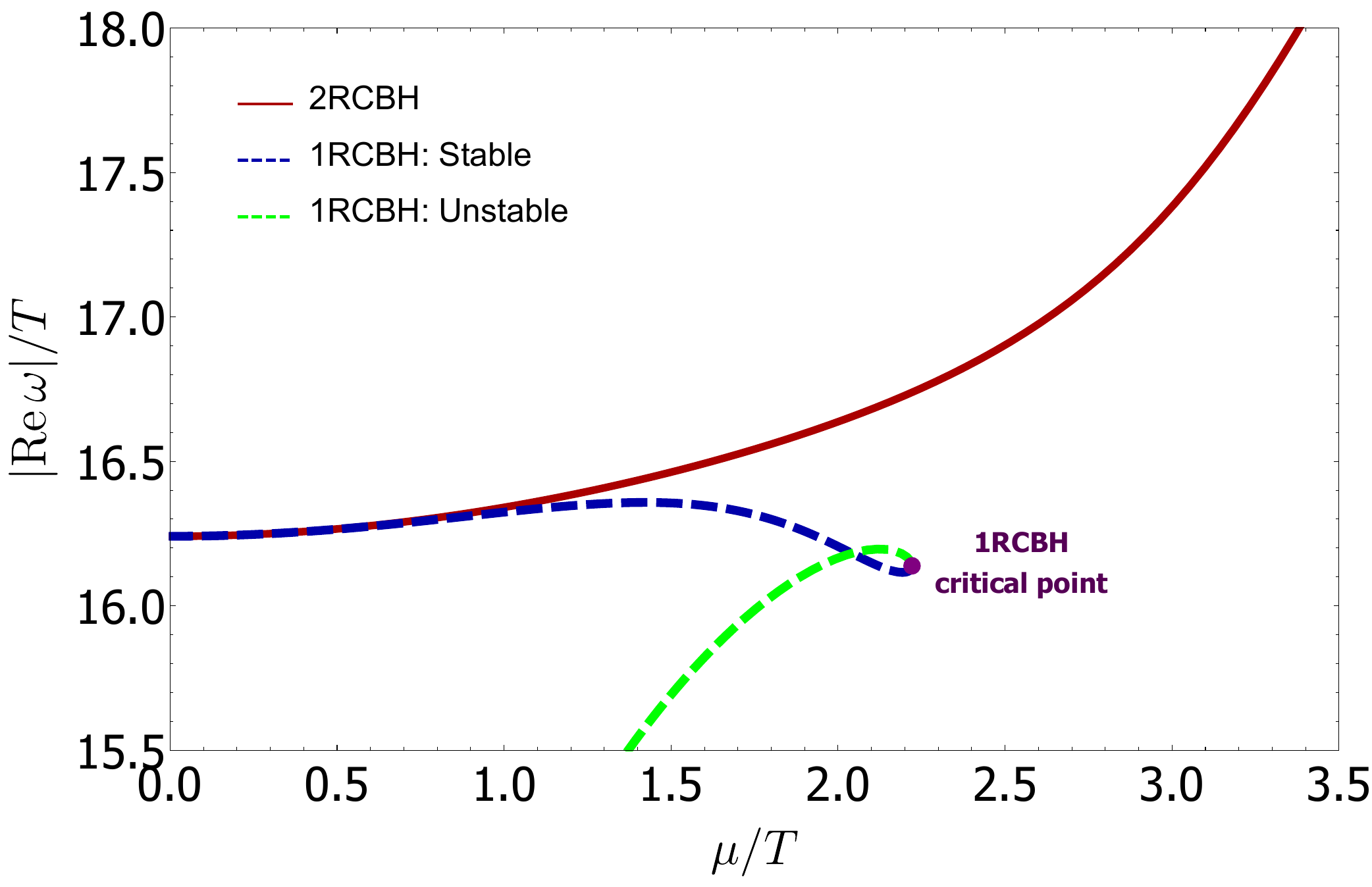}}
\subfigure[3rd mode]{\includegraphics[width=0.45\linewidth]{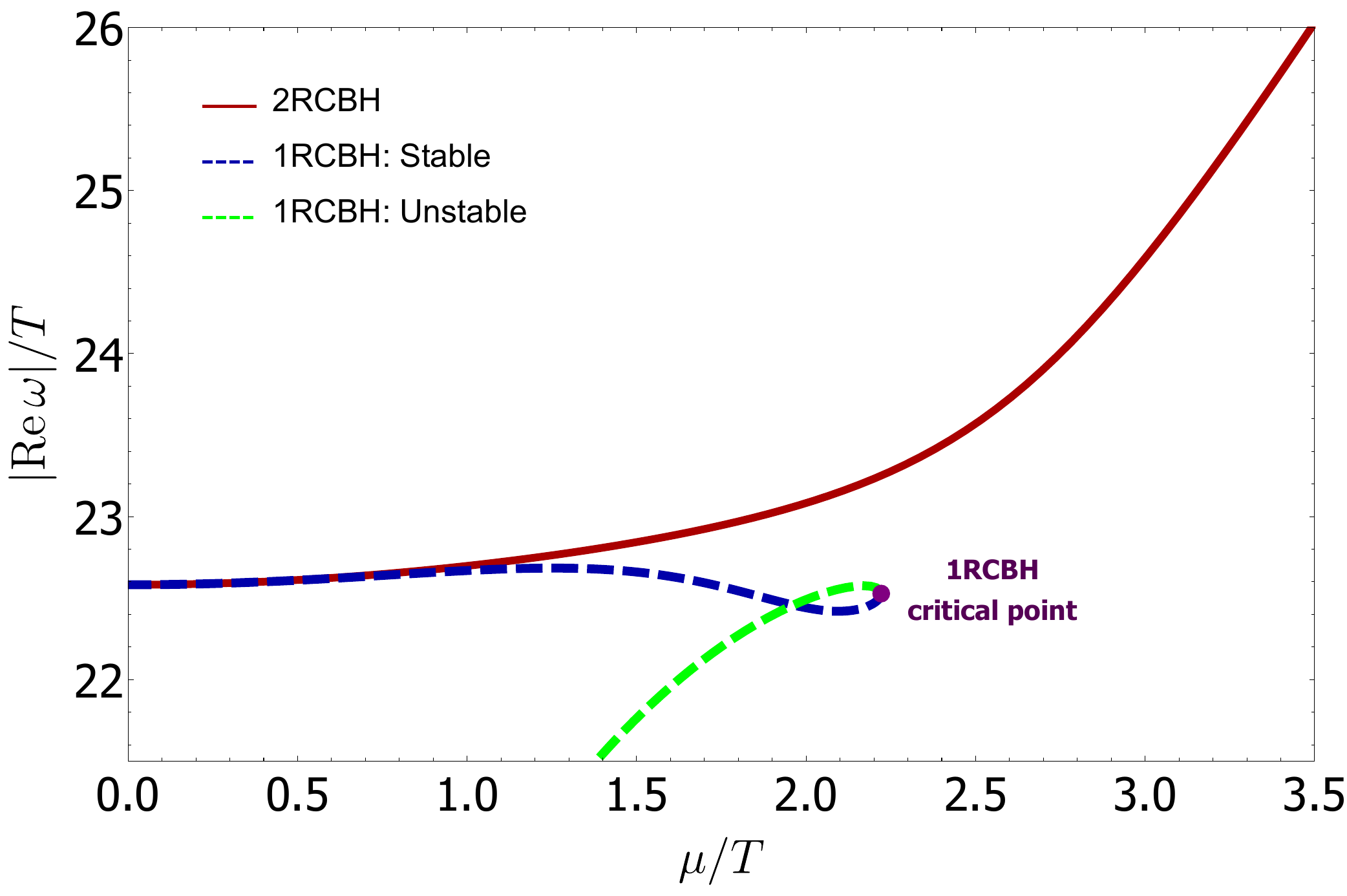}}
\subfigure[4th mode]{\includegraphics[width=0.45\linewidth]{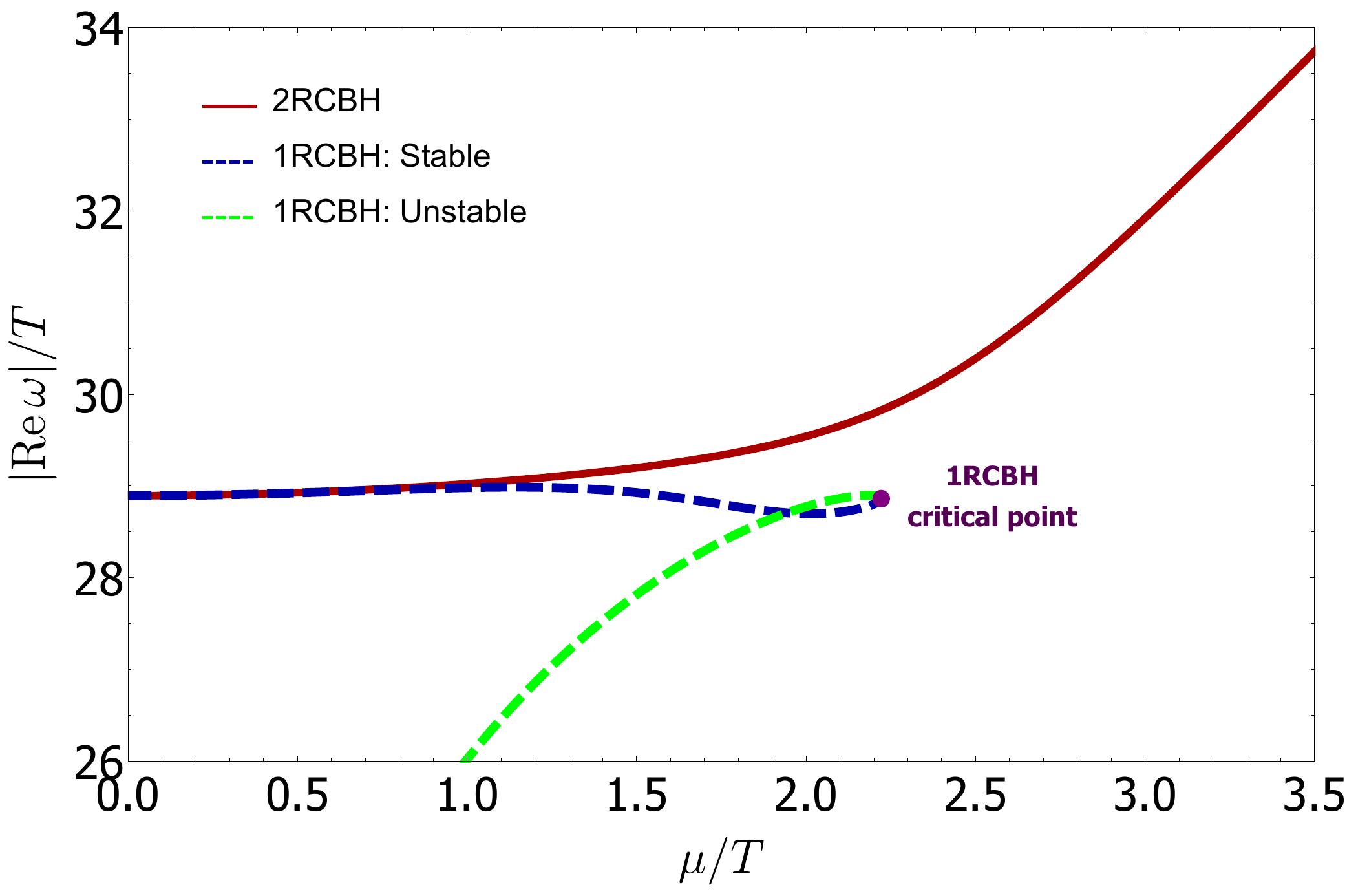}}
\caption{Real part for the first four quasinormal modes for the 1RCBH and 2RCBH models in the quintuplet channel.}
\label{fig:QNMQuiRe}
\end{figure}
\begin{figure}
\centering
\subfigure[1st mode]{\includegraphics[width=0.45\linewidth]{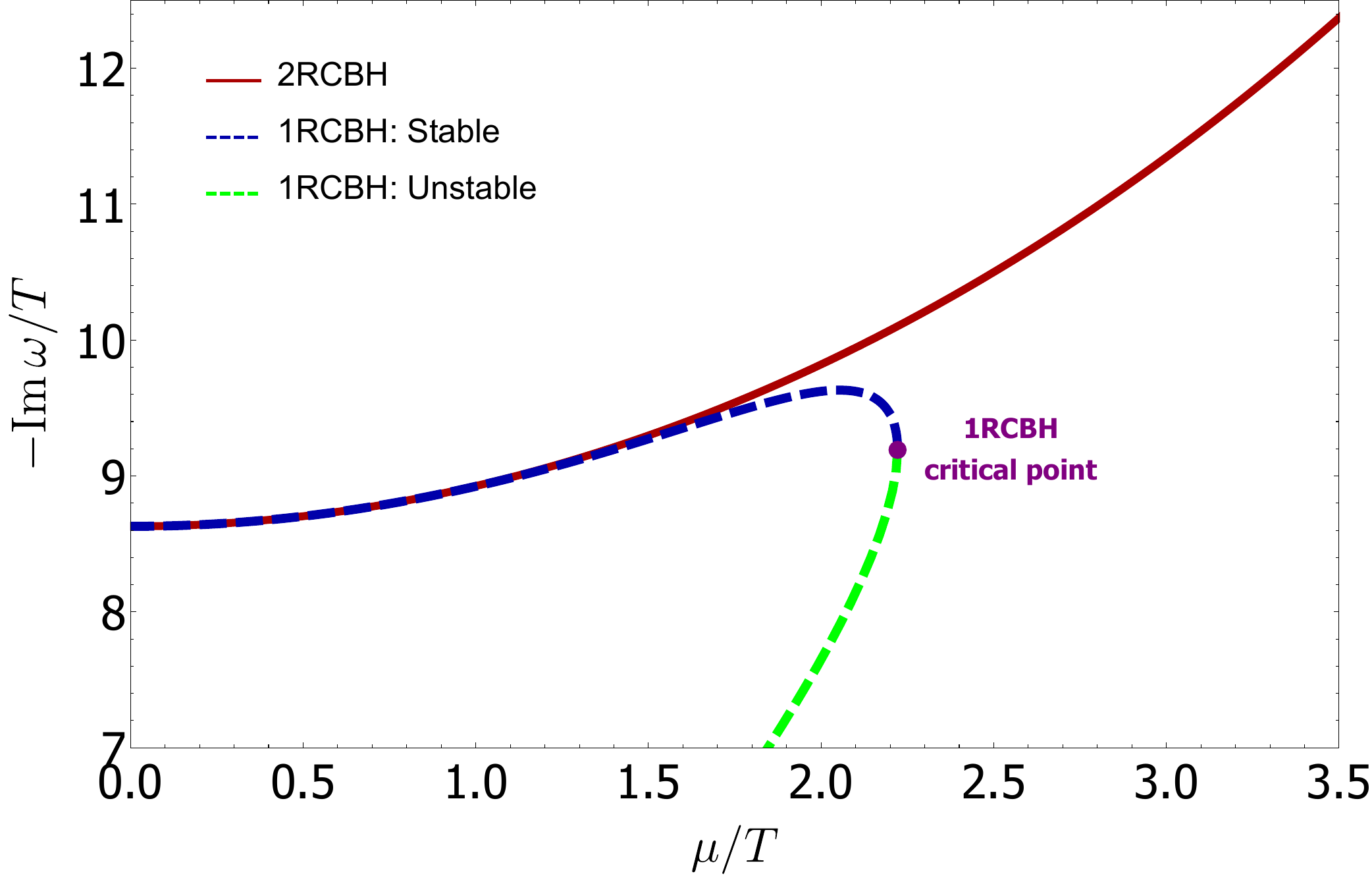}}
\subfigure[2nd mode]{\includegraphics[width=0.45\linewidth]{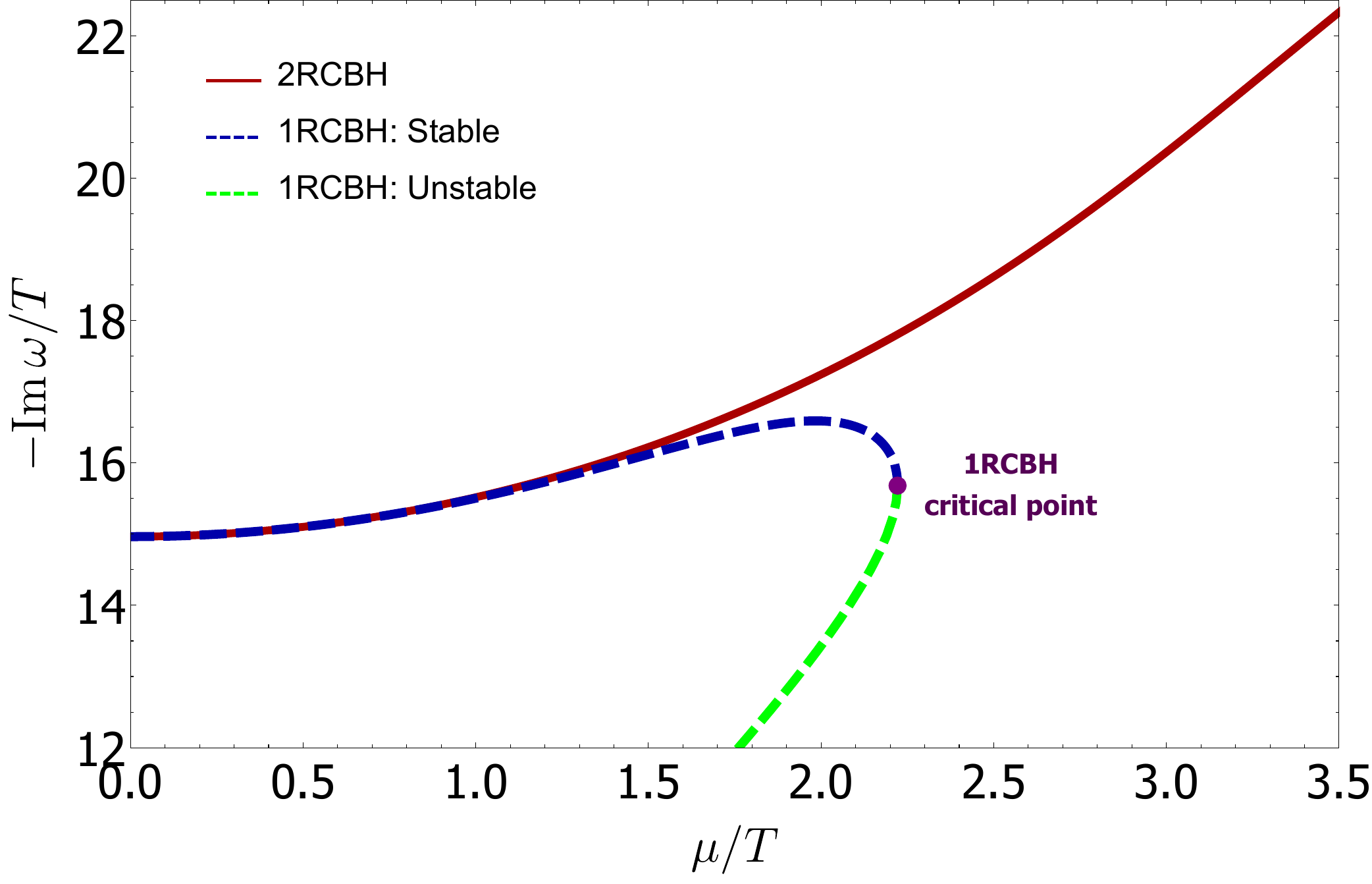}}
\subfigure[3rd mode]{\includegraphics[width=0.45\linewidth]{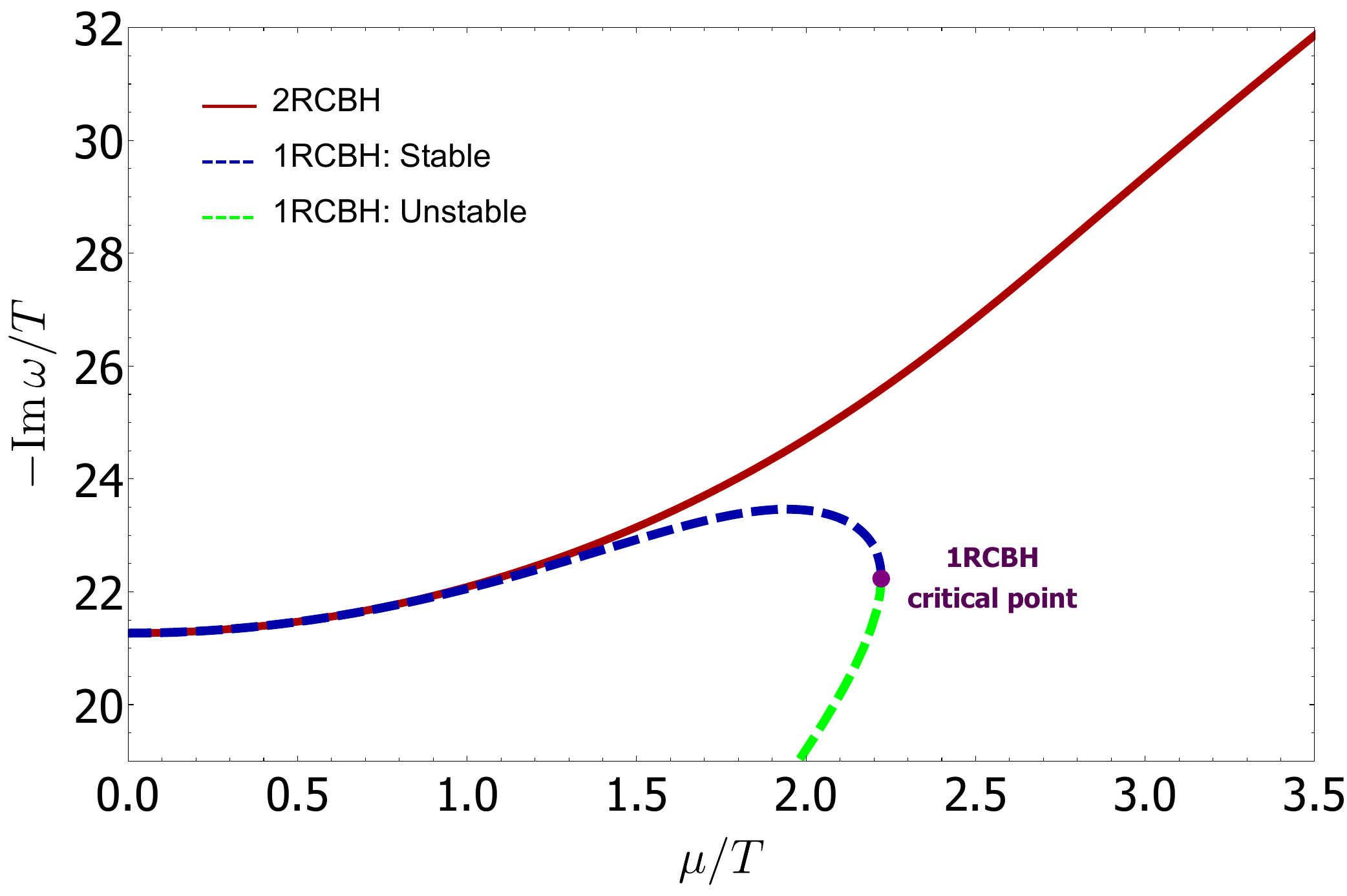}}
\subfigure[4th mode]{\includegraphics[width=0.45\linewidth]{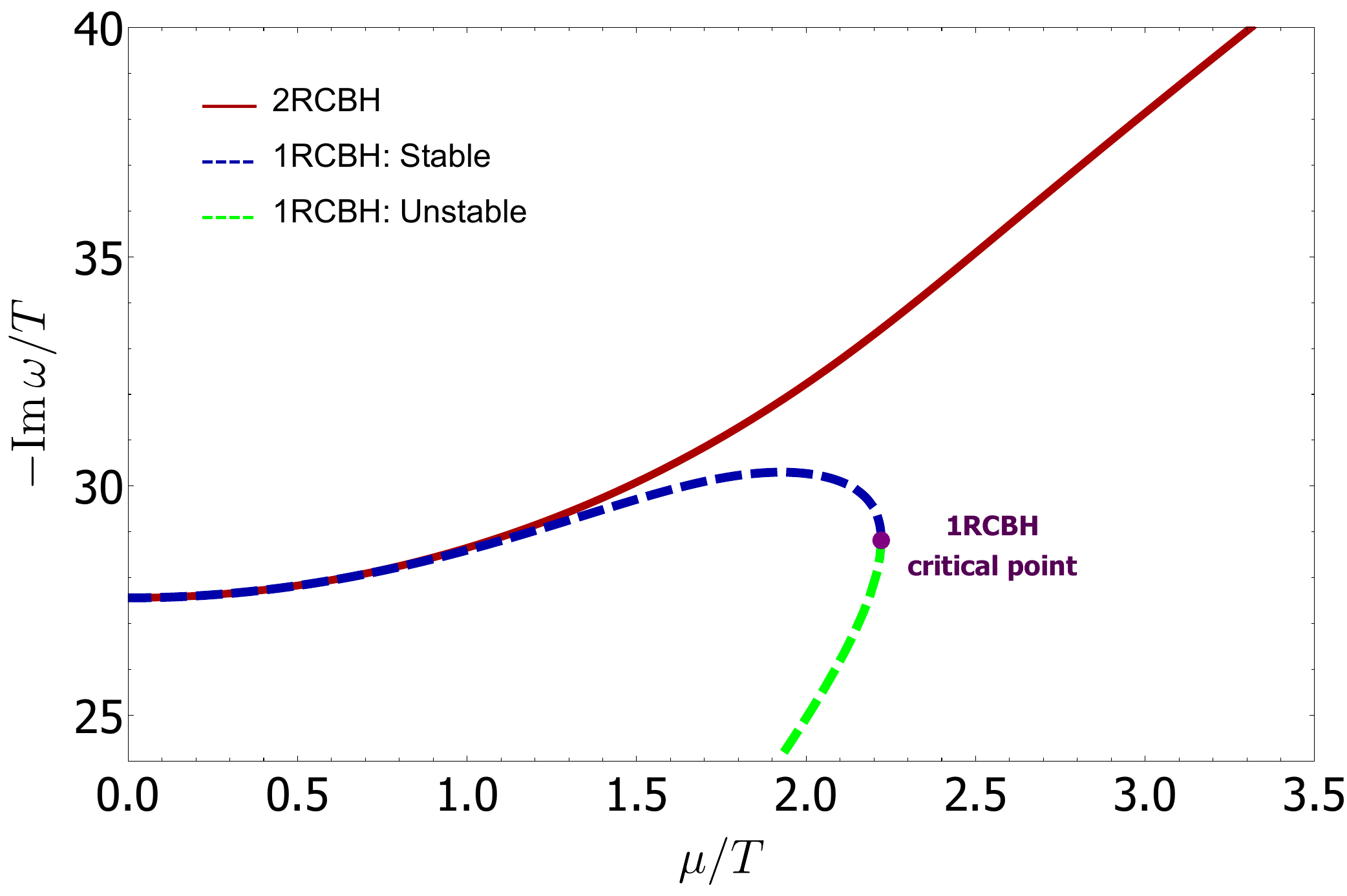}}
\caption{Imaginary part for the first four quasinormal modes for the 1RCBH and 2RCBH models in the quintuplet channel.}
\label{fig:QNMQuiIm}
\end{figure}
In terms of the EF coordinates~\eqref{eq:EFmetric}, the dynamics for the $\chi$ perturbation~\eqref{eq:chidiag} is rewritten as,\footnote{One considers the chain rules, $\partial_t=\partial_v$, and, $\dd/\dd r=(\partial v/\partial r)\partial_v+\partial_r=h^{-1}e^{B-A}\partial_v+\partial_r\,\dot{=}-i\omega h^{-1}e^{B-A}+\partial_r$, which also imply that, $\dd^2/\dd r^2\,\dot{=}\, \partial_r^2-2i\omega h^{-1}e^{B-A}\partial_r-\omega^2 h^{-2} e^{2(B-A)}+i\omega h^{-1}e^{B-A}(h'/h+A'-B')$.}
\begin{equation}
    \chi''+\left(4A'-B'+\frac{h'}{h}-\frac{2i\omega}{h}e^{B-A}\right)\chi'-\frac{3i\omega A'e^{2(B-A)}}{h}\chi=0.
\end{equation}

In order to numerically solve the equation of motion for the field fluctuation $\chi$ using the pseudospectral method \cite{boyd01}, as discussed in Appendix B of \cite{Finazzo:2016psx}, we transform the radial coordinate $r$, which is defined on the infinite interval $r\in[r_H,\infty)$, to a new dimensionless radial coordinate $u=r_H/r$, defined on the compact interval $u\in[0,1]$, which is very suited for numerical computations. In terms of these new coordinates, the equation of motion for the perturbation field $\chi$ in the case of the 1RCBH model becomes \cite{Finazzo:2016psx},
\begin{align}
    \nonumber\chi_1 ''&+ \left[\frac{Q^2 u^2 \left(u^2+1\right)+r_H^2 \left(u^4+3\right)}{u \left(u^2-1\right) \left(u^2 \left(Q^2+r_H^2\right)+r_H^2\right)}-\frac{2 i \omega_1 \sqrt{Q^2 u^2+r_H^2}}{\left(u^2-1\right) \left(u^2 \left(Q^2+r_H^2\right)+r_H^2\right)}\right]\chi_1 '\\
    \label{eq:EOM1RCBH}&+ \left[\frac{i \omega_1 \left(2 Q^2 u^2+3 r_H^2\right)}{u \left(u^2-1\right) \sqrt{Q^2 u^2+r_H^2} \left(u^2 \left(Q^2+r_H^2\right)+r_H^2\right)}\right]\chi_1 =0,
\end{align}
whereas for the 2RCBH model, one has,
\begin{align}
    \nonumber\chi_2 '' &+\left[\frac{2 Q^2 u^2 \left(u^2+1\right)+r_H^2 \left(u^4+3\right)}{u \left(u^2-1\right) \left(u^2 \left(2 Q^2+r_H^2\right)+r_H^2\right)}-\frac{2 i \omega_2 \left(Q^2 u^2+r_H^2\right)}{r_H \left(u^2-1\right) \left(u^2 \left(2 Q^2+r_H^2\right)+r_H^2\right)}\right]\chi_2 '\\
    &+ \label{eq:EOM2RCBH}\left[\frac{i \omega_2 \left(Q^2 u^2+3 r_H^2\right)}{r_H u \left(u^2-1\right) \left(u^2 \left(2 Q^2+r_H^2\right)+r_H^2\right)}\right]\chi_2 =0.
\end{align}

As aforementioned, the spectra of QNMs for both the 1RCBH and 2RCBH models can be obtained by treating Eqs.~\eqref{eq:EOM1RCBH} and \eqref{eq:EOM2RCBH} as eigenvalue problems subjected to the Dirichlet boundary condition that sets the respective fluctuation field, $\chi_k,\,k=1,2$, to zero at the boundary. The ultraviolet, near-boundary expansion for the fluctuation in the $SO(3)$ quintuplet channel is of the form, $\chi_k(u,\omega)=G_k(u,\omega)+u^4 F_k(u,\omega)$, with $G_k$ and $F_k$ representing non-normalizable and normalizable modes, respectively. Regarding the QFT operator $\hat{O}_k$ dual to the bulk fluctuating field $\chi_k$, the leading non-normalizable mode at the boundary, $G_k(u\to 0,\omega)=J_k(\omega)$, is interpreted as its external source in the generating functional of correlation functions, while the subleading normalizable mode at the boundary, $F_k(u\to 0,\omega)=\langle \hat{O}_k(\omega)\rangle$, gives its expectation value. In fact, from the real-time prescription proposed in~\cite{Son:2002sd}, near the boundary ($u\to 0$), the thermal retarded propagator of the QFT operator $\hat{O}_k$ is specified by the ratio between the normalizable and the non-renormalizable modes of its dual bulk field fluctuation, $\mathcal{G}_{\hat{O}\hat{O}}^R(\omega)=-\langle \hat{O}_k(\omega)\rangle/J_k(\omega)=-F_k(0,\omega)/G_k(0,\omega)$. By setting the Dirichlet boundary condition $G_k(0,\omega)=J_k(\omega)=0$ with $F_k(0,\omega)=\langle\hat{O}_k(\omega)\rangle\neq 0$, one implements the adequate boundary condition specifying the QNM eigenfrequency problem in asymptotically AdS spacetimes, with the set of complex eigenfrequencies $\{\omega\}$ corresponding to the poles of the thermal retarded propagator $\mathcal{G}_{\hat{O}\hat{O}}^R(\omega)$.

Therefore, we set $\chi_k(u)=u^4 F_k(u)$, with $F_k(0)\neq 0$, such that for the 1RCBH model one obtains,
\begin{align}
\nonumber F_1''&+ \left[\frac{5}{u}+2 u \left(\frac{Q^2+r_H^2}{u^2 \left(Q^2+r_H^2\right)+r_H^2}+\frac{1}{u^2-1}\right)-\frac{2 i \omega_1 \sqrt{Q^2 u^2+r_H^2}}{\left(u^2-1\right) \left(u^2 \left(Q^2+r_H^2\right)+r_H^2\right)}\right]F_1'\\
\label{eq:NormalizableMode1}&+\left[\frac{16 u^2 \left(Q^2+r_H^2\right)-8 Q^2}{ u^2 \left(Q^2+r_H^2\right)+r_H^2}-\frac{i \omega_1 \left(6 Q^2 u^2+5 r_H^2\right)}{u  \sqrt{Q^2 u^2+r_H^2} \left(u^2 \left(Q^2+r_H^2\right)+r_H^2\right)}\right]\frac{F_1}{u^2-1} =0,
\end{align}

whereas for the 2RCBH model, one gets,
\begin{align}
    \nonumber F_2''&+ \left[\frac{5}{u}+2 u \left(\frac{2 Q^2+r_H^2}{u^2 \left(2 Q^2+r_H^2\right)+r_H^2}+\frac{1}{u^2-1}\right)-\frac{2 i \omega_2 \left(Q^2 u^2+r_H^2\right)}{r_H \left(u^2-1\right) \left(u^2 \left(2 Q^2+r_H^2\right)+r_H^2\right)}\right]F_2'\\
    \label{eq:NormalizableMode2}&+ \left[8 \left(\frac{2 Q^2+r_H^2}{u^2 \left(2 Q^2+r_H^2\right)+r_H^2}+\frac{1}{u^2-1}\right)+\frac{i\omega_2}{u r_H}\frac{\left(7 Q^2 u^2+5  r_H^2\right)}{ u^2 [2 Q^2(1-u^2)-r_H^2 u^2]+r_H^2}\right]F_2=0.
\end{align}

\begin{figure}[t]
\centering  
{\includegraphics[width=0.45\linewidth]{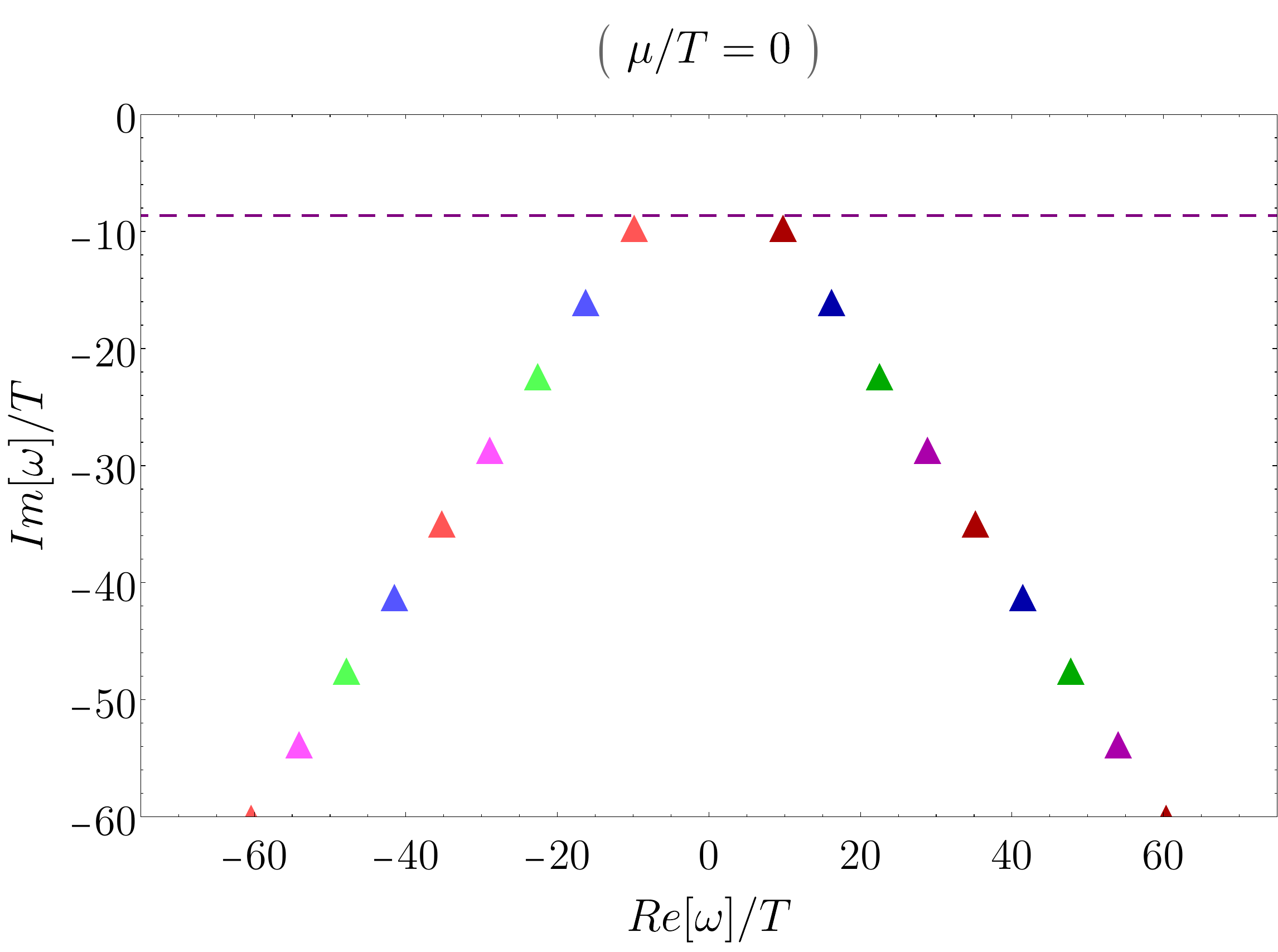}}
{\includegraphics[width=0.45\linewidth]{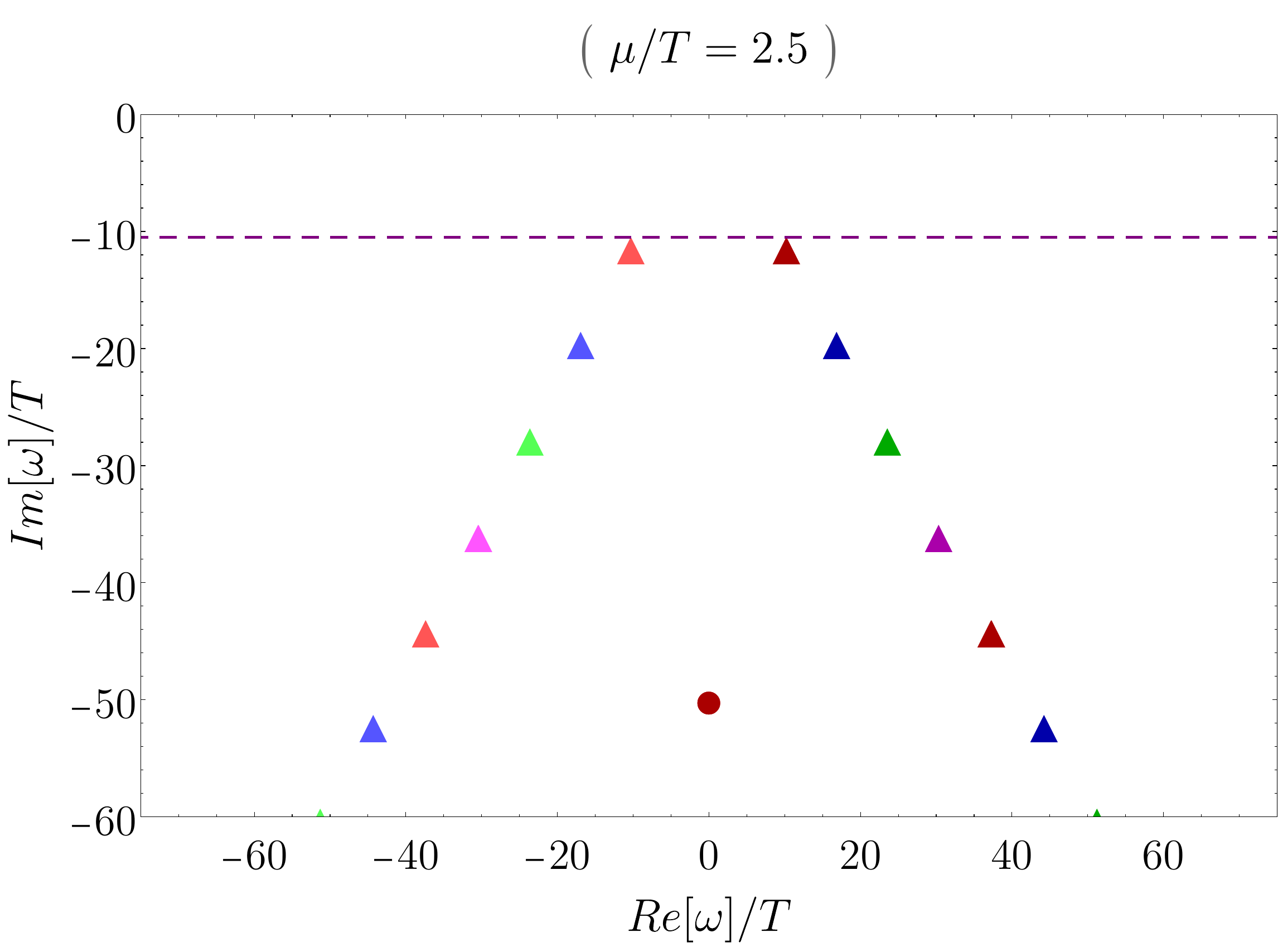}}
{\includegraphics[width=0.45\linewidth]{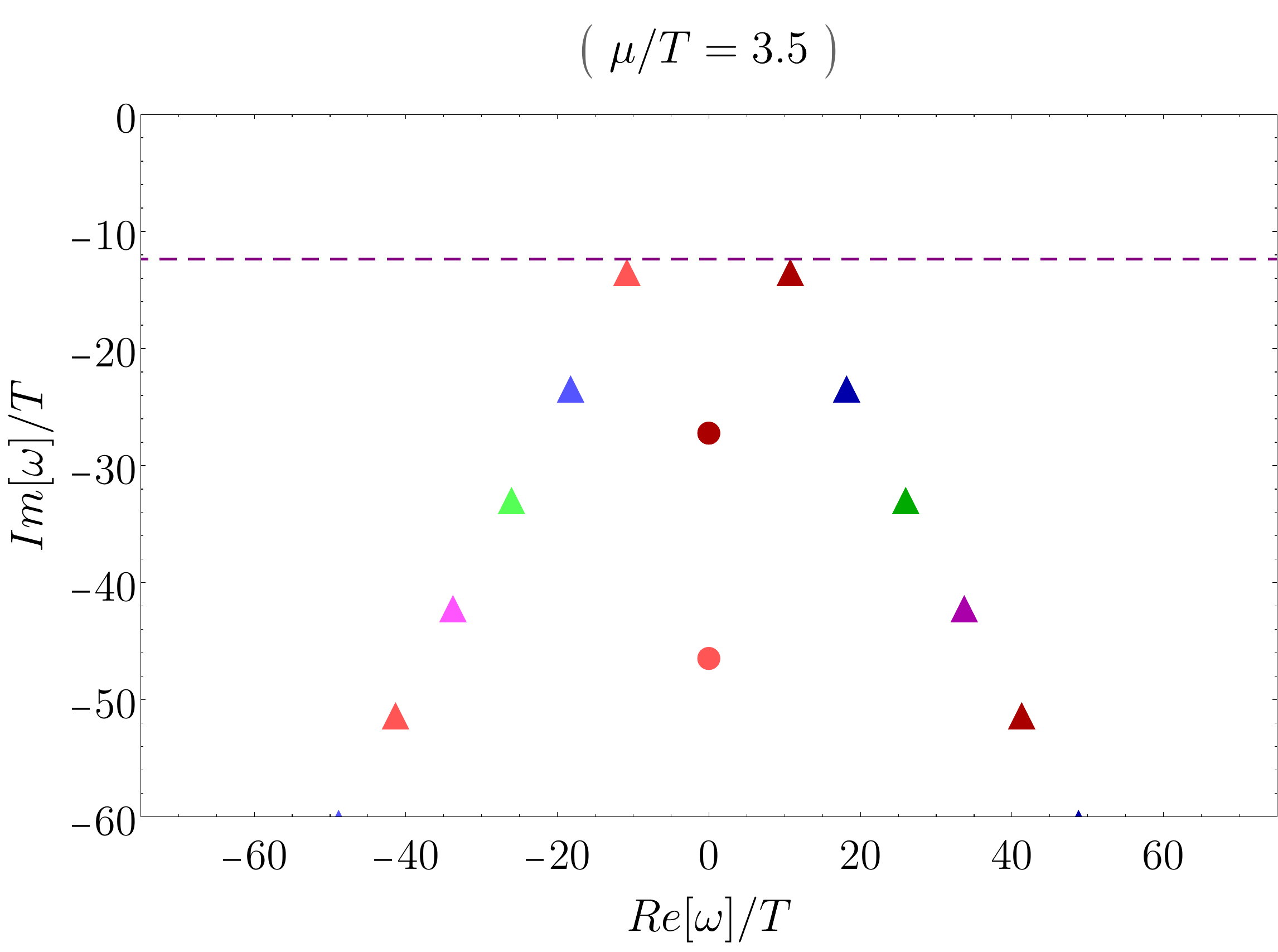}}
{\includegraphics[width=0.45\linewidth]{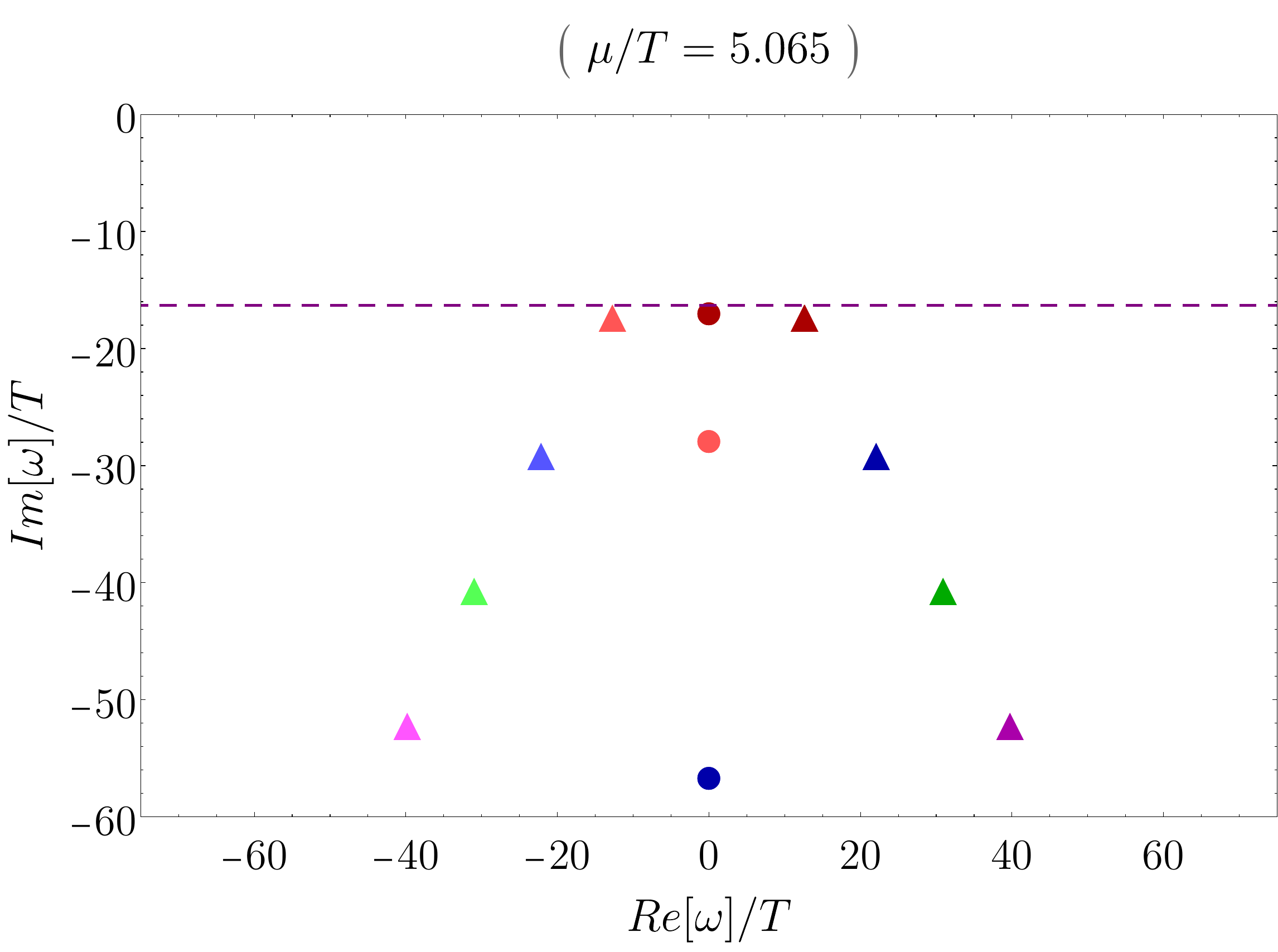}}
{\includegraphics[width=0.45\linewidth]{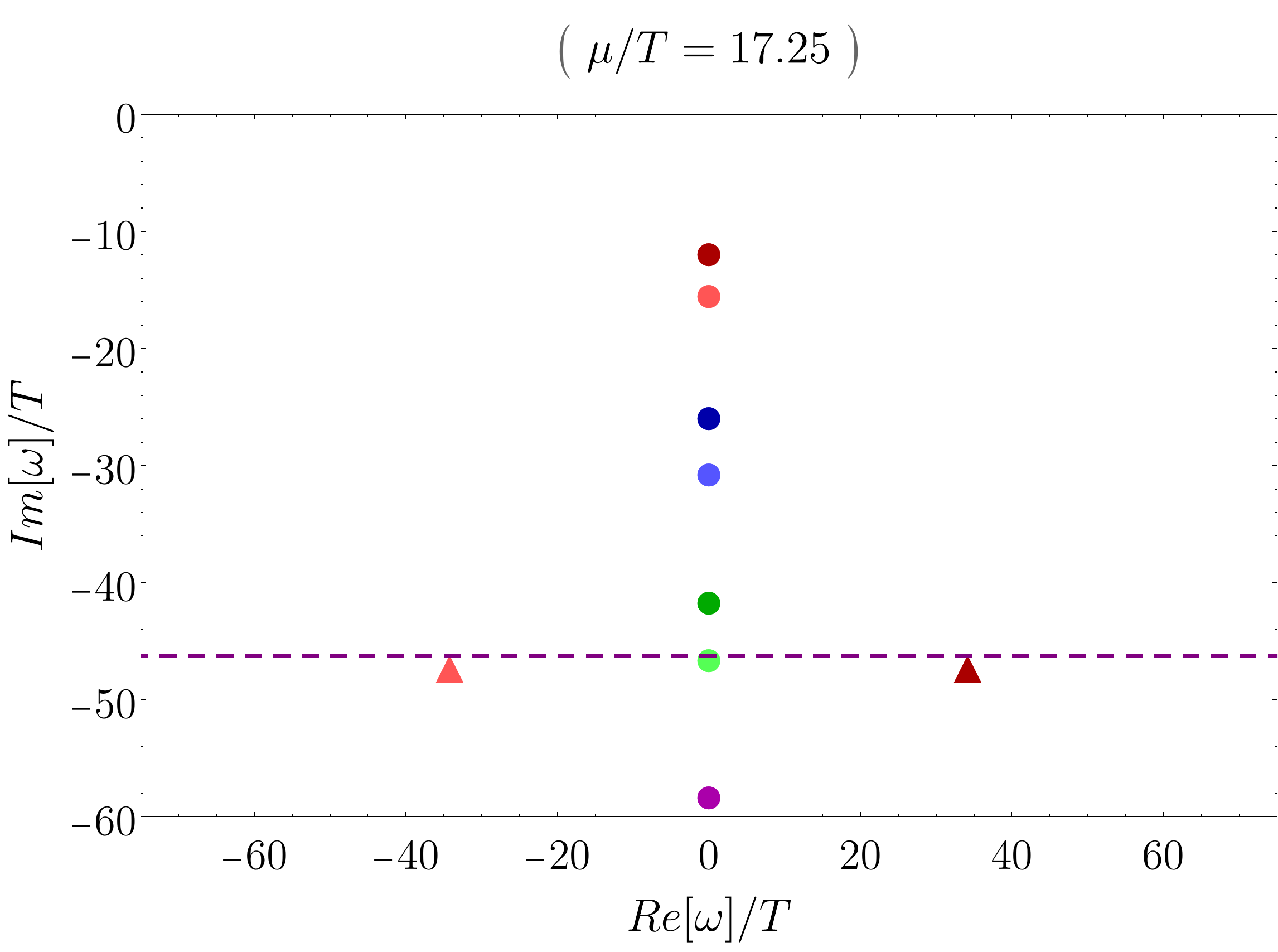}}
{\includegraphics[width=0.45\linewidth]{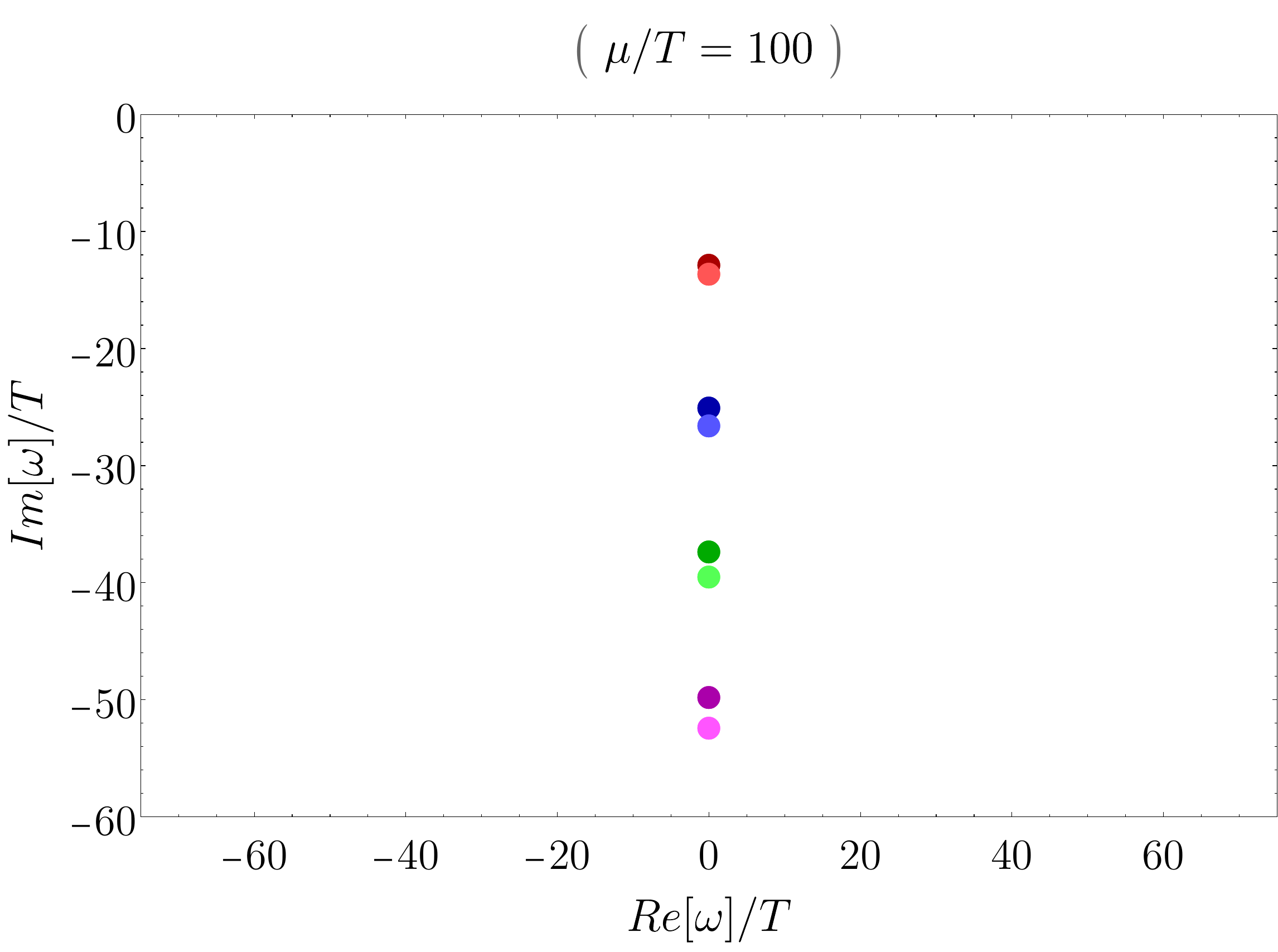}}
\caption{QNMs snapshots for 6 different values of $\mu/T$ in the $SO(3)$ quintuplet channel of the 2RCBH model. In the figure, colored triangles represent OQNMs (ordinary QNMs with nonzero real part), colored circles depict PIQNMs (purely imaginary QNMs) while the dashed horizontal line measures the imaginary part of the lowest OQNMs (i.e., the OQNMs with lowest imaginary part in modulus).}
\label{fig:StrQNMsQui}
\end{figure}

\subsection{QNM Spectra}

\begin{figure}
\centering  
\subfigure[Real part of the First 8 Ordinary QNMs]
{\includegraphics[width=0.45\linewidth]{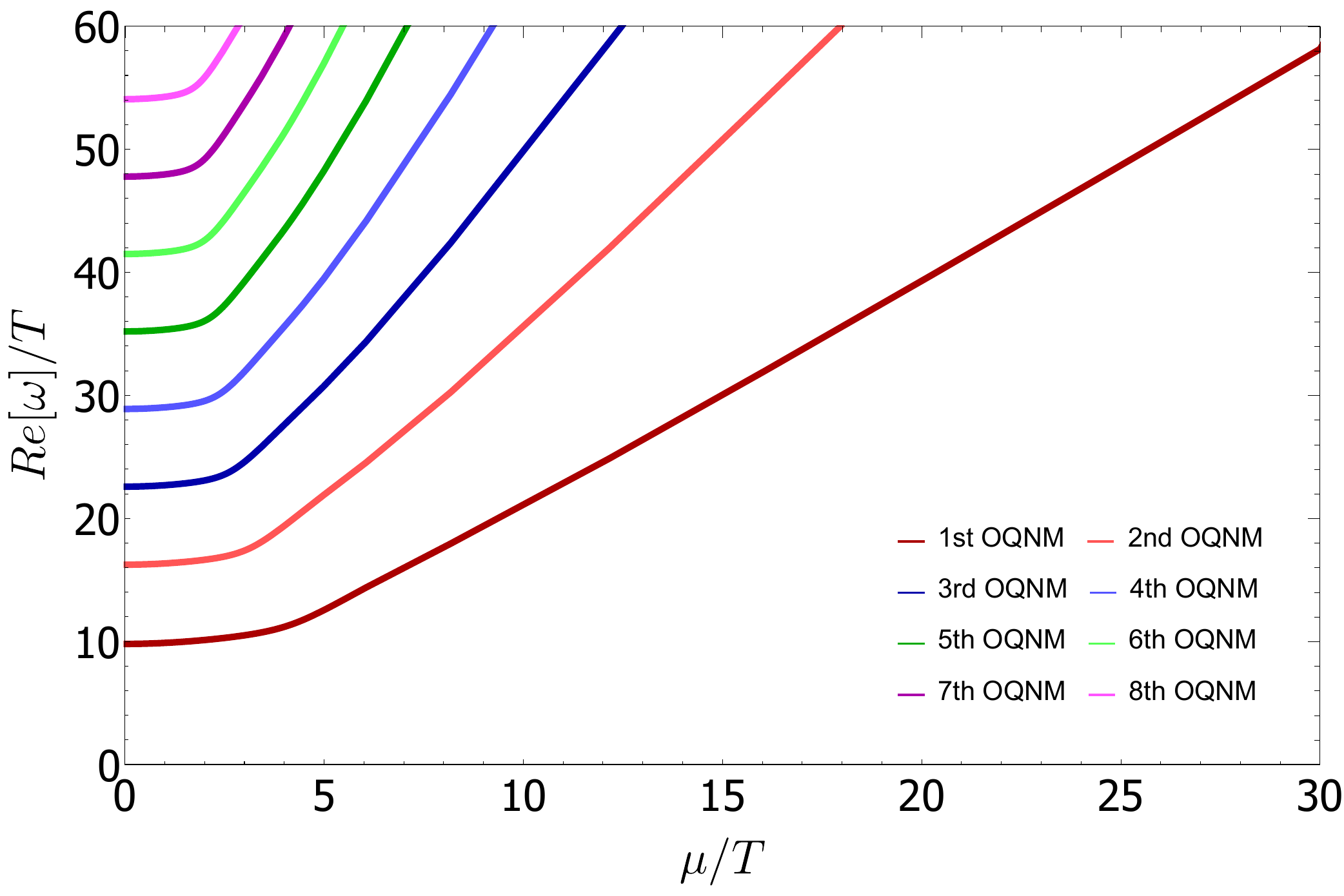}}
\label{fig:OQNMQuiRe2RCBH}
\subfigure[Imaginary part of the First 8 Ordinary QNMs]{\includegraphics[width=0.45\linewidth]{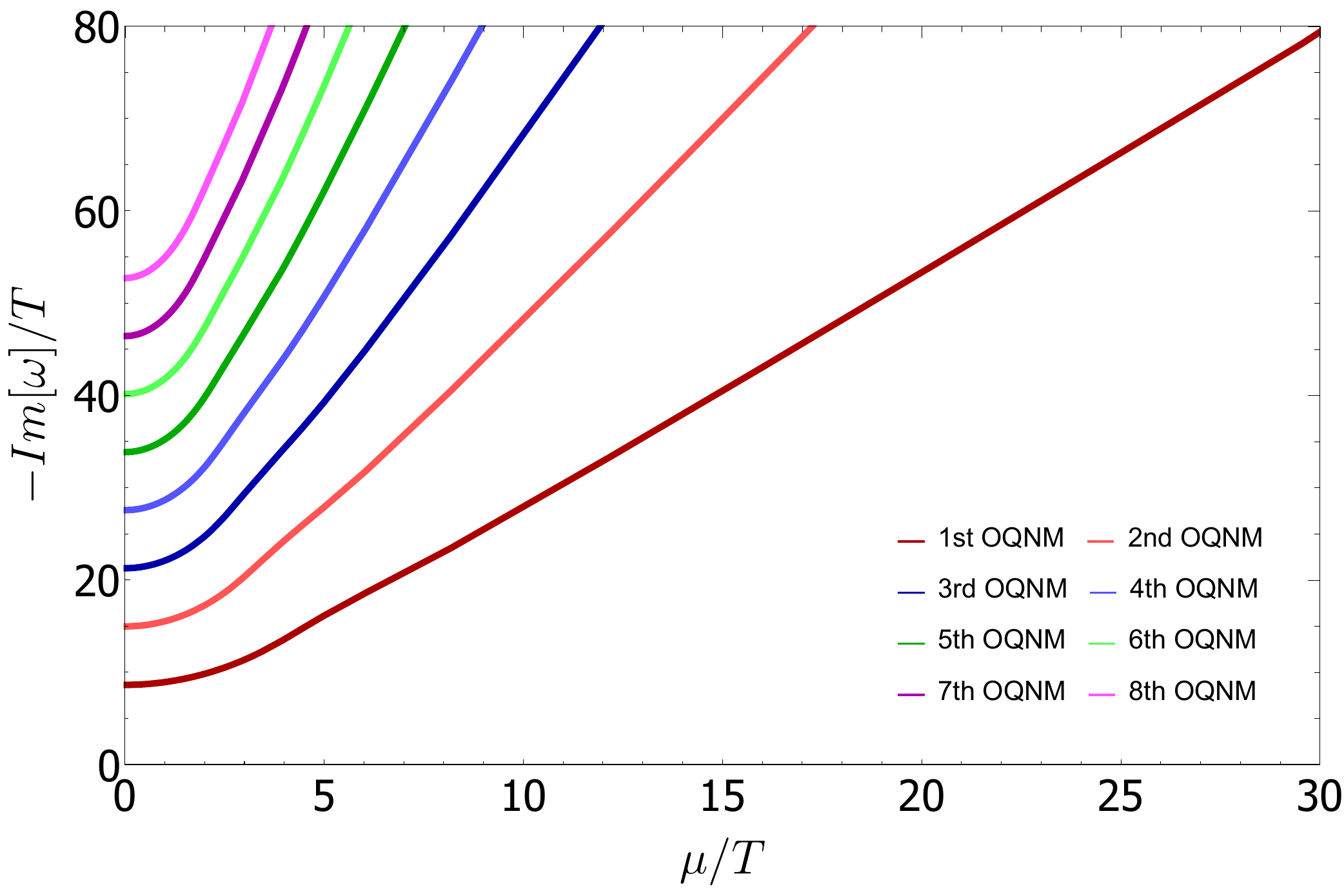}}
\label{fig:OQNMQuiIm2RCBH}
\subfigure[Imaginary part of the First 8 Purely Imaginary QNMs]
{\includegraphics[width=0.6\linewidth]{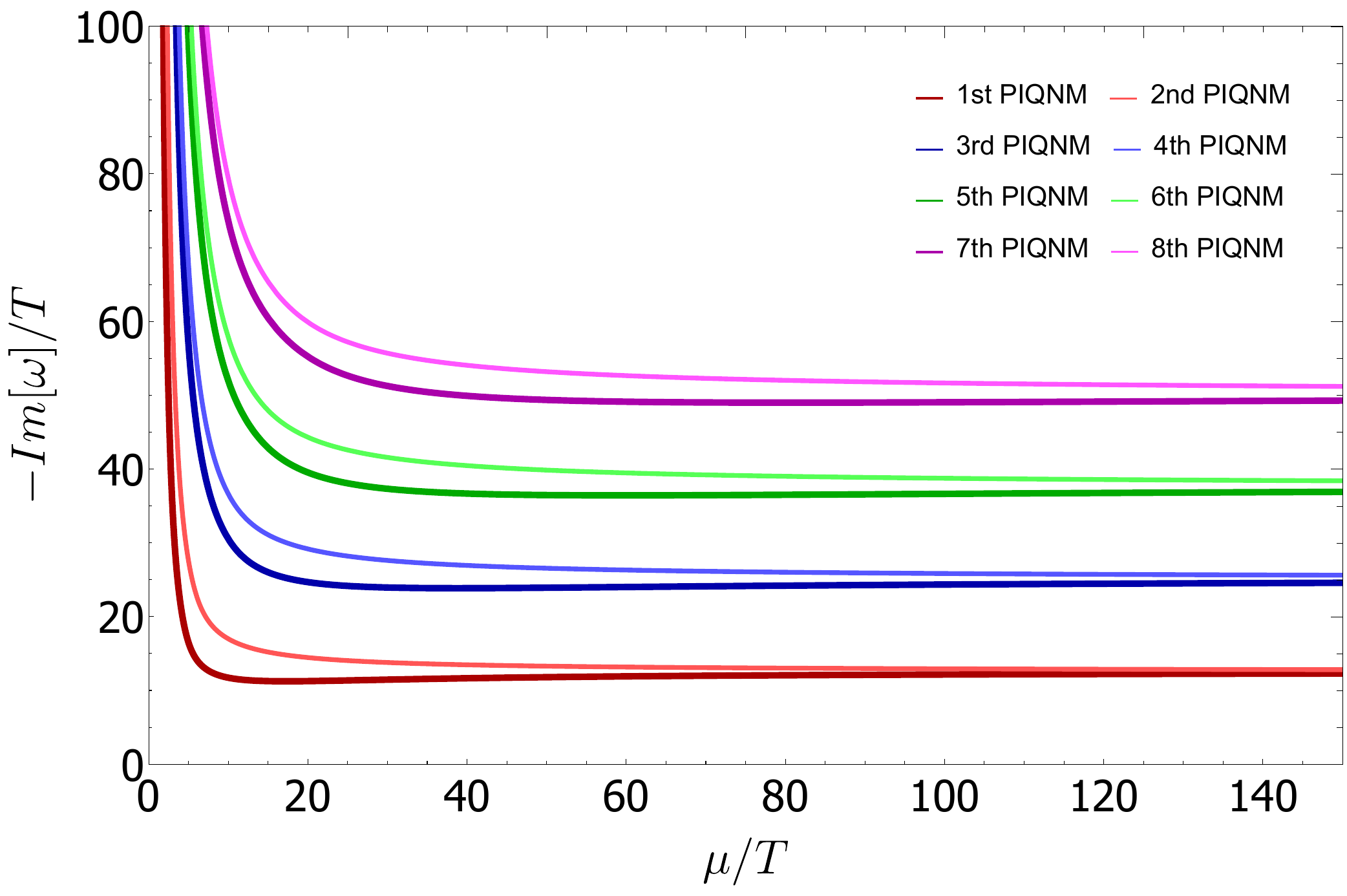}}
\label{fig:PIQNMQui2RCBH}
\caption{QNMs as functions of $\mu/T$ in the $SO(3)$ quintuplet channel of the 2RCBH model.}
\label{fig:PIQNMQui2RCBH}
\end{figure}

Figs. \ref{fig:QNMQuiRe} and \ref{fig:QNMQuiIm} provide a direct comparison for, respectively, the real and imaginary parts of the normalized dimensionless eigenfrequencies, $\omega/T$, corresponding to the first four \textit{ordinary quasinormal modes} (OQNMs --- i.e., QNMs with nonzero real part) in the $SO(3)$ quintuplet channel of the 1RCBH and 2RCBH models. As expected, the results for the 1RCBH and 2RCBH models agree at low values of $\mu/T$, since both models converge to the purely thermal SYM plasma in the limit $\mu/T\to 0$.\footnote{As discussed in section \ref{sec:thermo}, in the limit $\mu/T\to 0$ the 1RCBH model converges to the purely thermal SYM plasma within the stable branch of black hole solutions; in the same limit, within the unstable branch, one has instead the superstar background.} Notice also that, while the phase diagram of the 1RCBH model is limited within the compact range $\mu/T\in[0,\pi/\sqrt{2}]$, with the stable and unstable branches of black hole solutions merging at the critical point $\mu/T=\pi/\sqrt{2}$, the phase diagram of the 2RCBH model probes the infinite range $\mu/T\in[0,\infty)$ with a only a single branch of black hole solutions. As discussed in \cite{Finazzo:2016psx}, at the critical point of the 1RCBH model, the vertical merging of the two branches of black hole solutions displayed in Figs. \ref{fig:QNMQuiRe} and \ref{fig:QNMQuiIm} can be characterized by associated critical exponents.

The behavior of both real and imaginary parts of the 2RCBH OQNMs for a much broader range of values of $\mu/T$ can be seen in Figs.~\ref{fig:PIQNMQui2RCBH}(a) and \ref{fig:PIQNMQui2RCBH}(b), where we plot the first few OQNMs with positive real part (there is a mirrored structure for the OQNMs, with the same values of the imaginary part but negative values of the real part, as illustrated in Fig.~\ref{fig:StrQNMsQui}). One observes that the initial growths for the real and imaginary parts of the OQNMs turn into linearly increasing asymptotic behaviors for large values of $\mu/T$.

The main result for the 2RCBH model is the discovery of a new and intricate structure of pairs of\textit{ purely imaginary quasinormal modes} (PIQNMs), as shown in Fig.~\ref{fig:PIQNMQui2RCBH}(c) --- see also Fig.~\ref{fig:StrQNMsQui}.  Analyses of the behaviors of these PIQNMs indicate that the moduli of such modes start at infinity, decreasing in magnitude as $\mu/T$ increases until reaching a minimum, and beyond that it slightly increase tending to asymptotically stabilize at some constant values. Fairly interestingly, in the $SO(3)$ quintuplet channel of the 2RCBH model, these purely imaginary QNMs come in pairs, with each PIQNM within a given pair seemingly to converge to the same asymptotic value, while the different pairs of PIQNMs are themselves evenly spaced apart, as shown in Fig.~\ref{fig:PIQNMQui2RCBH}(c) --- see also Fig.~\ref{fig:StrQNMsQui}. In striking contrast, for the the 1RCBH model there are no purely imaginary QNMs in the $SO(3)$ quintuplet channel \cite{Finazzo:2016psx}, which indicate a very different structure of QNMs in both models, even at the qualitative level. We shall further comment in section \ref{sec:conc} on possible consequences of those differences for the isotropization and thermalization processes in far-from-equilibrium dynamics of both models, to be explored in future works.

In order to better visualize how the overall structure of QNMs is arranged in the $SO(3)$ quintuplet channel of the 2RCBH model, we plotted their real parts versus their imaginary parts in Fig.~\ref{fig:StrQNMsQui} for six different key values of $\mu/T$. In this figure, colored triangles represent OQNMs, colored circles depict PIQNMs while the dashed horizontal line measures the imaginary part of the lowest OQNM (i.e., the OQNM with the lowest magnitude of its imaginary part). From the sequence of snapshots in Fig.~\ref{fig:StrQNMsQui}, one can observe that a pair of OQNMs starts as the fundamental QNMs of the system, corresponding to the QNMs with the lowest imaginary part in modulus, until they are overtaken at $\mu/T\approx 5.065$ by the first PIQNM that rapidly came from $-i\infty $. After that, the PIQNMs continue to decrease in magnitude until they reach a minimum value (at $\mu/T\approx17.25$ for the fundamental QNM). Subsequently, the now fundamental PIQNM begins to increase in magnitude and tends toward a constant value that appears to be the same as the next PIQNM. The structure of PIQNMs is then clearly organized, for large values of $\mu/T$, in pairs of modes with the same asymptotic imaginary value, all evenly spaced apart. Regarding the OQNMs, both their real and imaginary parts increase in magnitude as $\mu/T$ increases, following diagonal trajectories in the plane of complex eigenfrequencies illustrated in Fig.~\ref{fig:StrQNMsQui}.

As expected, by increasing the number of collocation points used in the Gauss-Lobato grid employed in the pseudospectral method to numerically solve the eigenfrequency problem (see Appendix B of~\cite{Finazzo:2016psx} for details on the numerical method), one may observe more highly excited pairs of PIQNMs at larger values of $\omega/T$. Of course, the computational cost of such numerical operations also increase with the number of collocation points used in the grid. Regarding the plots displayed in the present work for the $SO(3)$ quintuplet channel of the 2RCBH model, we used different data point densities for each $\mu/T$ interval as specified in Table~\ref{TabQuint}. The continuous curves shown in the plots were obtained by numerical interpolations through the generated data points. Moreover, in order to obtain a more accurate numerical estimate for the spacing $\delta_\textrm{5et}$ between two consecutive asymptotically fused pairs of PIQNMs at high values of chemical potential to temperature ratio in the quintuplet channel, we also evaluated the PIQNMs at $\mu/T=1500$, employing for this specific calculation $2500$ collocation points in the Gauss-Lobato grid, which allowed to determine the approximate value $\delta_\textrm{5et}\approx 12.565$ for the separation between the first two pairs of asymptotically merged PIQNMs. In the refereeing process of this work, the anonymous referee pointed out that this value is an approximation to $4\pi \approx 12.56637\dots$, which is indeed also observed (with different numerical accuracy) for the triplet and singlet channels, as we are going to discuss.

\begin{table}[]
\begin{tabular}{|c|c|c|c}
\cline{1-3}
$\mu/T$ & \# of collocation points & \# of data points &  \\ \cline{1-3}
{[}0{,} 7.5{]} & 150 & 250 &  \\
(7.5{,} 21{]} & 200 & 105 &  \\
{(}21{,} 120] & 500 & 40 &  \\
{(}120{,} 150{]} & 650 & 7 &  \\ \cline{1-3}
\end{tabular}
\caption{Some numerical details used to calculate the QNMs of the $SO(3)$ quintuplet channel.}
\label{TabQuint}
\end{table}

\begin{figure}
\centering  
\subfigure[Equilibration time.]{\includegraphics[width=0.45\linewidth]{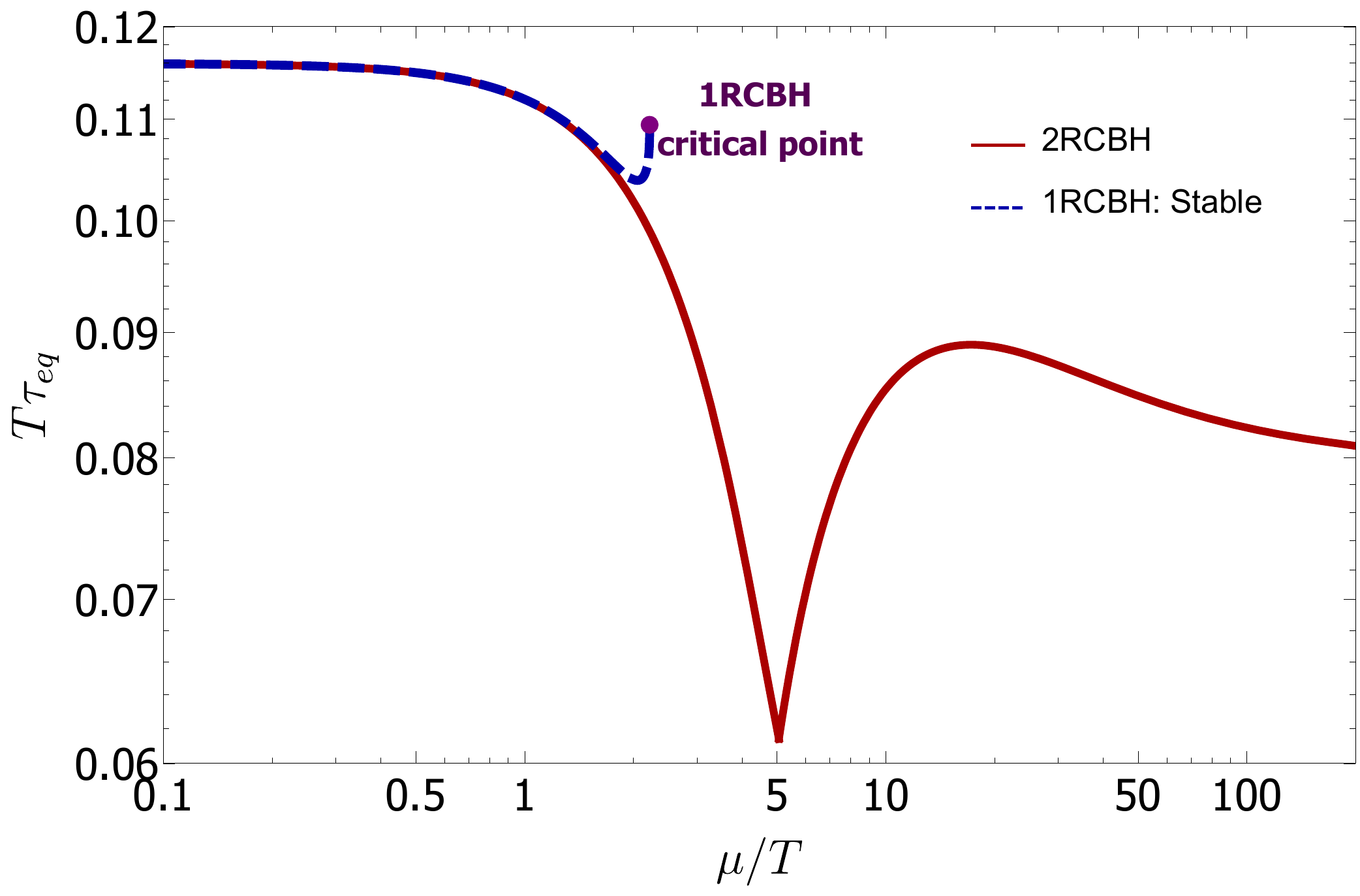}}
\subfigure[First $(\mu/T)$-derivative of the equilibration time.]{\includegraphics[width=0.45\linewidth]{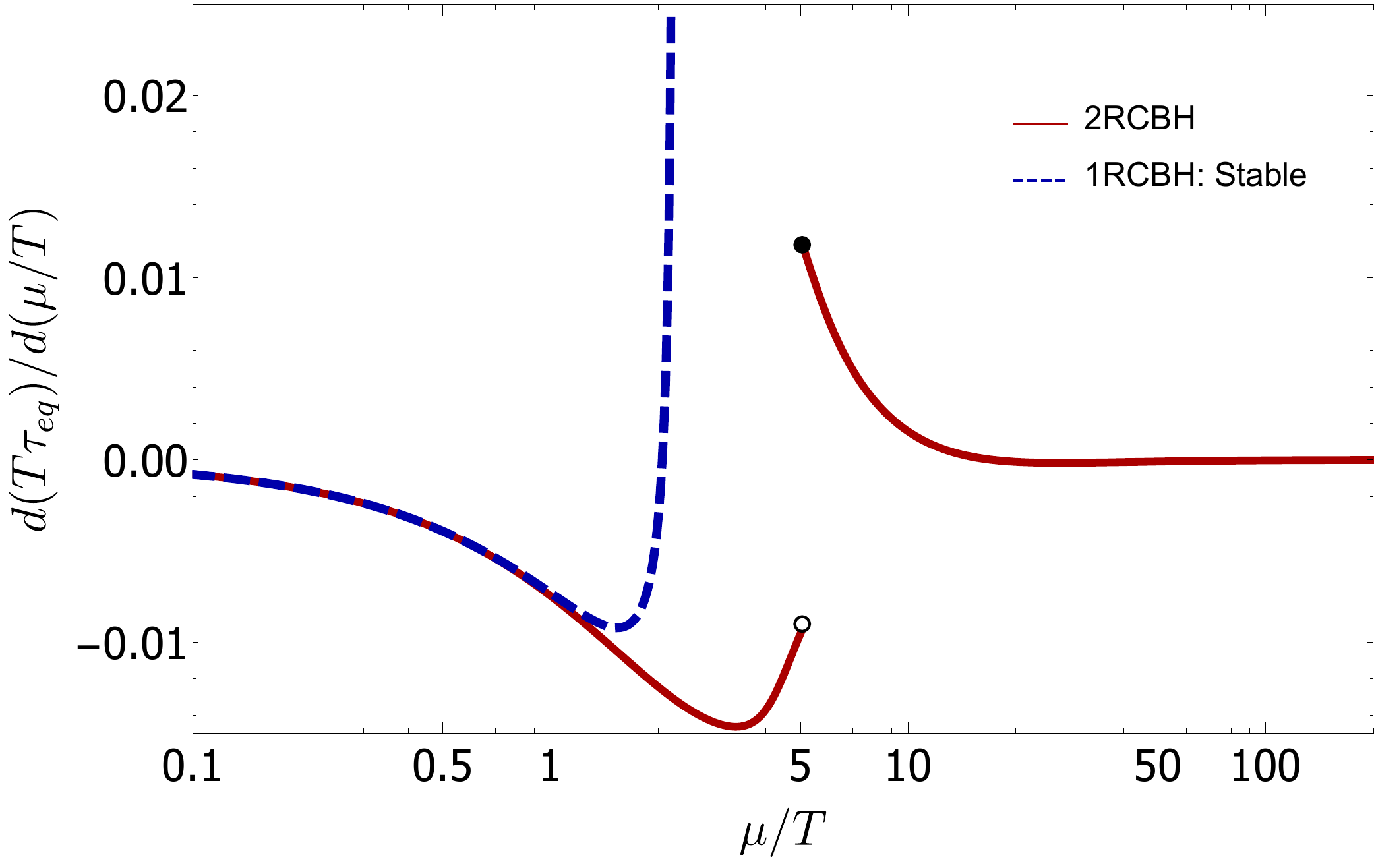}}
\caption{Characteristic equilibration time and its first derivative with respect to $\mu/T$ for the $SO(3)$ quintuplet channel of the 1RCBH and 2RCBH models.}
\label{fig:eqtime&der}
\end{figure}

By following the general reasoning discussed in Ref.~\cite{Horowitz:1999jd}, an upper bound for the characteristic equilibration time of the $SO(3)$ quintuplet channel of the 1RCBH and 2RCBH plasmas, $\tau_{\text{eq}}$, can be estimated as minus the inverse of the imaginary part of the lowest/fundamental non-hydrodynamic QNM of that channel, since it is the mode with the slowest decay in the considered channel. Fig.~\ref{fig:eqtime&der} shows a comparison between the equilibration time and its first derivative with respect to $\mu/T$, both for the 1RCBH and the 2RCBH models in the quintuplet channel. As expected, the curves agree for small values of $\mu/T$. In the case of the 1RCBH model, far from its critical point the equilibration time of the plasma decreases with the increase of $\mu/T$, while close to the critical point it starts to increase and acquires an infinite slope at the critical point, where its derivative diverges with a characteristic critical exponent of $\theta=1/2$ \cite{Finazzo:2016psx}. For the 2RCBH model, one observes a continual decrease of the equilibration time until $\mu/T\approx 5.065$, when the first PIQNM abruptly overtakes the position as the fundamental QNM of the quintuplet channel --- see also Fig.~\ref{fig:StrQNMsQui}. This abrupt change is depicted in the graph for the first derivative of the equilibration time with respect to $\mu/T$, where a discontinuity is present. By further increasing the value of $\mu/T$ beyond that point, the equilibration time increases again, reaching a maximum for the lowest PIQNM, before decreasing and eventually stabilizing at an asymptotic constant value.

\section{Triplet Channel}
\label{sec:5}

\subsection{Field Equation for the Perturbation}

\begin{figure}
\centering  
\subfigure[1st mode]{\includegraphics[width=0.45\linewidth]{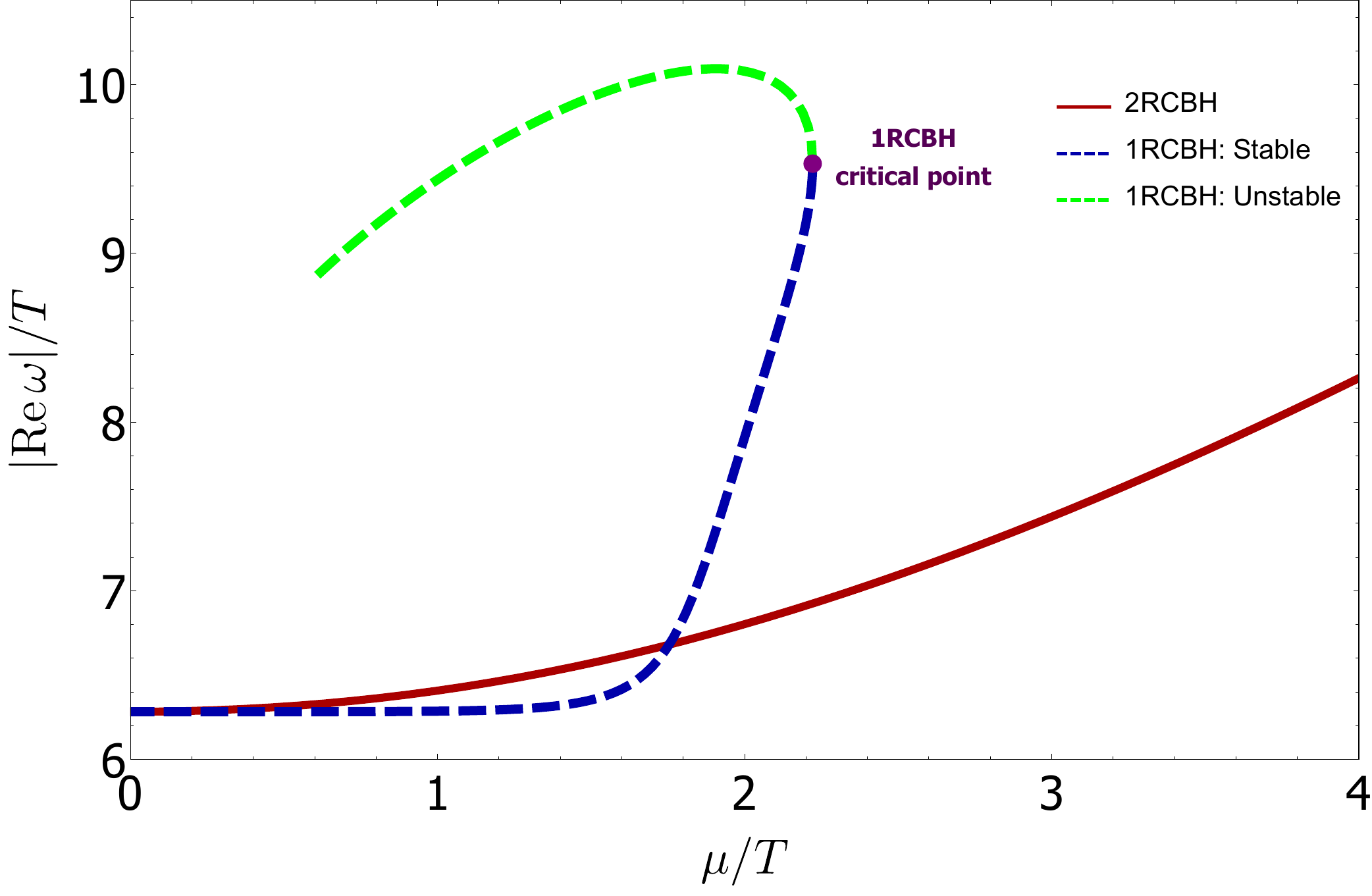}}
\subfigure[2nd mode]{\includegraphics[width=0.45\linewidth]{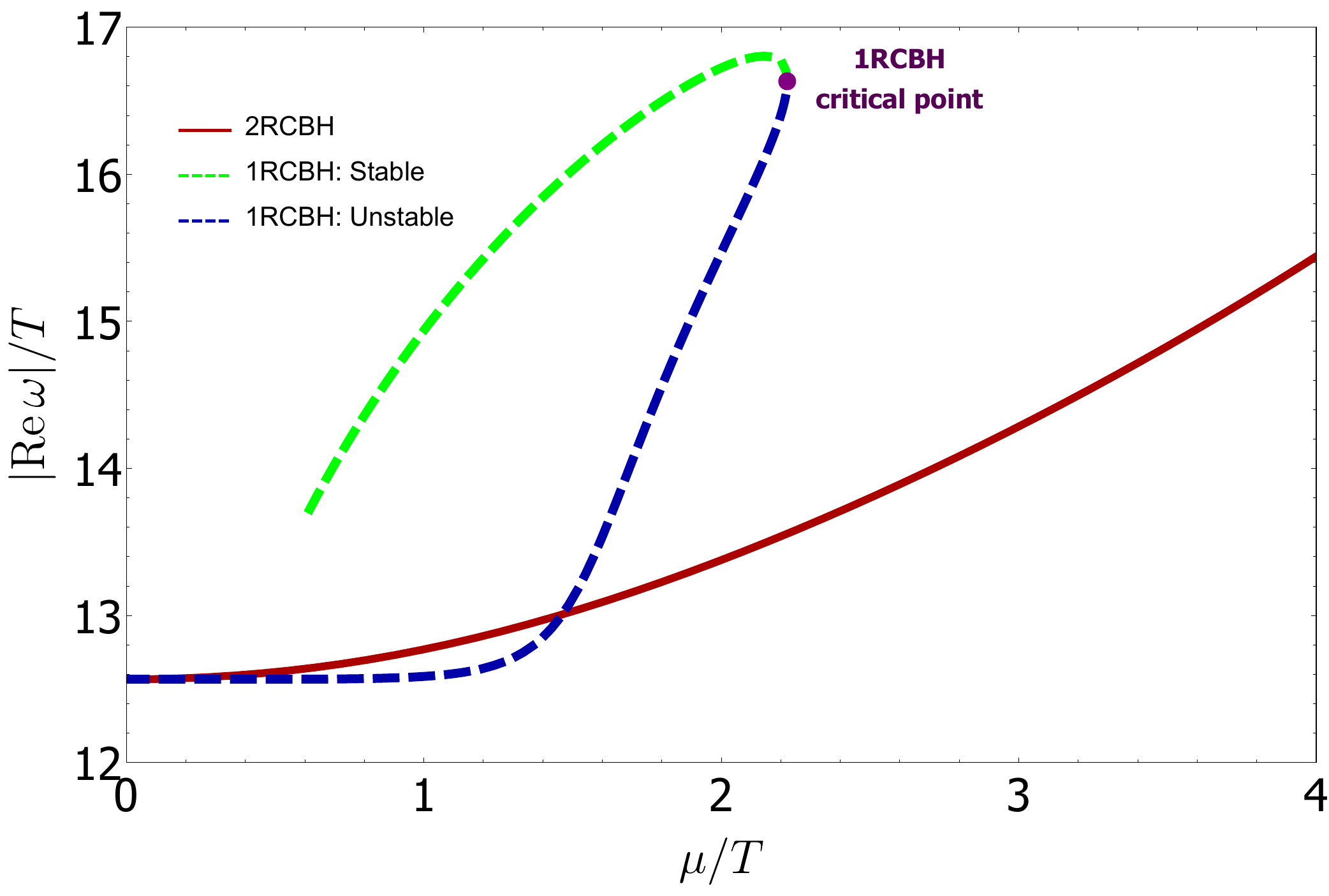}}
\subfigure[3rd mode]{\includegraphics[width=0.45\linewidth]{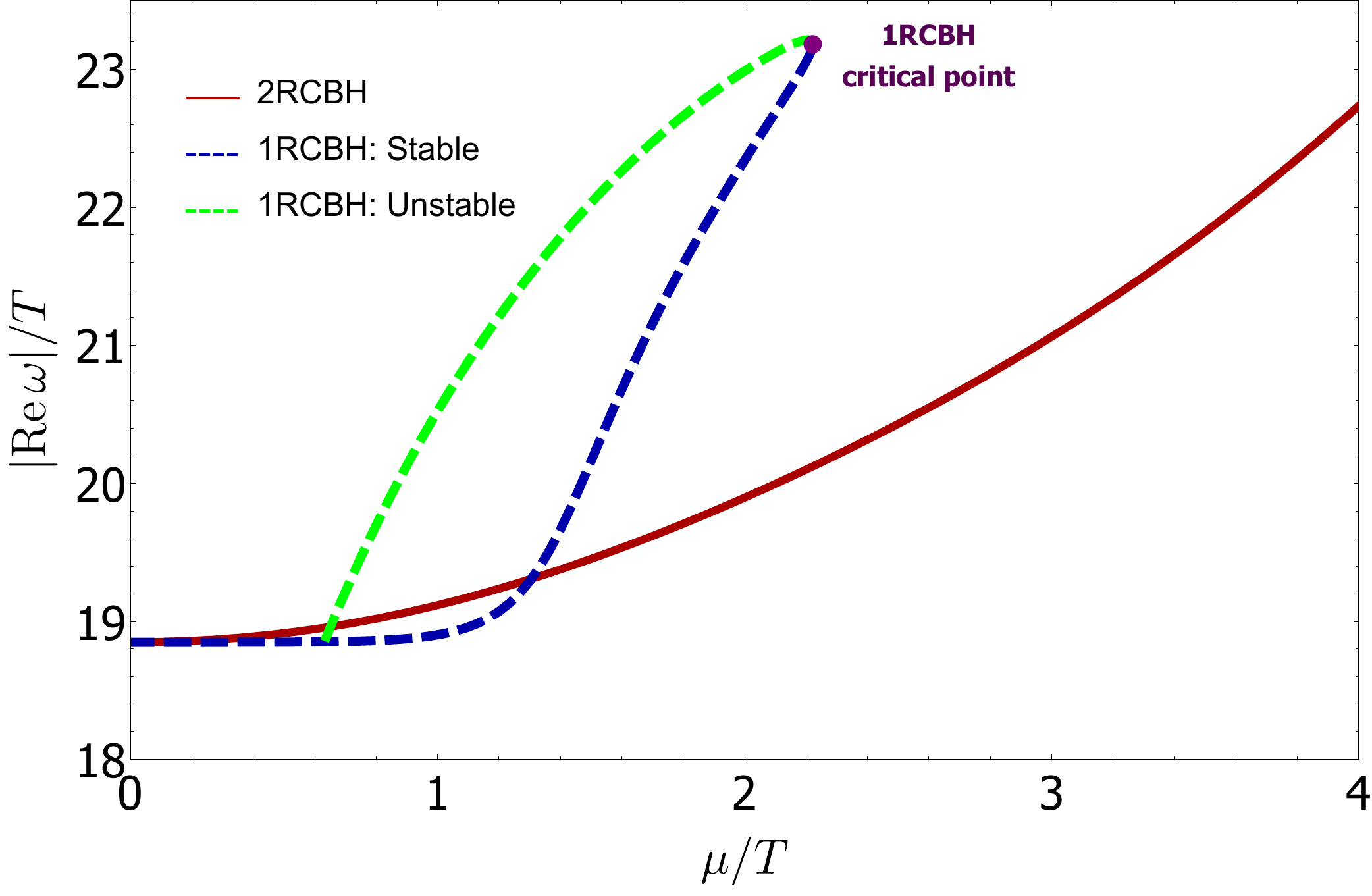}}
\subfigure[4th mode]{\includegraphics[width=0.45\linewidth]{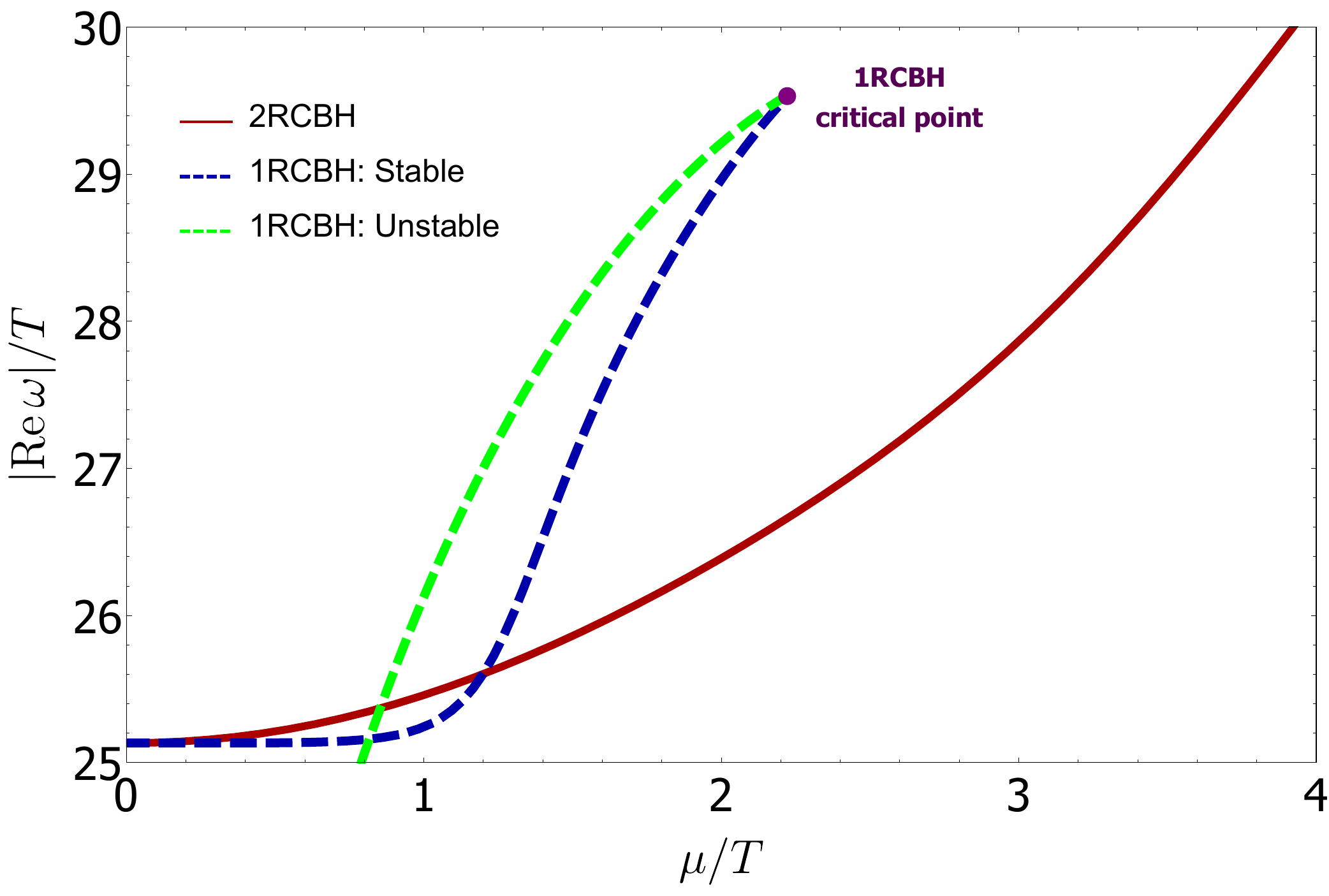}}
\caption{Real part for the first four quasinormal modes for the 1RCBH and 2RCBH models in the triplet channel.}
\label{fig:QNMTriRe}
\end{figure}
\begin{figure}
\centering
\subfigure[1st mode]{\includegraphics[width=0.45\linewidth]{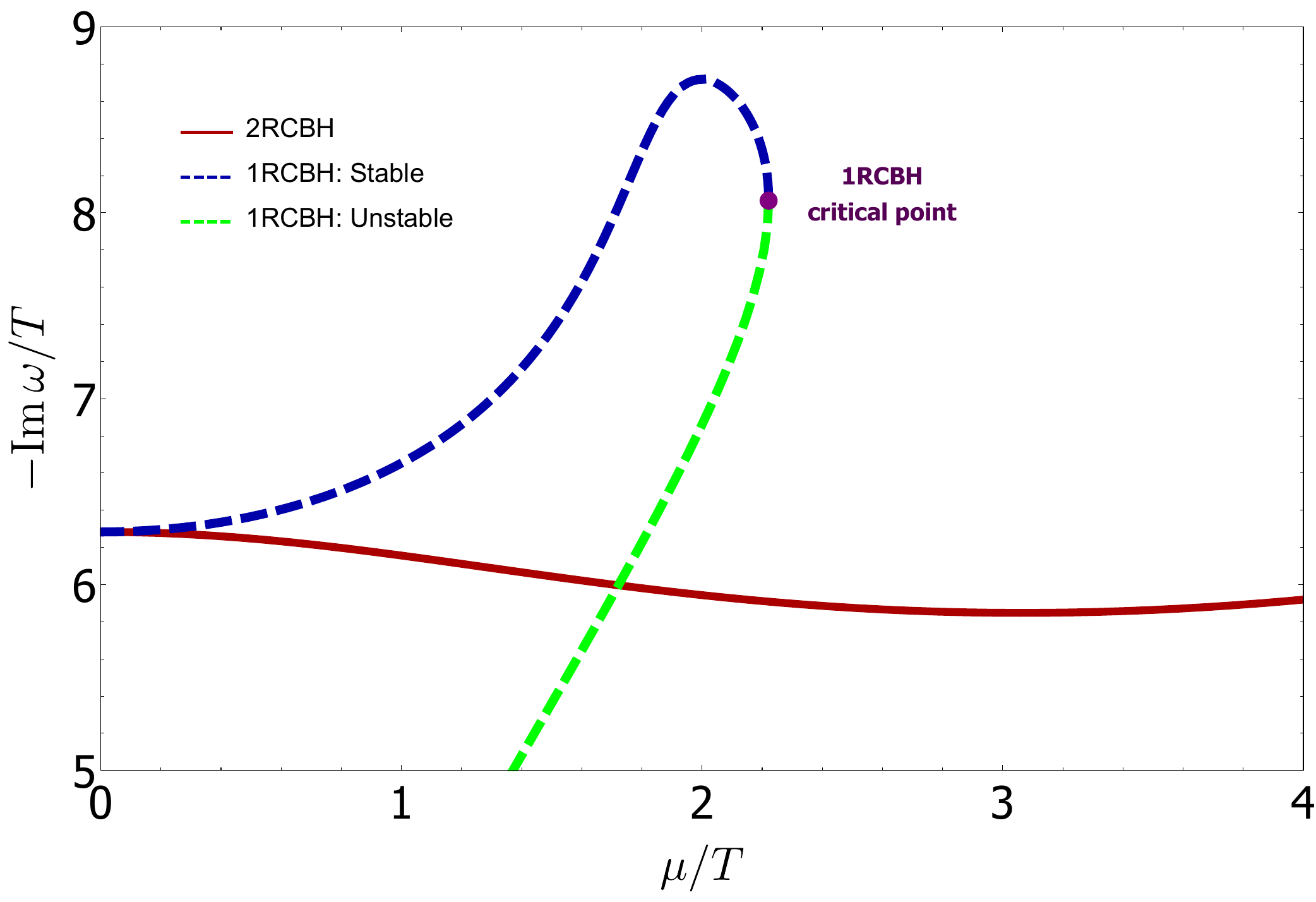}}
\subfigure[2nd mode]{\includegraphics[width=0.45\linewidth]{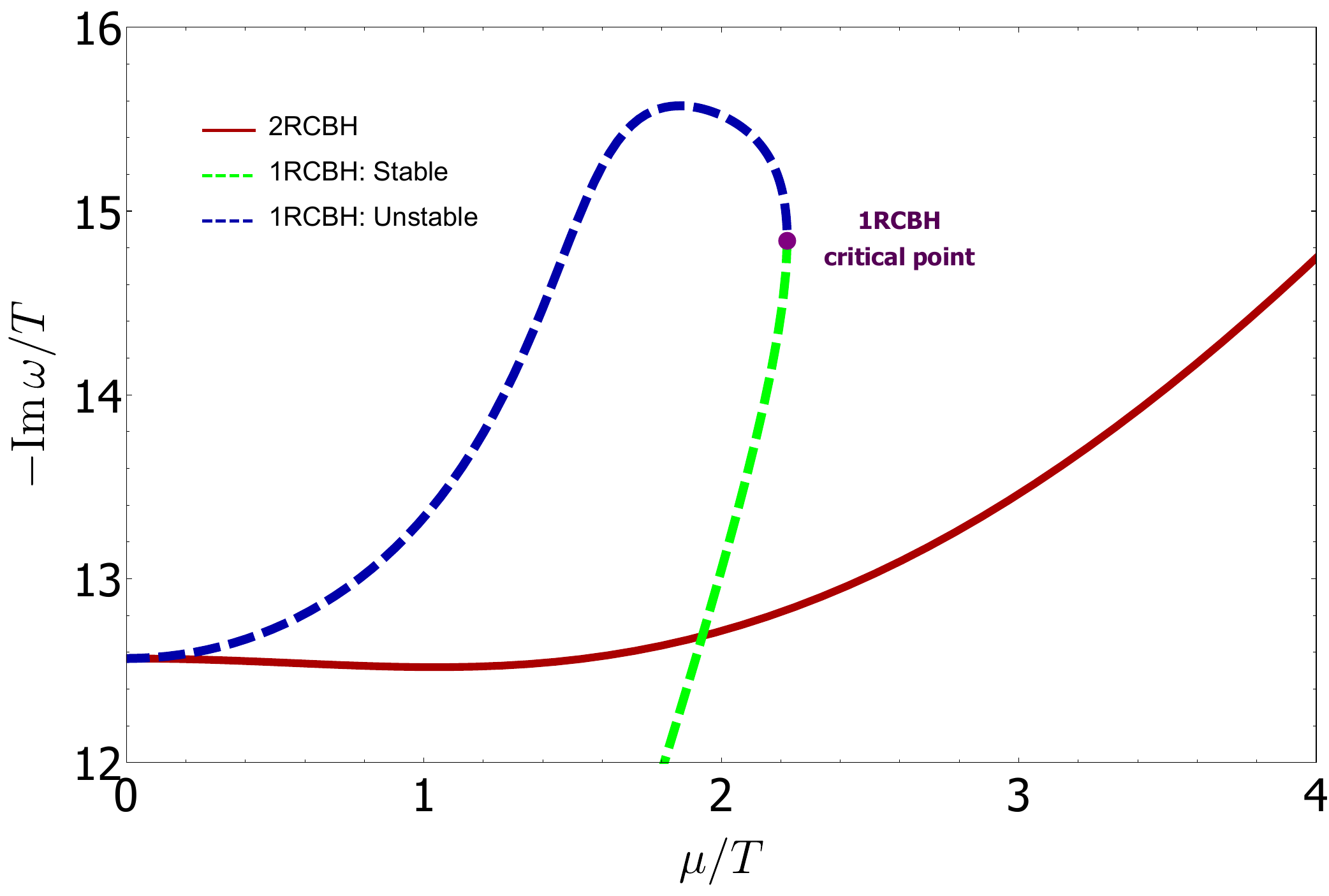}}
\subfigure[3rd mode]{\includegraphics[width=0.45\linewidth]{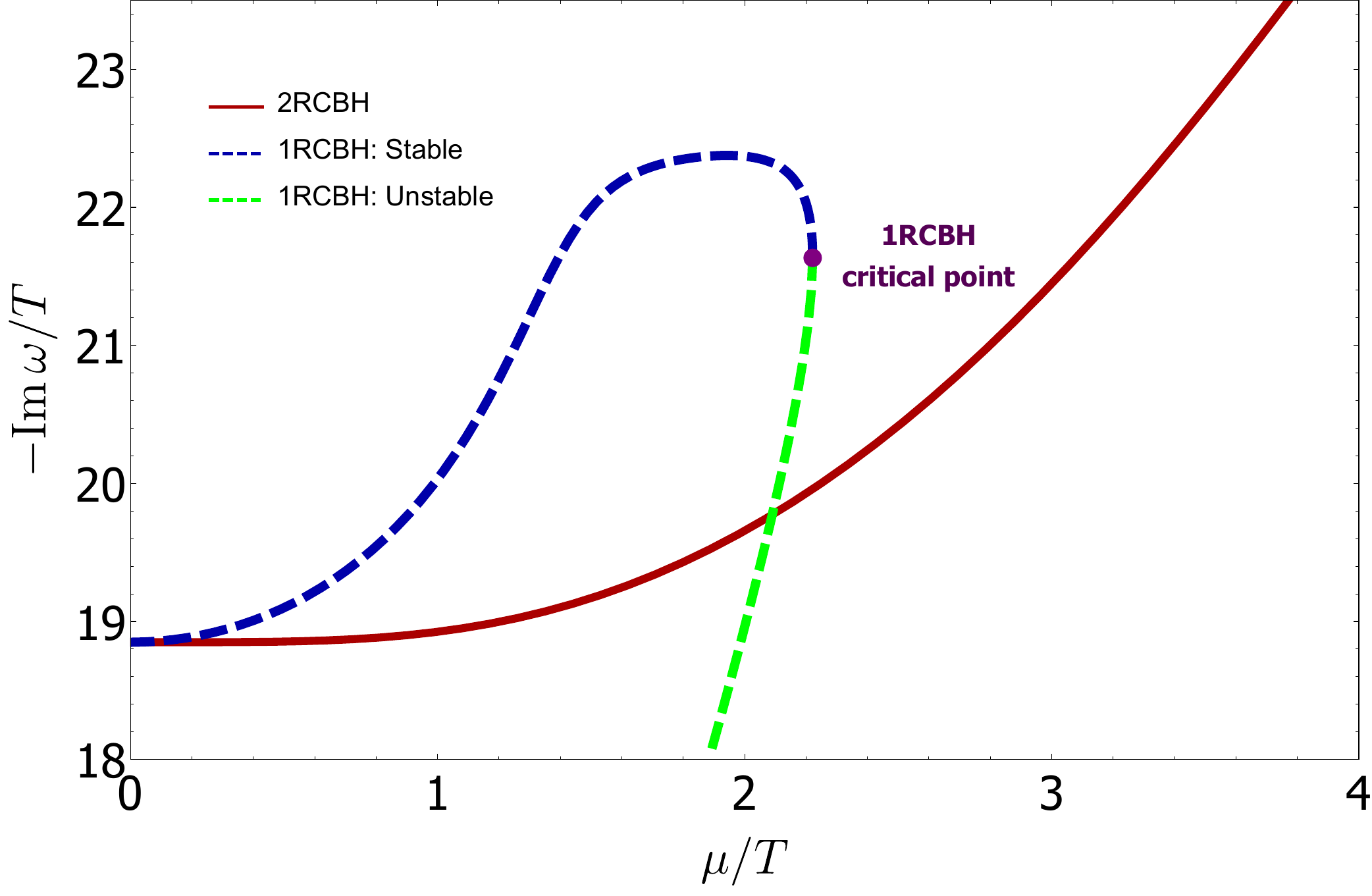}}
\subfigure[4th mode]{\includegraphics[width=0.45\linewidth]{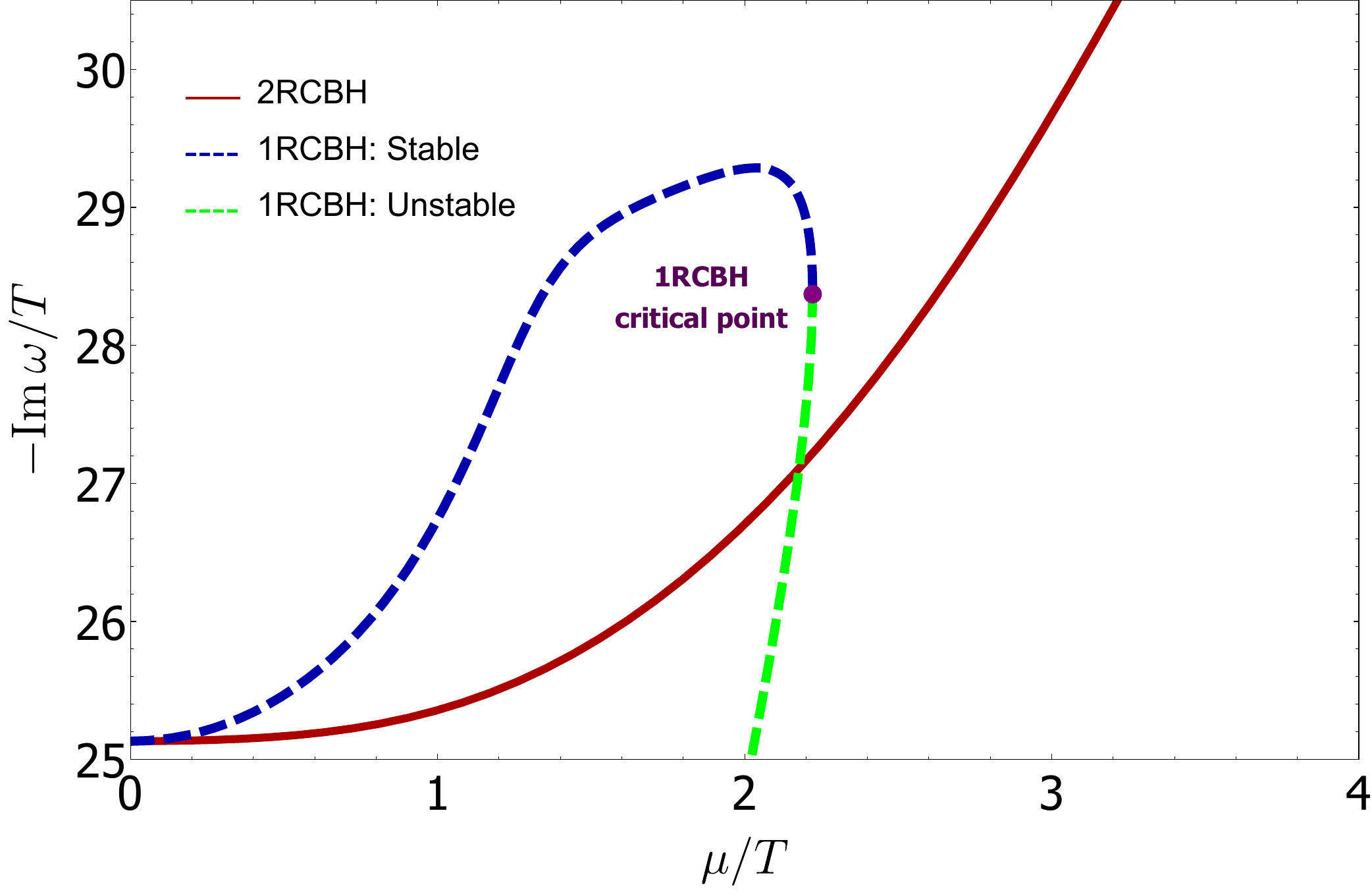}}
\caption{Imaginary part for the first four quasinormal modes for the 1RCBH and 2RCBH models in the triplet channel.}
\label{fig:QNMTriIm}
\end{figure}

By substituting the ansatz for the background field \eqref{AnsatzEqs} along with the fluctuating fields \eqref{FluctuationEqs} into the Maxwell's equation \eqref{eq:Maxwell}, it becomes evident that any spatial component of the gauge field $a\equiv a_i$, mixes at linear order only with the metric field fluctuation component $h_0^i$ \cite{Finazzo:2016psx}. This results in the following equation of motion for the fluctuating gauge field,
\begin{align}
    a''&+ \left(2 A'-B'+\frac{h'}{h}+\frac{\phi ' \partial_\phi f(\phi )}{f(\phi )}\right)a'+\frac{\omega ^2 e^{2 B-2 A}}{h^2}a\nonumber\\
    &+\frac{\Phi '}{h} h_0^i\,\!'+\left[\frac{ \left(f(\phi ) \left(\Phi ' \left(2 A'-B'\right)+\Phi ''\right)+\partial_\phi f(\phi )\phi '\Phi '  \right)}{h f(\phi )}\right]h_0^i=0,
\end{align}
together with the Maxwell's equation for the background field given by Eq.~\eqref{eq:MaxwellAnsatz}.

Employing the infalling EF coordinates defined in Eq.~\eqref{eq:EEcoordinate}, one can rewrite the last equation as \cite{Finazzo:2016psx},
    \begin{align}
      \nonumber a''&+\left(2A'-B'+\frac{h'}{h}-2i\omega \frac{e^{A-B}}{h}+\frac{\partial_\phi f(\phi)\phi'}{f(\phi)}\right)a'-i\omega \frac{f(\phi)A'+\partial_\phi f(\phi)\phi'}{f(\phi)h}e^{B-A}a\\
       \label{eq:tripletEE}&+\frac{\Phi'}{2h'}h_0^i\,\!'+\left[\frac{f(\phi)\left(\left(2A'-B'\right)\Phi'+\Phi''\right)+\partial_\phi f(\phi)\phi'\Phi'}{2f(\phi)h}\right]h_0^i=0,
    \end{align}

Using Einstein's equations for the metric field given by Eq.~\eqref{eq:Einstein}, it is possible to decouple the perturbation $h_0^i$ from the equation of motion for the perturbation $a$. This can be achieved by subtracting the $vz$-component of Eq.~\eqref{eq:Einstein} from the corresponding $rz$-component, resulting in the following constraint \cite{Finazzo:2016psx},
\begin{equation}
\label{eq:constraint}
    h_0^i\,\!'=-f(\phi)\Phi'e^{-2A}a.
\end{equation}

Then, substituting Eq.~\eqref{eq:constraint} together with the zeroth order Maxwell's Eq.~\eqref{eq:MaxwellAnsatz} into Eq.~\eqref{eq:tripletEE}, one is able to decouple the equation of motion for the gauge field perturbation as follows \cite{Finazzo:2016psx},
\begin{align}
    a''&+ \left[2 A'-\frac{2 i \omega  e^{B-A}}{h}-B'+\frac{\phi ' \partial_\phi f( \phi )}{f(\phi )}+\frac{h'}{h}\right]a'\nonumber\\
&+ \left[-\frac{e^{-2 A} f(\phi ) \Phi '^2}{h}-\frac{i \omega  e^{B-A} \left(A' f(\phi )+\phi ' \partial_\phi f(\phi )\right)}{h f(\phi )}\right]a=0.
\end{align}

Applying the coordinate transformation $r\to r_H/u$, as done in the quintuplet channel, and substituting the background solutions~\eqref{eq:AnsatzAll}, we obtain for the 1RCBH model the following expression,
\begin{align}
   \nonumber&a_1''+\left[2 u \left(\frac{Q^2}{Q^2 u^2+r_H^2}+\frac{Q^2+r_H^2}{u^2 \left(Q^2+r_H^2\right)+r_H^2}+\frac{1}{u^2-1}\right)-\frac{2 i \omega_1 \sqrt{Q^2 u^2+r_H^2}}{\left(u^2-1\right) \left(u^2 \left(Q^2+r_H^2\right)+r_H^2\right)}-\frac{1}{u}\right]a_1'\\
    &+\left[\frac{4 Q^2 r_H^2 u^4 \left(Q^2+r_H^2\right)}{\left(u^2-1\right) \left(Q^2 u^2+r_H^2\right)^2 \left(u^2 \left(Q^2+r_H^2\right)+r_H^2\right)}+\frac{i \omega_1 \left(r_H^2-2 Q^2 u^2\right)}{u \left(u^2-1\right) \sqrt{Q^2 u^2+r_H^2} \left(u^2 \left(Q^2+r_H^2\right)+r_H^2\right)}\right]a_1=0,
\end{align}
whereas for the 2RCBH model, we have,
\begin{align}
    \nonumber&a_2''+\left[2 u \left(\frac{2 Q^2+r_H^2}{u^2 \left(2 Q^2+r_H^2\right)+r_H^2}+\frac{1}{u^2-1}\right)-\frac{2 i \omega_2 \left(Q^2 u^2+r_H^2\right)}{r_H \left(u^2-1\right) \left(u^2 \left(2 Q^2+r_H^2\right)+r_H^2\right)}-\frac{1}{u}\right]a_2' \\
    &+ \left[\frac{8 Q^2 u^4 \left(Q^2+r_H^2\right)^2}{\left(u^2-1\right) \left(Q^2 u^2+r_H^2\right)^2 \left(u^2 \left(2 Q^2+r_H^2\right)+r_H^2\right)}+\frac{i \omega_2 (Q u+r_H) (r_H-Q u)}{r_H u \left(u^2-1\right) \left(u^2 \left(2 Q^2+r_H^2\right)+r_H^2\right)}\right]a_2=0.
\end{align}

In the case of the gauge field perturbation in the $SO(3)$ triplet channel, the normalizable mode at the boundary is obtained by setting $a(u)=u^2 F(u)$, with $F(0)\neq 0$ \cite{DeWolfe:2011ts,Finazzo:2016psx}. Implementing this last prescription, one obtains for the 1RCBH model the following eigenvalue problem,
\begin{align}
    \nonumber u \left(u^2-1\right) F_1''+ \Bigg[&\left(u^2-1\right) \left(9+\frac{2}{u^2-1}-\frac{2r_H^2}{u^2 \left(Q^2+r_H^2\right)+r_H^2}-\frac{2r_H^2}{Q^2 u^2+r_H^2}\right)-\frac{2 i u \omega_1 \sqrt{Q^2 u^2+r_H^2}}{u^2 \left(Q^2+r_H^2\right)+r_H^2}\Bigg]F_1'\\
    \nonumber +\Bigg[&\frac{4 Q^6 u^5 \left(3 u^2-2\right)+4 Q^4 r_H^2 u^3 \left(3 u^4+6 u^2-4\right)+8 Q^2 r_H^4 u \left(3 u^4+u^2-1\right)+8 r_H^6 u^3}{\left(Q^2 u^2+r_H^2\right)^2 \left(u^2 \left(Q^2+r_H^2\right)+r_H^2\right)}\\
    &-\frac{3 i \omega_1 \left(2 Q^2 u^2+r_H^2\right)}{\sqrt{Q^2 u^2+r_H^2} \left(u^2 \left(Q^2+r_H^2\right)+r_H^2\right)}\Bigg]F_1=0,
    \label{eq:TriplF1}
\end{align}
while in the case of the 2RCBH model, one has,
\begin{align}
    \nonumber u \left(u^2-1\right) F_2''&+\left[\frac{2 Q^2 u^2 \left(7 u^2-5\right)+r_H^2 \left(7 u^4-3\right)}{u^2 \left(2 Q^2+r_H^2\right)+r_H^2}-\frac{2 i u \omega_2 \left(Q^2 u^2+r_H^2\right)}{r_H u^2 \left(2 Q^2+r_H^2\right)+r_H^3}\right]F_2' \\
    &+\left[8 u \left(1-\frac{r_H^2 \left(Q^2+r_H^2\right) \left(r_H^2-Q^2 u^2 \left(u^2-2\right)\right)}{\left(Q^2 u^2+r_H^2\right)^2 \left(u^2 \left(2 Q^2+r_H^2\right)+r_H^2\right)}\right)-\frac{i \omega_2 \left(5 Q^2 u^2+3 r_H^2\right)}{r_H u^2 \left(2 Q^2+r_H^2\right)+r_H^3}\right]F_2 =0.
\label{eq:TriplF2}
\end{align}

\begin{figure}[t]
\centering  
{\includegraphics[width=0.45\linewidth]{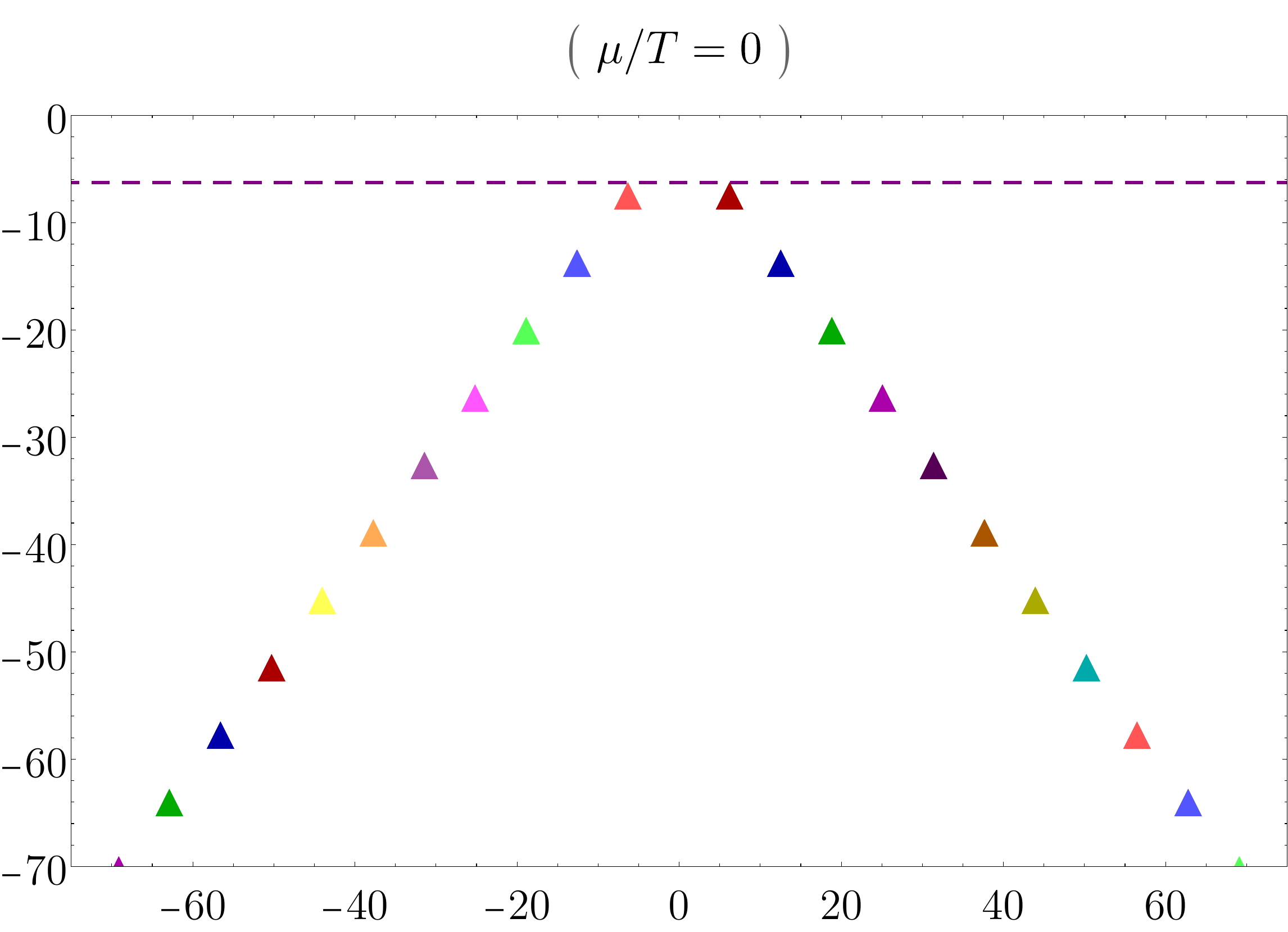}}
{\includegraphics[width=0.45\linewidth]{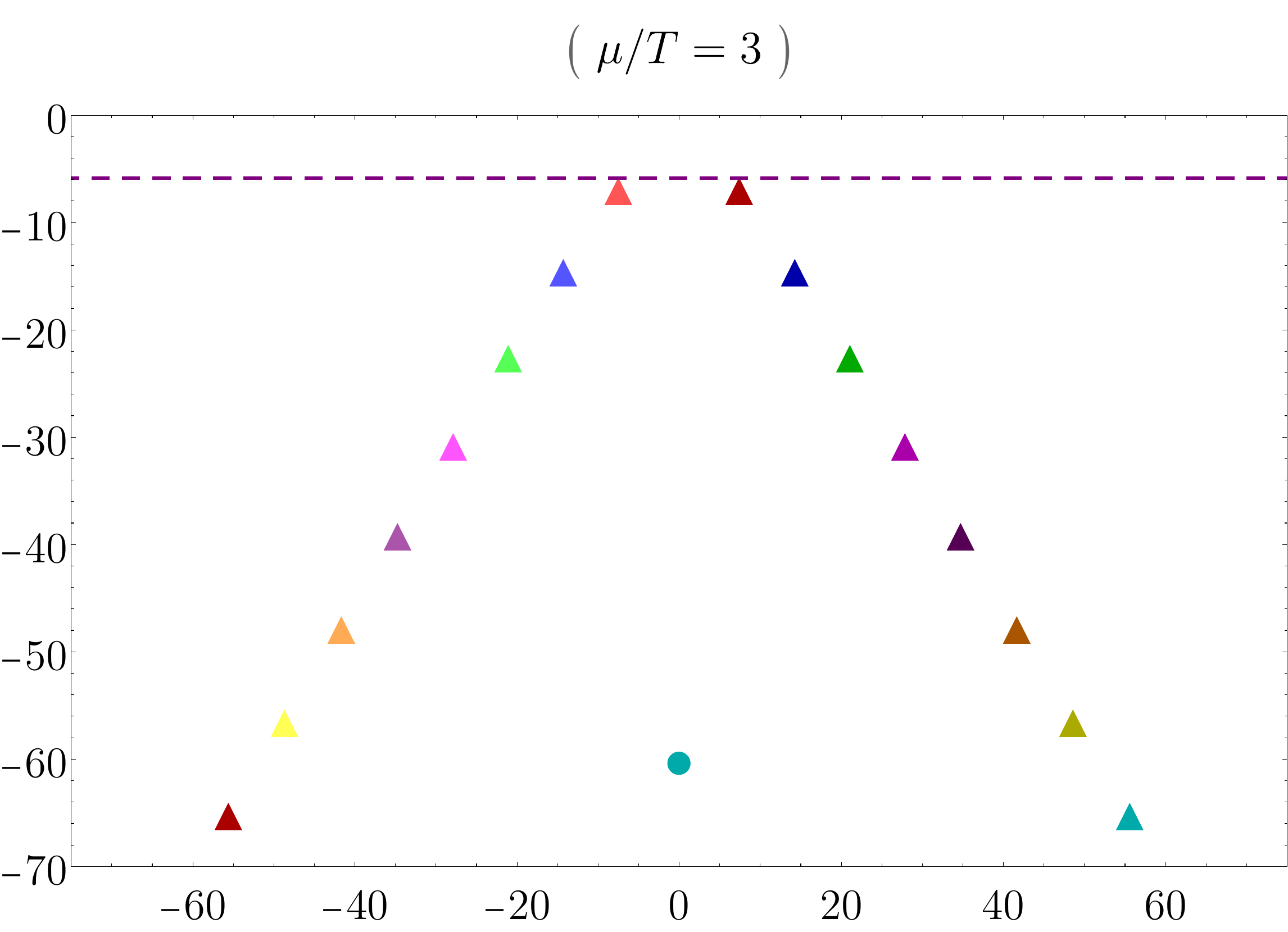}}
{\includegraphics[width=0.45\linewidth]{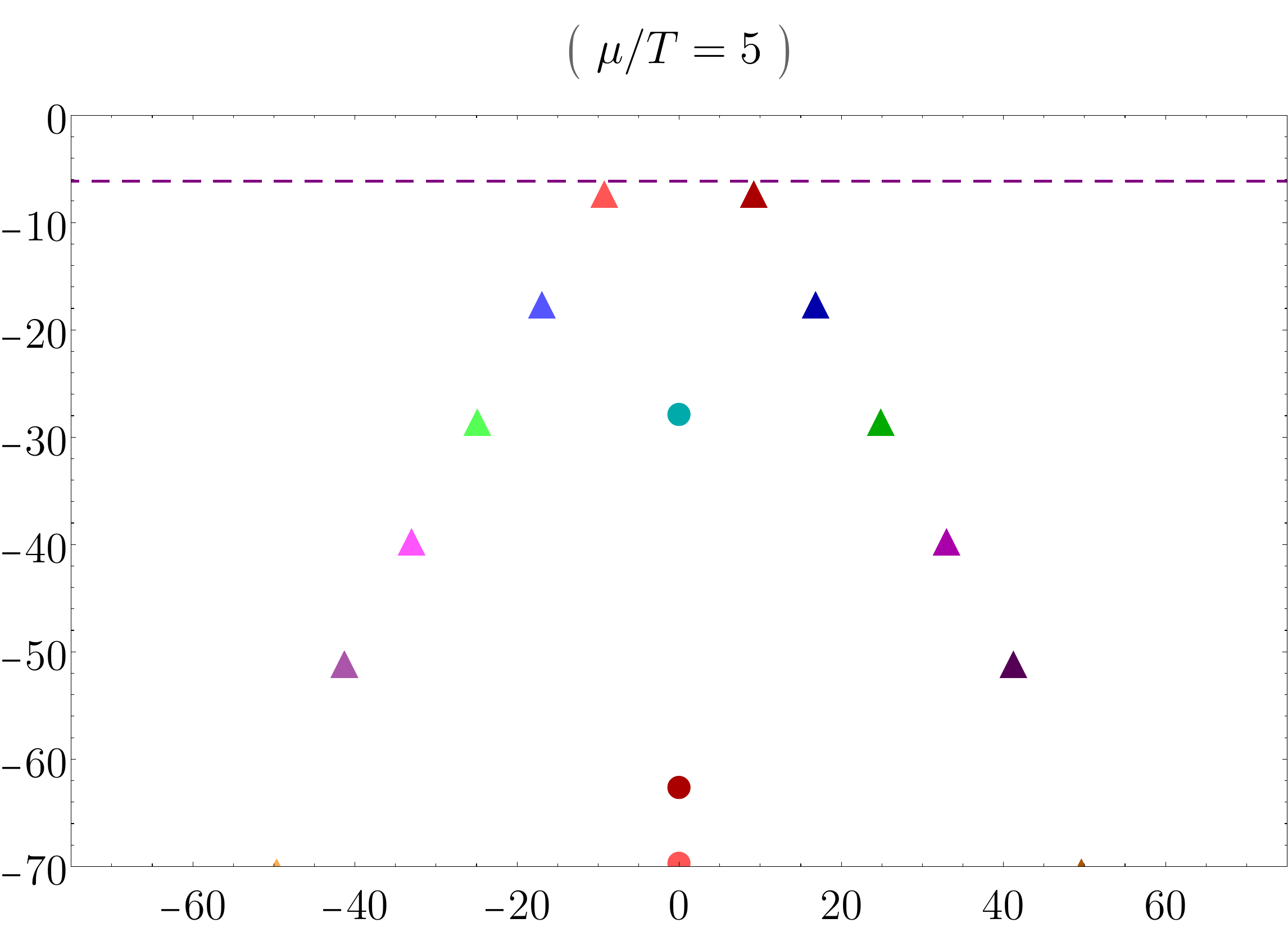}}
{\includegraphics[width=0.45\linewidth]{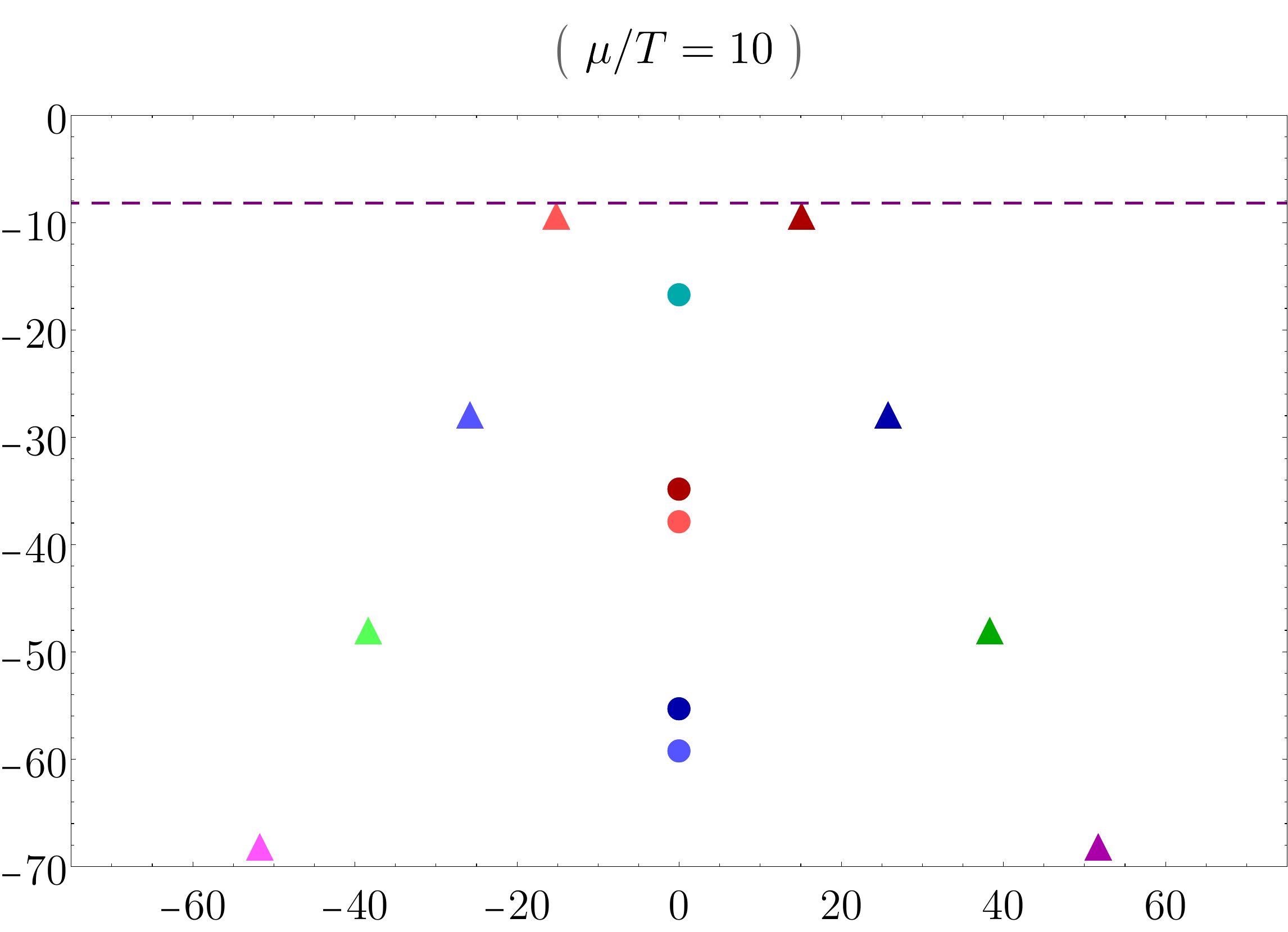}}
{\includegraphics[width=0.45\linewidth]{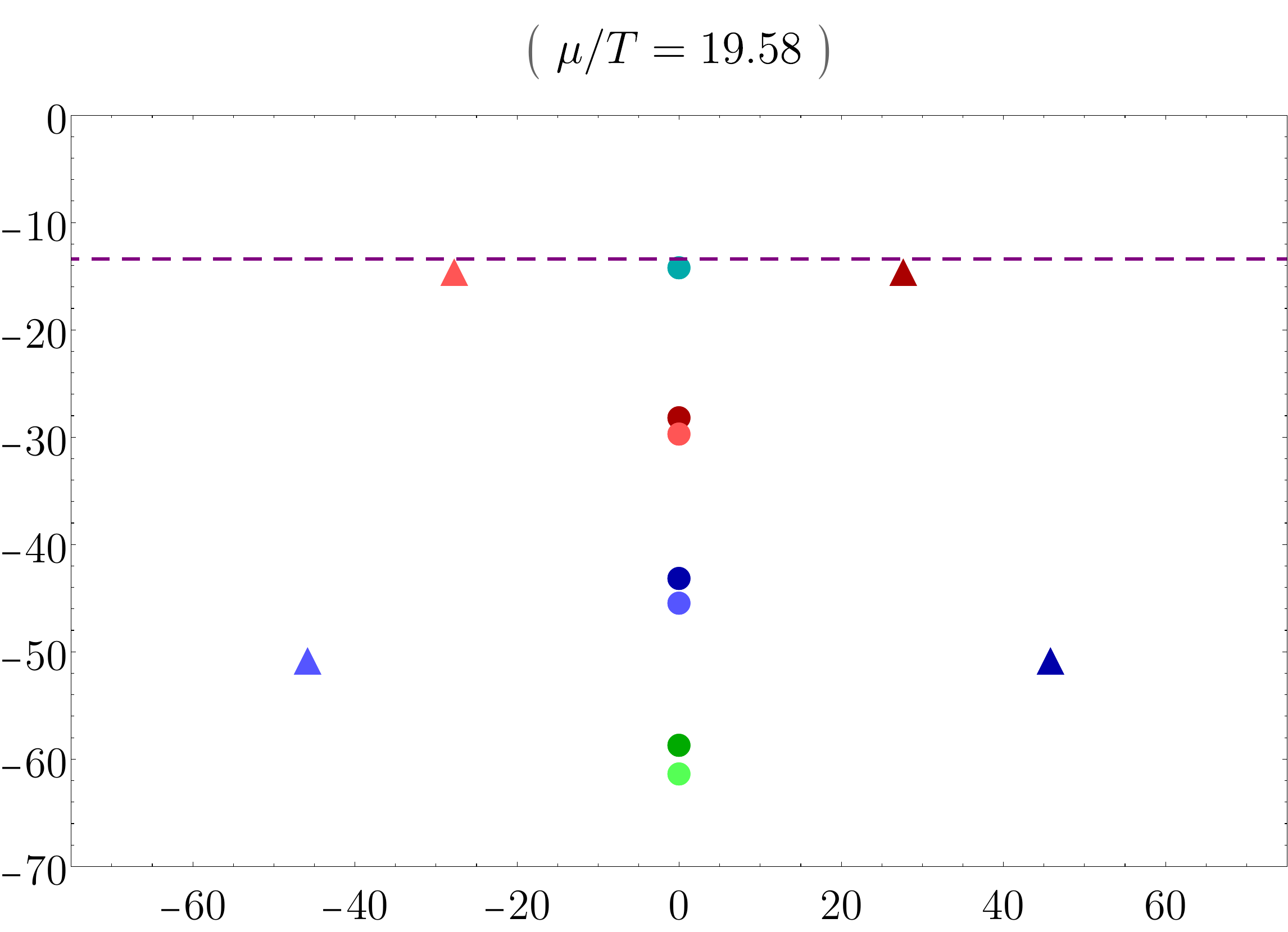}}
{\includegraphics[width=0.45\linewidth]{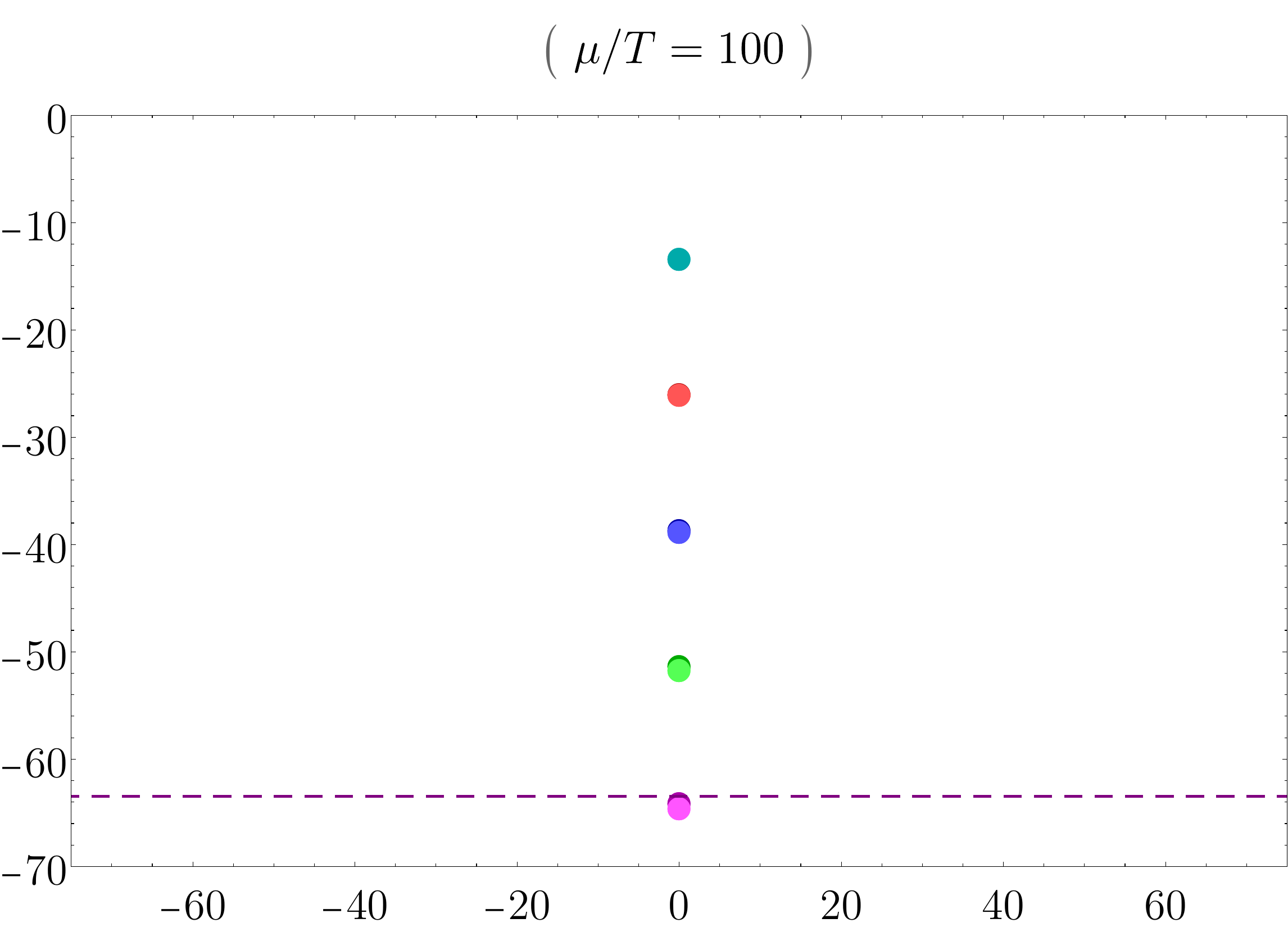}}
\caption{QNMs snapshots for 6 different values of $\mu/T$ in the $SO(3)$ triplet channel of the 2RCBH model. In the figure, colored triangles represent OQNMs (ordinary QNMs with nonzero real part), colored circles depict PIQNMs (purely imaginary QNMs) while the dashed horizontal line measures the imaginary part of the lowest OQNMs (i.e., the OQNMs with lowest imaginary part in modulus).}
\label{fig:StrQNMsTri}
\end{figure}

\subsection{QNM Spectra}

Figs. \ref{fig:QNMTriRe} and \ref{fig:QNMTriIm} provide a direct comparison for, respectively, the real and imaginary parts of the first four OQNMs in the $SO(3)$ triplet channel of the 1RCBH and 2RCBH models, plotted as functions of $\mu/T$.

\begin{figure}[h]
\centering  
\subfigure[Real part of the First 8 Ordinary QNMs]{\includegraphics[width=0.45\linewidth]{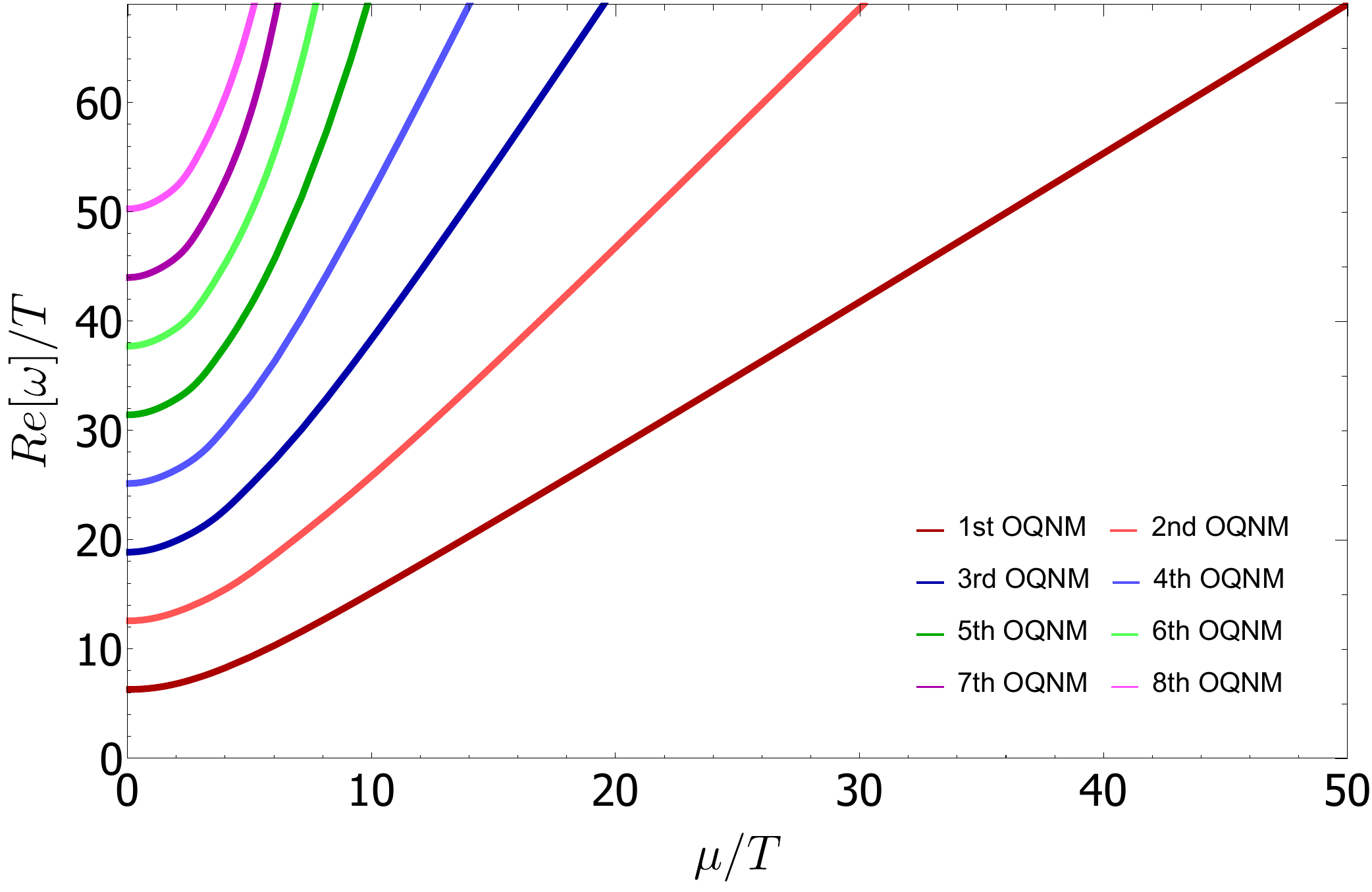}}
\subfigure[Imaginary part of the First 8 Ordinary QNMs]{\includegraphics[width=0.45\linewidth]{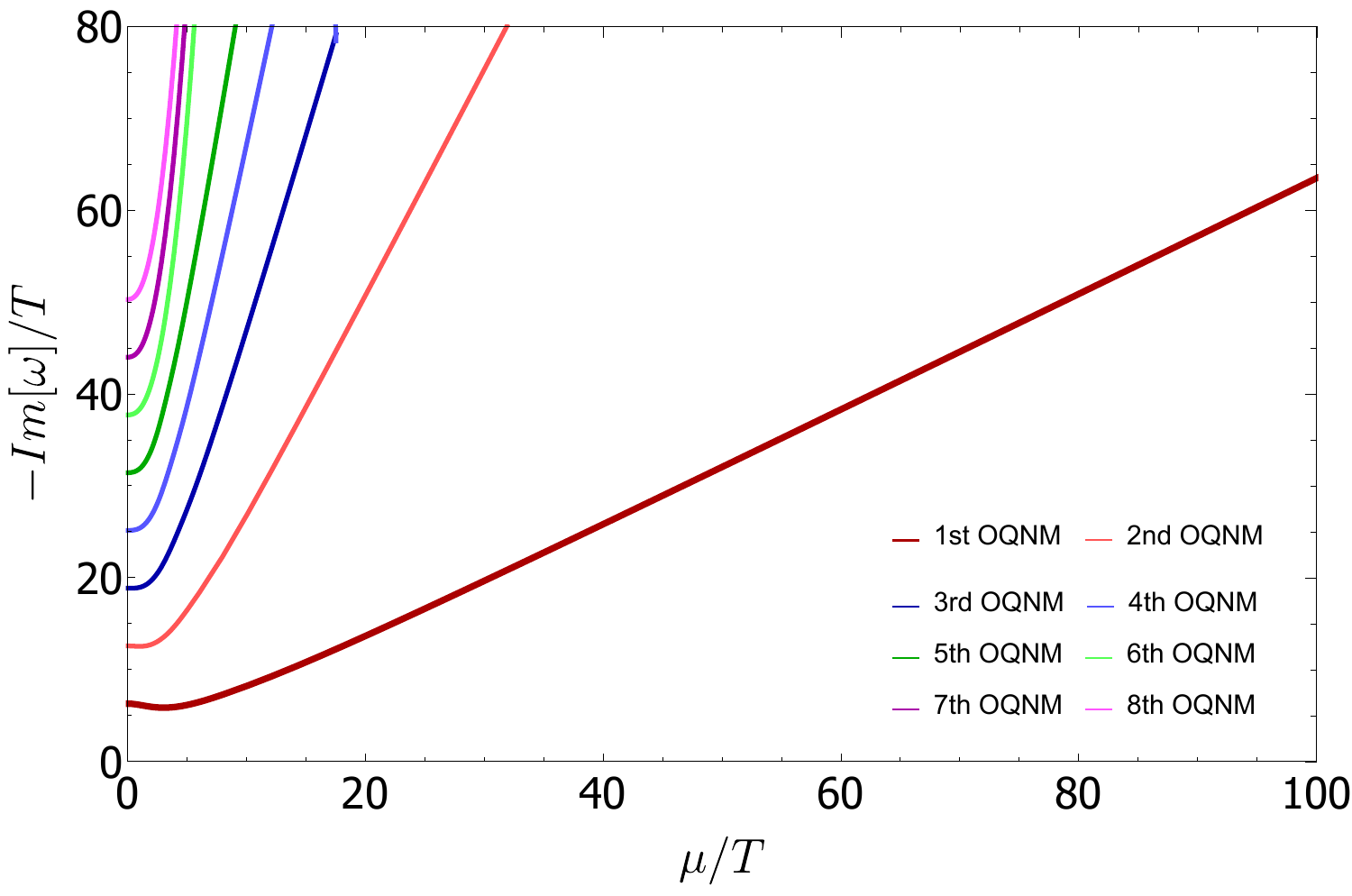}}
\subfigure[Imaginary part of the First 9 Purely Imaginary QNMs]
{\includegraphics[width=0.45\linewidth]{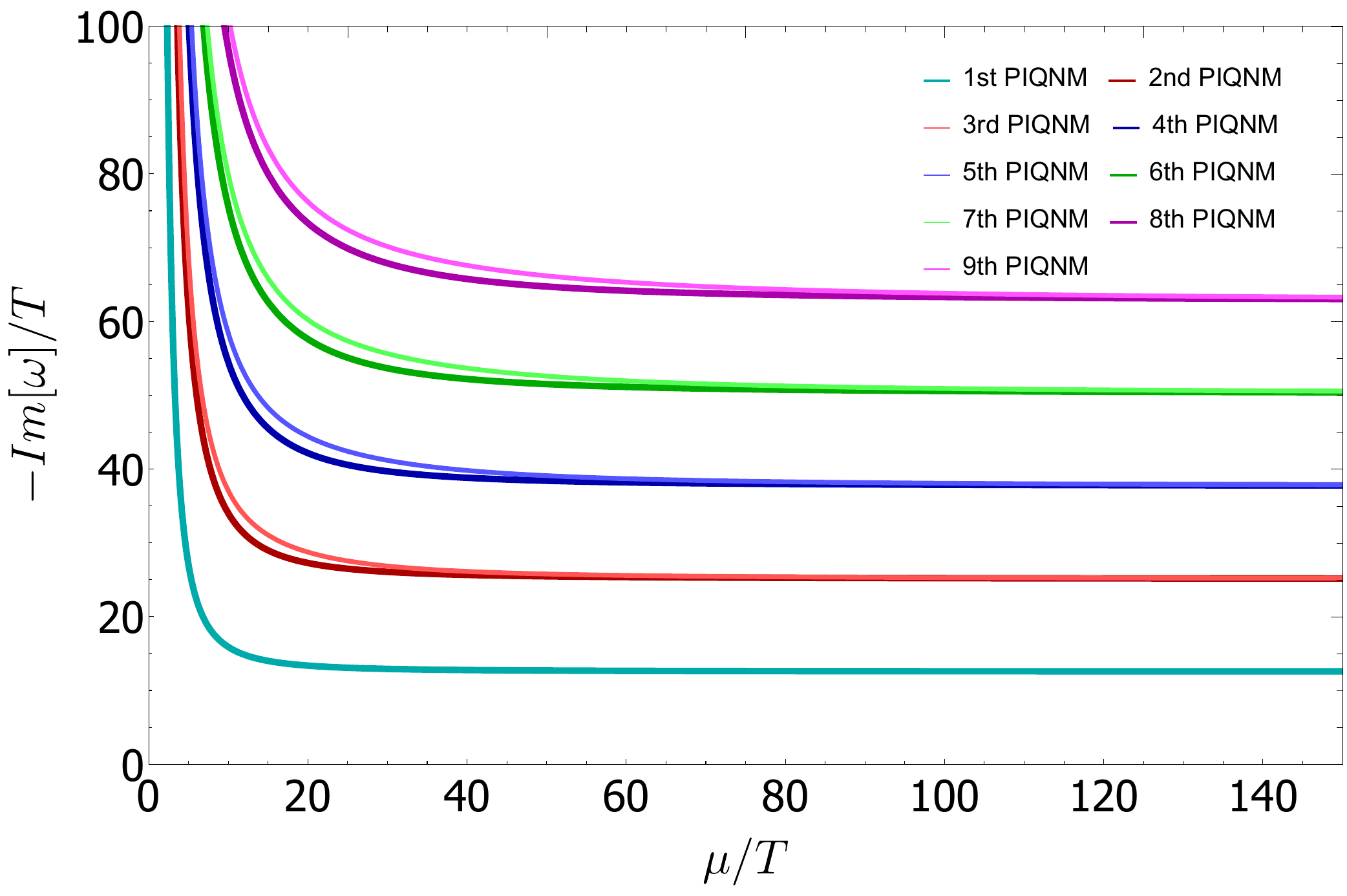}}
\subfigure[Comparison betwenn the 1RCBH and 2RCBH isolated PIQNMs.]
{\includegraphics[width=0.45\linewidth]{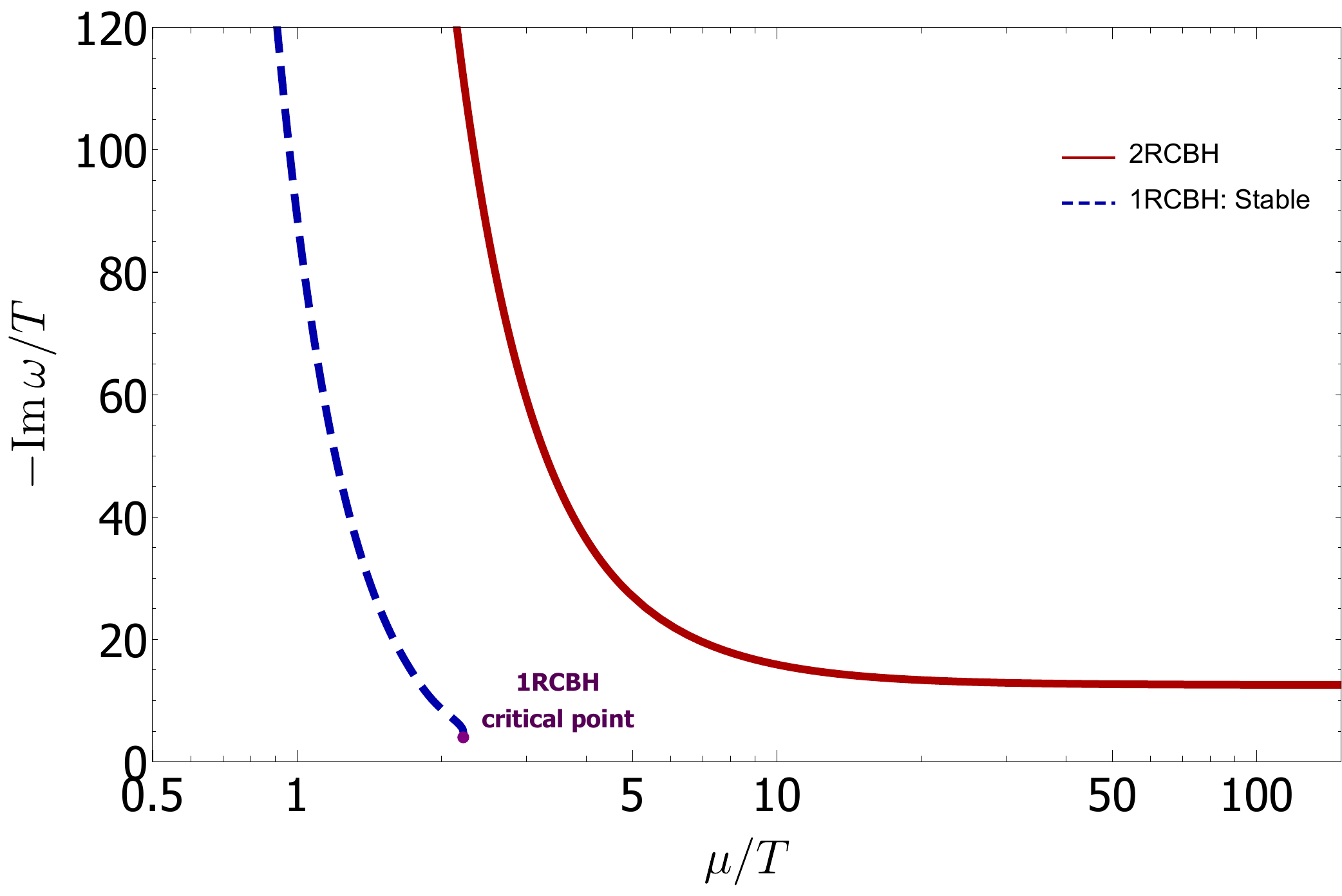}}
\caption{QNMs as functions of $\mu/T$ in the $SO(3)$ triplet channel of the 2RCBH model.}
\label{fig:PIQNMTri2RCBH}
\end{figure}

The behavior of both real and imaginary parts of the 2RCBH OQNMs for a much broader range of values of $\mu/T$ is shown in Figs.~\ref{fig:PIQNMTri2RCBH}(a) and \ref{fig:PIQNMTri2RCBH}(b), where, as before, we plot the first few OQNMs with positive real part. As in the quintuplet channel, the initial growths for the real and imaginary parts of the OQNMs turn into linearly increasing asymptotic behaviors for large values of $\mu/T$.

As in the quintuplet channel of the 2RCBH model, a similar structure of pairs of PIQNMs is also present in the triplet channel of the same model, as shown in Fig.~\ref{fig:PIQNMTri2RCBH}(c) --- see also Fig.~\ref{fig:StrQNMsTri}. However, differently from the quintuplet channel, the first PIQNM in the triplet channel of the 2RCBH model appears isolated, without a companion pole, what is more akin to the triplet channel of the 1RCBH model; on the other hand, in the triplet channel of the 1RCBH model, there is just a single PIQNM (besides the usual structure of mirrored OQNMs) \cite{Finazzo:2016psx}, differently from the further structure of pairs of PIQNMs observed here for the 2RCBH model. Interestingly, the first and isolated PIQNM is a distinctive feature of the triplet channel, not observed in the QNM spectra of the quintuplet and singlet channels.\footnote{It is also interesting to note that the emergence of purely imaginary quasinormal modes was also reported in the context of the holographic AdS$_4$-Reissner-Nordstrom model in \cite{Arean:2020eus}.}

Another difference with respect to the quintuplet channel of the 2RCBH model, is that in the triplet channel the moduli of the PIQNMs start at infinity, decreasing monotonically in magnitude as $\mu/T$ increases until asymptotically stabilizing at some constant values, without ever receding. Each purely imaginary QNM within a given pair of modes (apart from the first lone PIQNM) seems to converge to the same asymptotic value, while the different pairs are themselves evenly spaced apart. By calculating the PIQNMs of the triplet channel at $\mu/T=1500$ with $2500$ collocation points, it was possible to estimate a separation value of approximately $\delta^{(1)}_\textrm{3et}\approx 12.5665$ between the fundamental lone mode and the first excited pair of merged PIQNMs, and a value of approximately $\delta_\textrm{3et}\approx 12.567$ between the first and the second excited pairs of asymptotically merged PIQNMs. Again, the obtained separations are remarkably close to $4\pi$.

As before, in order to better visualize how the overall structure of QNMs is organized in the $SO(3)$ triplet channel of the 2RCBH model, we plotted their real parts versus their imaginary parts in Fig.~\ref{fig:StrQNMsTri} for six different key values of $\mu/T$. In this figure, colored triangles represent OQNMs, colored circles depict PIQNMs while the dashed horizontal line measures the imaginary value of the lowest OQNMs. From the sequence of snapshots in Fig.~\ref{fig:StrQNMsTri}, one can observe that a pair of OQNMs starts as the fundamental QNMs of the system, until they are overtaken at $\mu/T\approx 19.58$ by the first PIQNM that rapidly came from $-i\infty $. Subsequently, the now fundamental PIQNM tends to stabilize at a constant value. The structure of PIQNMs is organized, for large values of $\mu/T$, as a fundamental and isolated purely imaginary pole, plus pairs of modes with the same asymptotic imaginary value, all evenly spaced apart. Concerning the OQNMs, both their real and imaginary parts increase in magnitude as $\mu/T$ increases, following diagonal trajectories in the plane of complex eigenfrequencies illustrated in Fig.~\ref{fig:StrQNMsTri}.

Concerning the plots displayed in the present work for the $SO(3)$ triplet channel of the 2RCBH model, the basic configuration used in our numerical pseudospectral routine is made available in Table~\ref{TabTrip}.
\begin{table}[]
\begin{tabular}{|c|c|c|c}
\cline{1-3}
$\mu/T$ & \# of collocation points & \# of data points &  \\ \cline{1-3}
{[}0, 5{]} & 150 & 200 &  \\
(5, 20{]} & 200 & 50 &  \\
{(}20,50] & 300 & 25 & \multicolumn{1}{l}{} \\
{(}50, 100{]} & 450 & 20 & \multicolumn{1}{l}{} \\
{(}100, 150{]} & 650 & 15 &  \\ \cline{1-3}
\end{tabular}
\caption{Some numerical details used to calculate the QNMs of the $SO(3)$ triplet channel.}
\label{TabTrip}
\end{table}

\begin{figure}
\centering  
\subfigure[Equilibration time]{\includegraphics[width=0.45\linewidth]{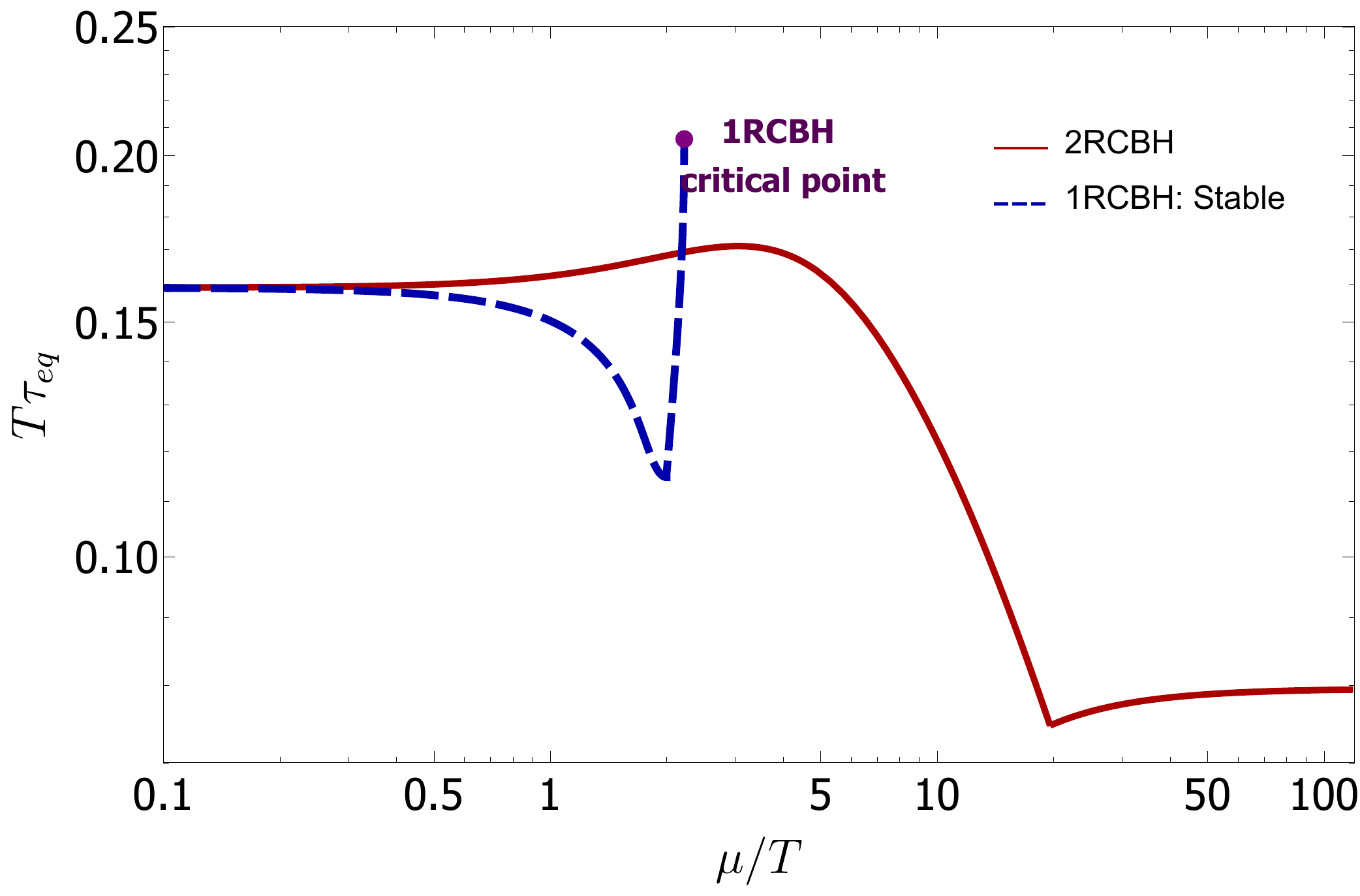}}
\subfigure[First $(\mu/T)$-derivative of the equilibration time]{\includegraphics[width=0.45\linewidth]{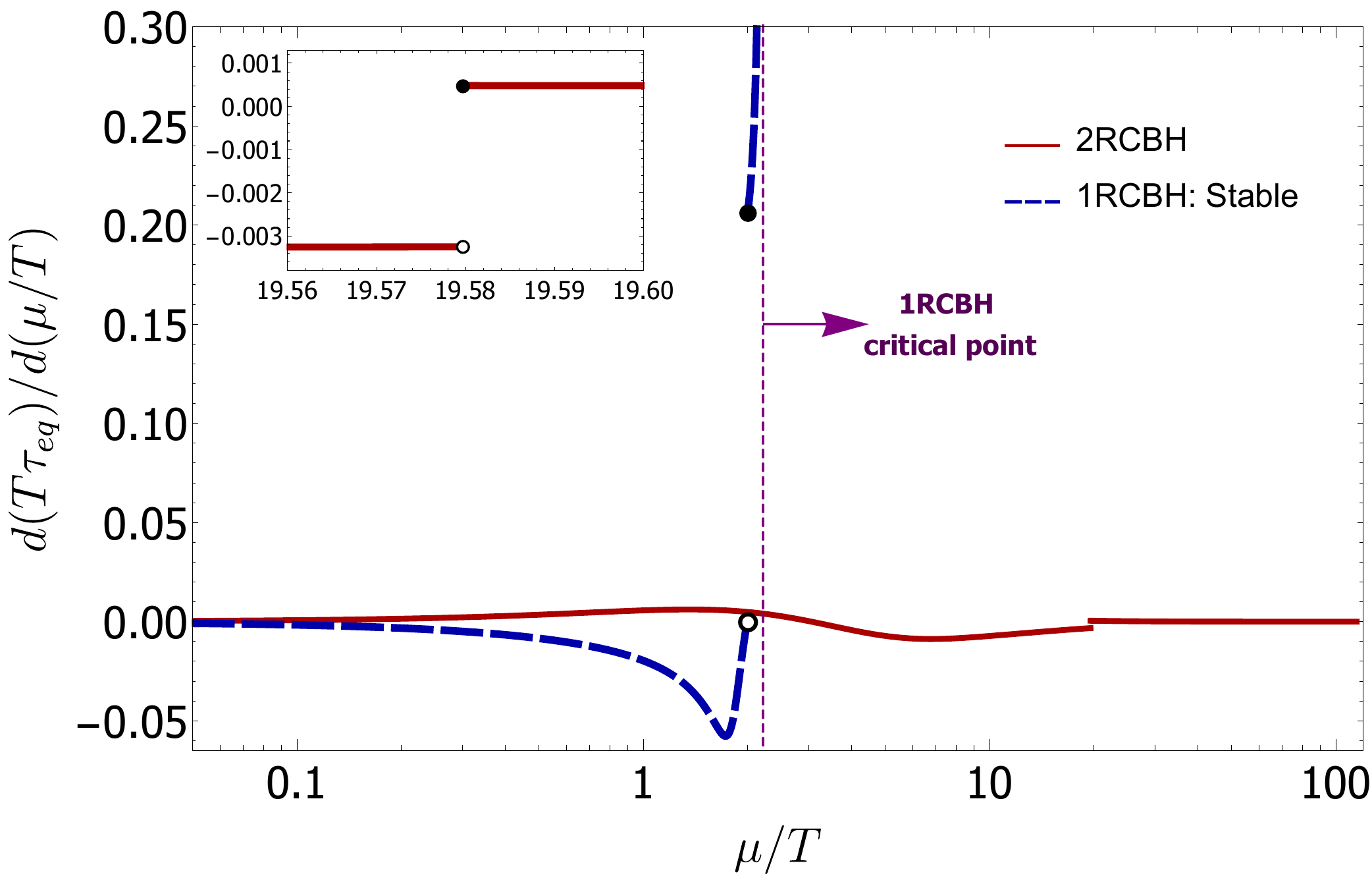}}
\caption{Characteristic equilibration time and its first derivative with respect to $\mu/T$ for the $SO(3)$ triplet channel of the 1RCBH and 2RCBH models. The inset plot on the right shows the discontinuity in the first $(\mu/T)$-derivative of the equilibration time of the 2RCBH model around $\mu/T\approx 19.58$, where the first PIQNM becomes the lowest QNM of the triplet channel.}
\label{fig:eqtime&derTri}
\end{figure}

In regard to the equilibration time and its first $(\mu/T)$-derivative for the QNMs in the $SO(3)$ triplet channel, Fig.~\ref{fig:eqtime&derTri} displays a comparison between the 1RCBH and the 2RCBH models. For the 1RCBH model, the main qualitative difference with respect to the quintuplet channel is the discontinuity in the first $(\mu/T)$-derivative of the equilibration time of the triplet channel due to the lone PIQNM becoming the fundamental QNM of this channel, while, as in the quintuplet channel, at the critical point of the 1RCBH model the first $(\mu/T)$-derivative of the equilibration time of the triplet channel diverges with a critical exponent of $\theta=1/2$ \cite{Finazzo:2016psx}. In what concerns the 2RCBH model, the equilibration time has an abrupt change at $\mu/T\approx 19.58$, where the first and isolated PIQNM becomes the lowest QNM of the triplet channel, with a corresponding discontinuity in the first $(\mu/T)$-derivative of the equilibration time of this channel --- see also Fig.~\ref{fig:StrQNMsTri}.

\section{Singlet Channel}
\label{sec:6}

\subsection{Field Equation for the Perturbation}

\begin{figure}
\centering  
\subfigure[1st mode]{\includegraphics[width=0.43\linewidth]{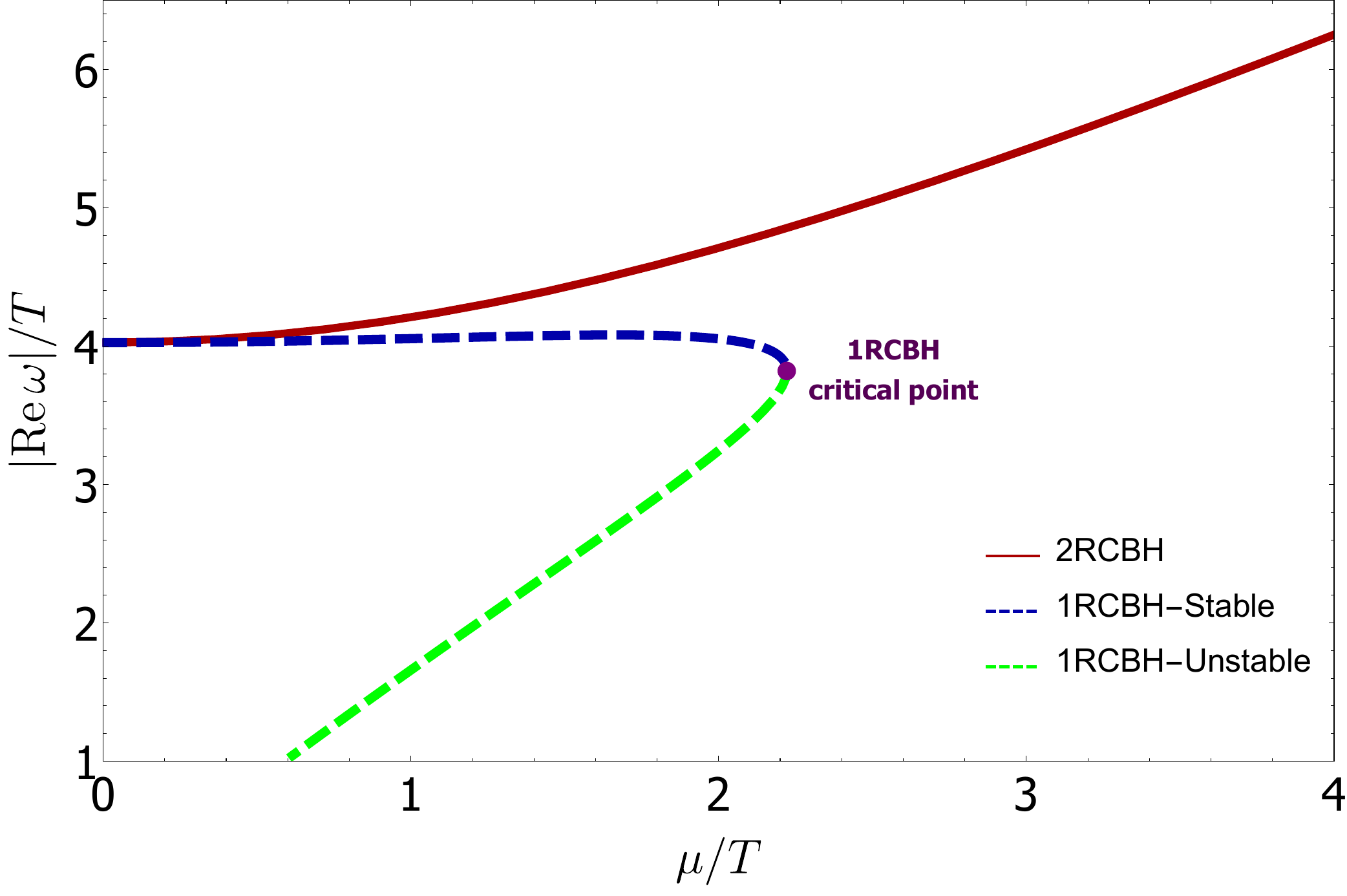}}
\subfigure[2nd mode]{\includegraphics[width=0.45\linewidth]{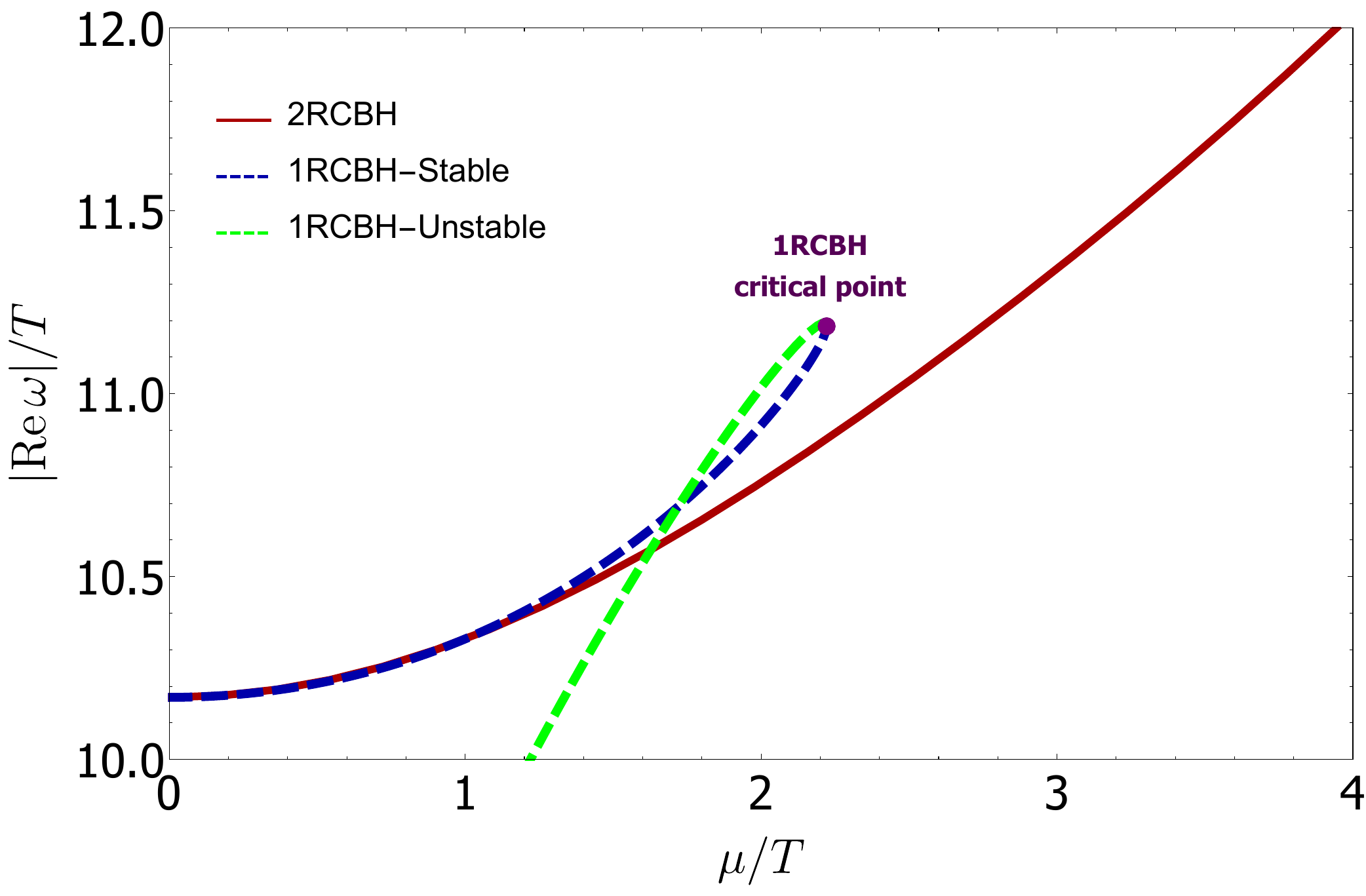}}
\subfigure[3rd mode]{\includegraphics[width=0.45\linewidth]{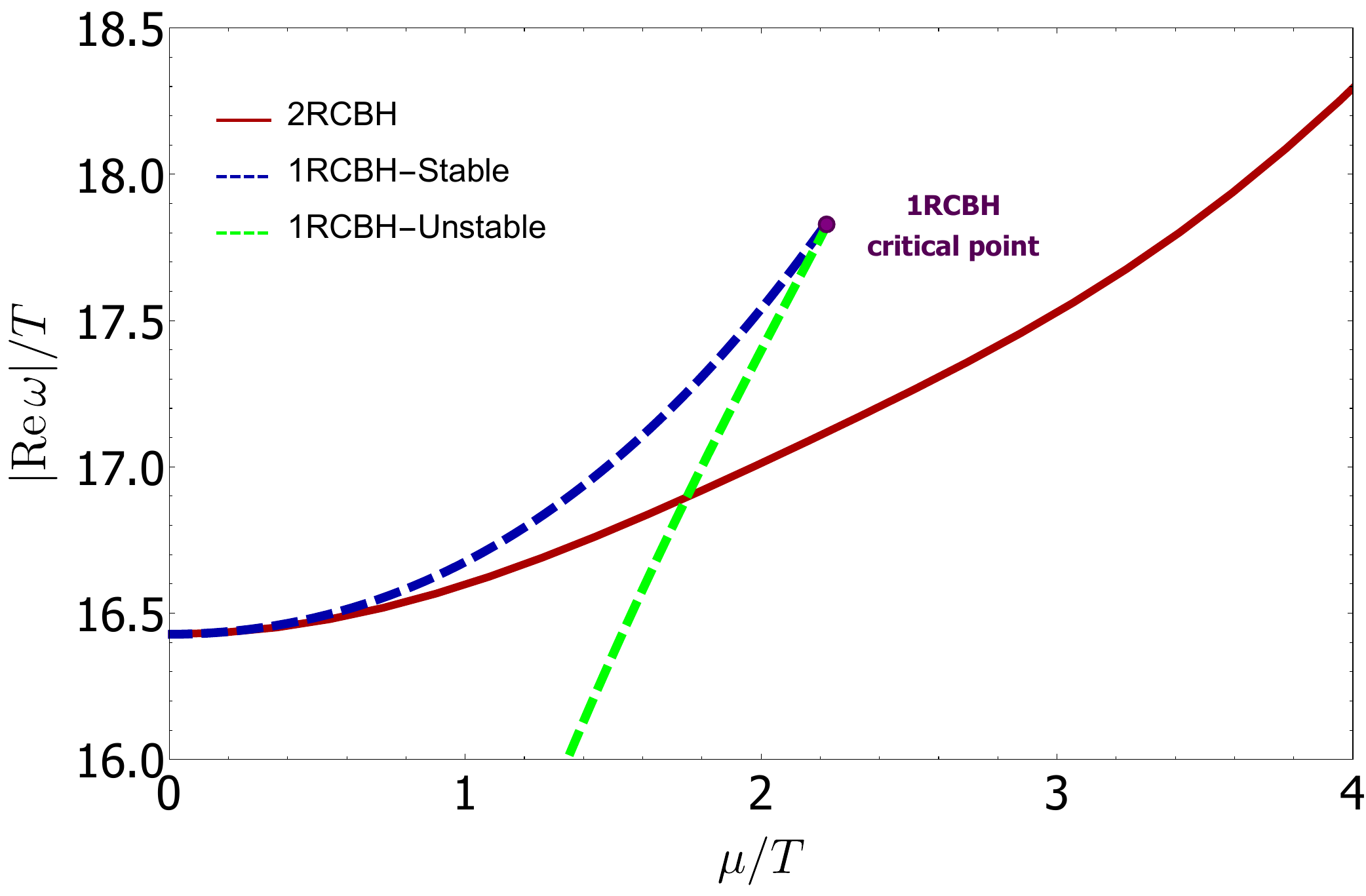}}
\subfigure[4th mode]{\includegraphics[width=0.45\linewidth]{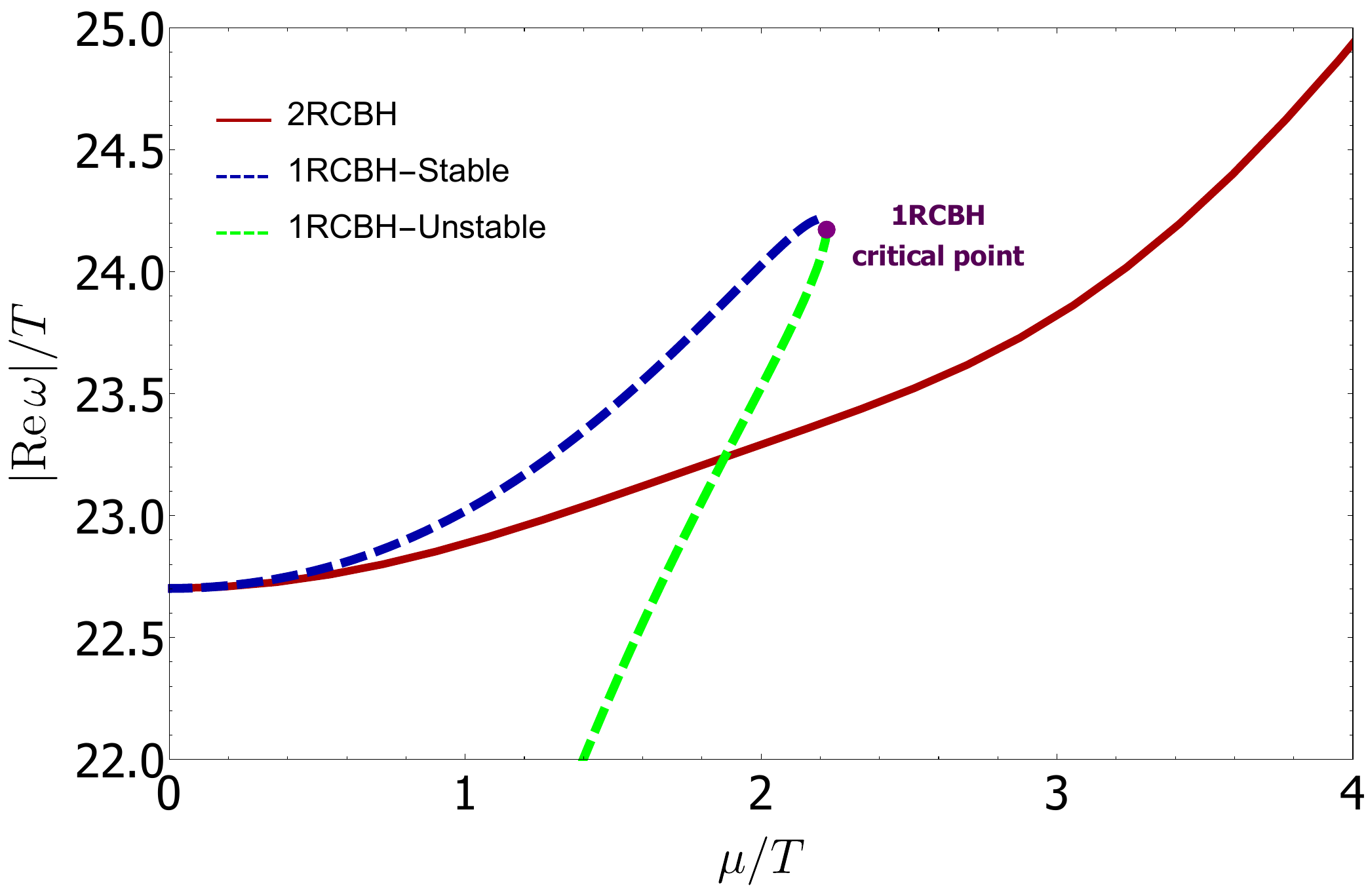}}
\caption{Real part for the first four quasinormal modes for the 1RCBH and 2RCBH models in the singlet channel.} 
\label{fig:QNMSinRe}
\end{figure}
\begin{figure}
\centering
\subfigure[1st mode]{\includegraphics[width=0.45\linewidth]{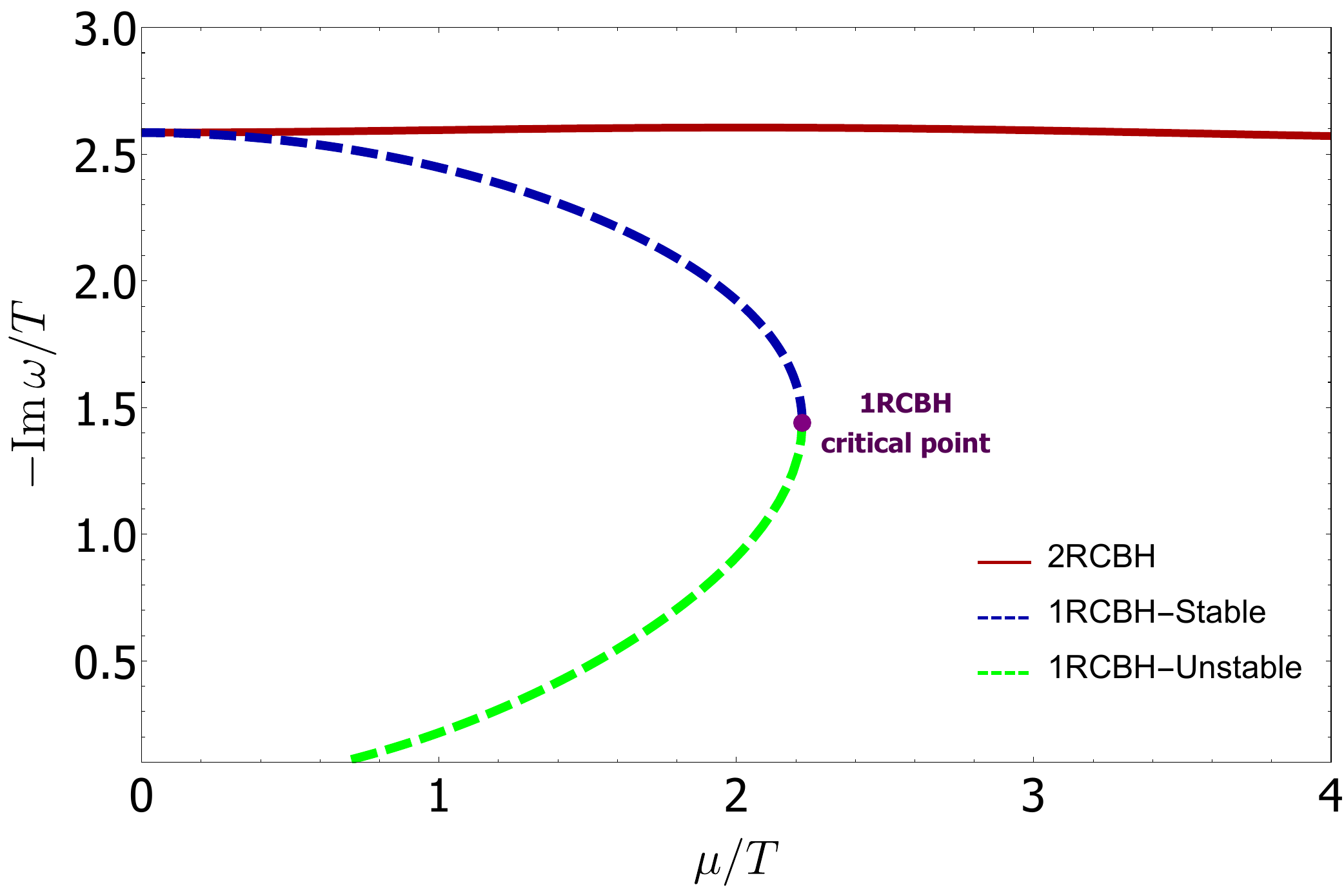}}
\subfigure[2nd mode]{\includegraphics[width=0.45\linewidth]{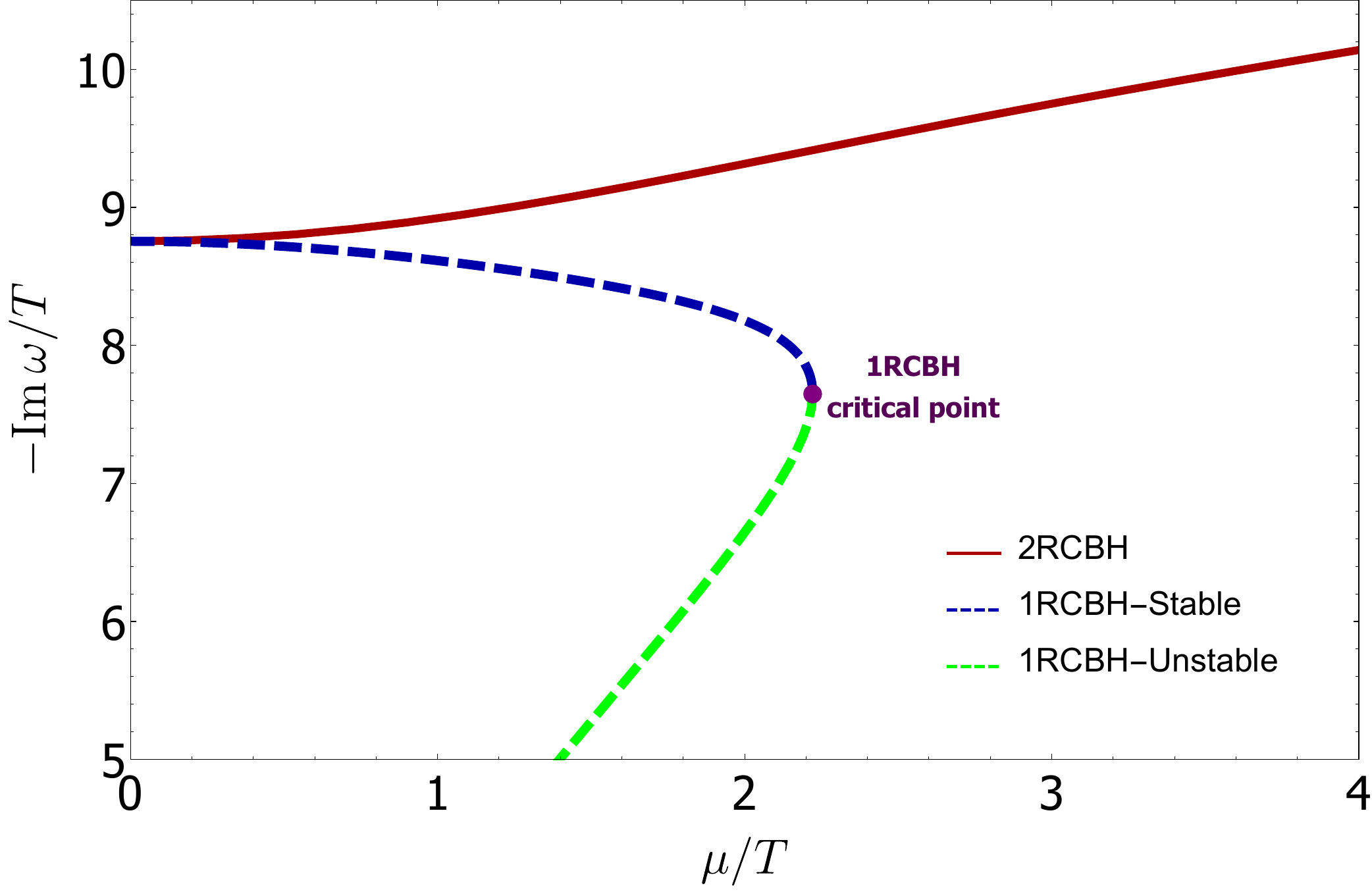}}
\subfigure[3rd mode]{\includegraphics[width=0.45\linewidth]{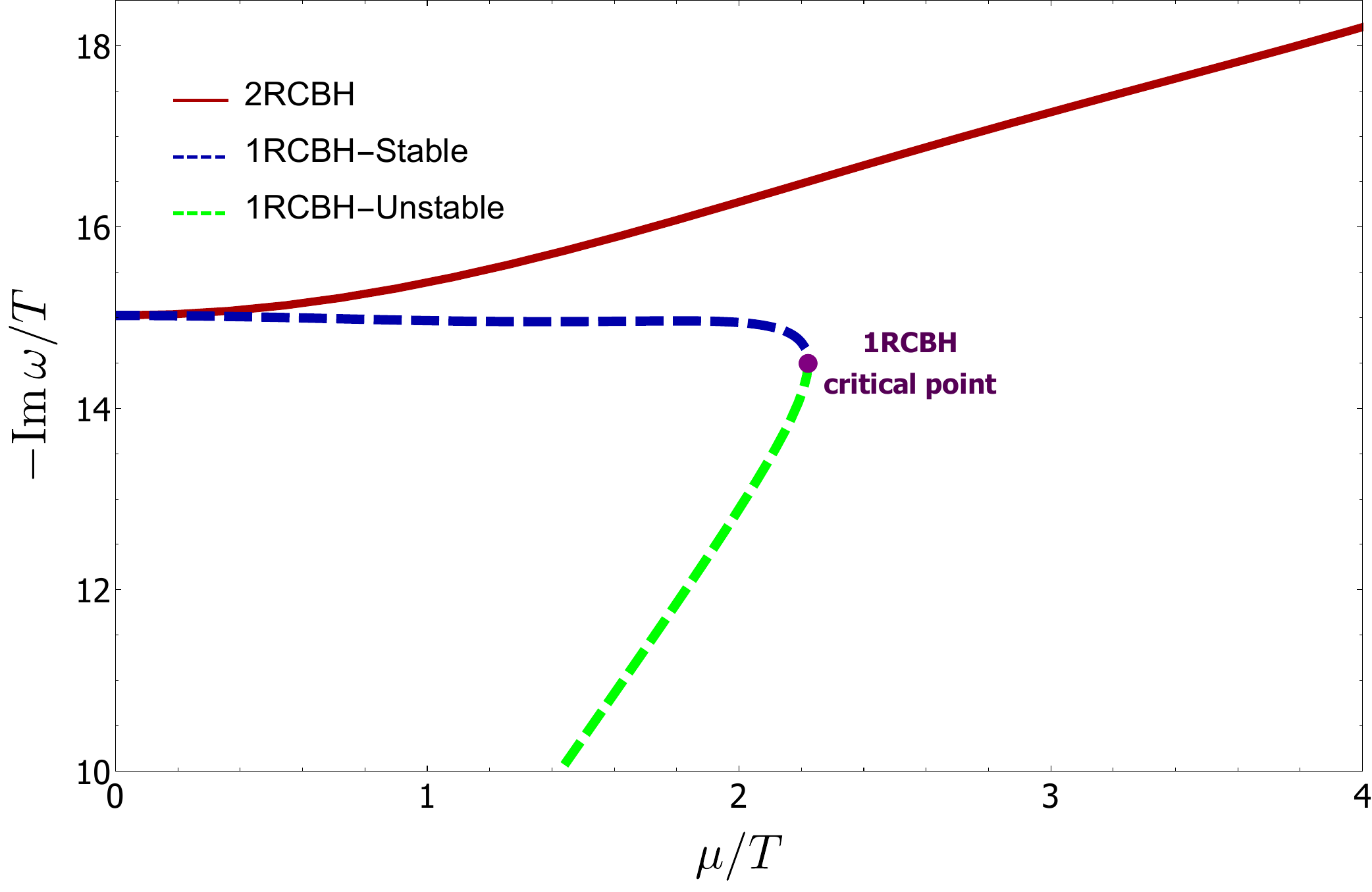}}
\subfigure[4th mode]{\includegraphics[width=0.45\linewidth]{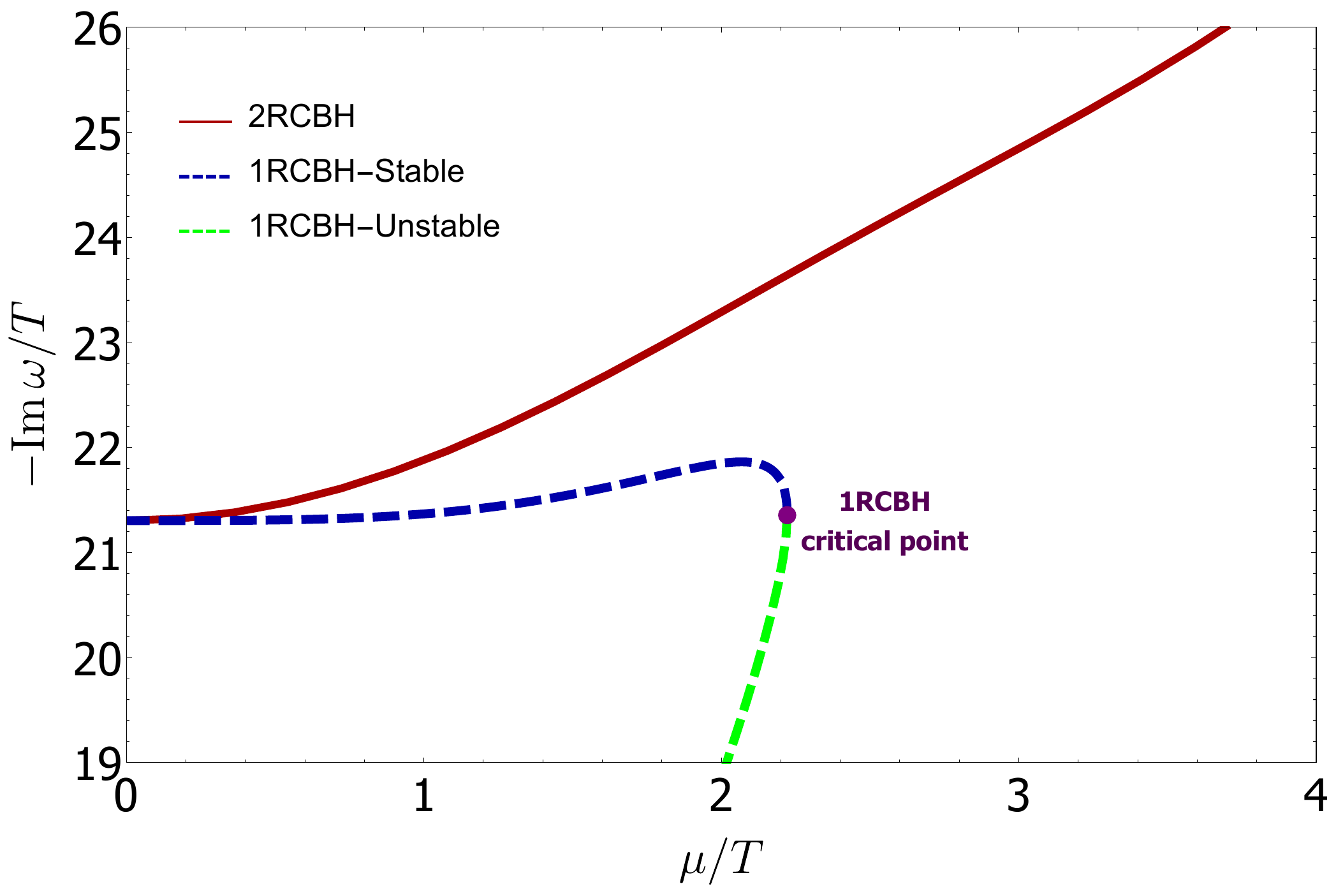}}
\caption{Imaginary part for the first four quasinormal modes for the 1RCBH and 2RCBH models in the singlet channel.}
\label{fig:QNMSinIm}
\end{figure}

As discussed before, in the case of the $SO(3)$ singlet channel, the relevant gauge and diffeomorphism invariant perturbation corresponds to a combination involving the dilaton perturbation $\varphi$ and the trace of the spatial part of the metric field perturbation $h_{ij}$,
\begin{equation}
    \mathcal{S}=\varphi-\frac{\phi'}{2A'}\frac{1}{3}\left(h_{xx}+h_{yy}+h_{zz}\right).
\end{equation}

The linearized equation of motion for the $\mathcal{S}$-perturbation was derived in Ref.~\cite{DeWolfe:2011ts} and takes the form,
\begin{equation}
    \mathcal{S}''+\left(4A'-B'+\frac{h'}{h}\right)\mathcal{S}'+\left(\frac{e^{2(B-A)}}{h^2}\omega^2+\Sigma(r)\right)\mathcal{S}=0,
\end{equation}
where $\Sigma(r)$ is given by,
\begin{align}
    \nonumber \Sigma(r)\equiv \frac{e^{-2A}}{18f(\phi)h^2A'^2}\Bigg[&-18hA'^2\partial_\phi f(\phi)^2\Phi'^2-e^{2A}f(\phi)h^2\phi'^4+6f(\phi)hA'\phi'\Bigg(-2e^{2(A+B)}\partial_\phi V(\phi)+e^{2A}h'\phi'+\partial_\phi f(\phi)\Phi'^2\Bigg)\\
    &+3f(\phi)A'^2\Bigg(8e^{2A}h^2\phi'^2+3h\Phi'^2\partial_\phi^2 f(\phi)-6e^{2(A+B)}h \partial_\phi^2 V(\phi)\Bigg)\Bigg].
\end{align}

In terms of the infalling EF coordinates defined in Eq.~\eqref{eq:EEcoordinate}, one can rewrite the last equation as \cite{Critelli:2017euk},
\begin{align}
 \nonumber S''+ &\left[ 4 A'-B'-\frac{2 i \omega}{h}  e^{B-A}+\frac{h'}{h}\right]S'+\frac{e^{-2 A}}{18 h A'^2 f(\phi )}  \Bigg\{-18 A'^2 \Phi '^2 \partial_\phi f(\phi )^2\\
 \nonumber &+f(\phi ) \Bigg[3 A'^2 \Bigg(-6 e^{2 (A+B)} \partial_\phi^2 V(\phi )+8 e^{2 A} h \phi '^2+3 \Phi '^2 \partial_\phi^2 f(\phi )\Bigg)\\
  \label{eq:EOMSingletEE}&+6 A' \phi ' \Bigg(e^{2 A} \left(h' \phi '-2 e^{2 B} \partial_\phi V(\phi )\right)+\Phi '^2 \partial_\phi f(\phi )\Bigg)-54 i \omega  e^{A+B} A'^3-e^{2 A} h \phi '^4\Bigg]\Bigg\}S=0.
\end{align}

\begin{figure}[t]
\centering  
\subfigure[$\mu/T=0$]{\includegraphics[width=0.45\linewidth]{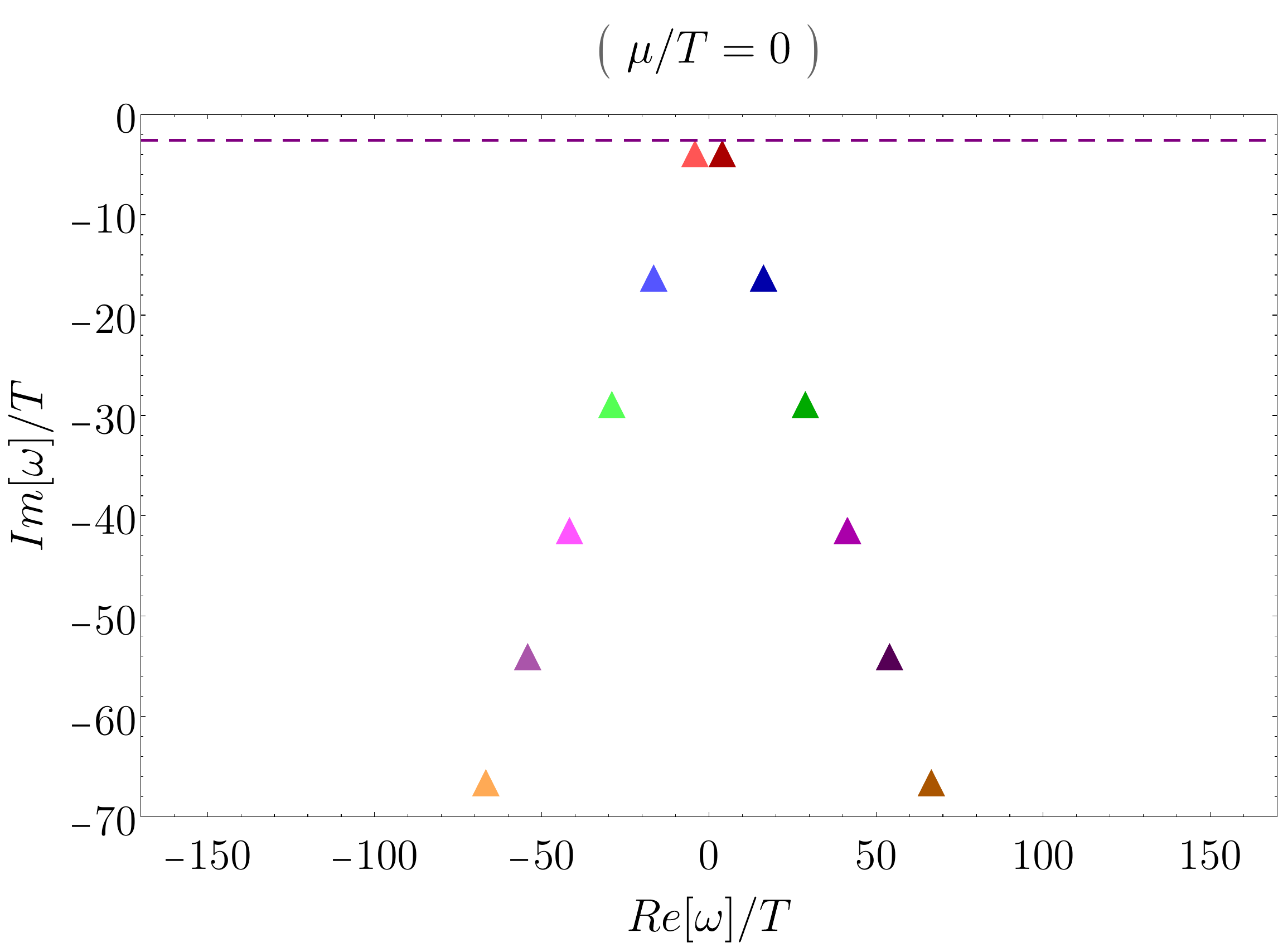}}
\subfigure[$\mu/T=5.0$]{\includegraphics[width=0.45\linewidth]{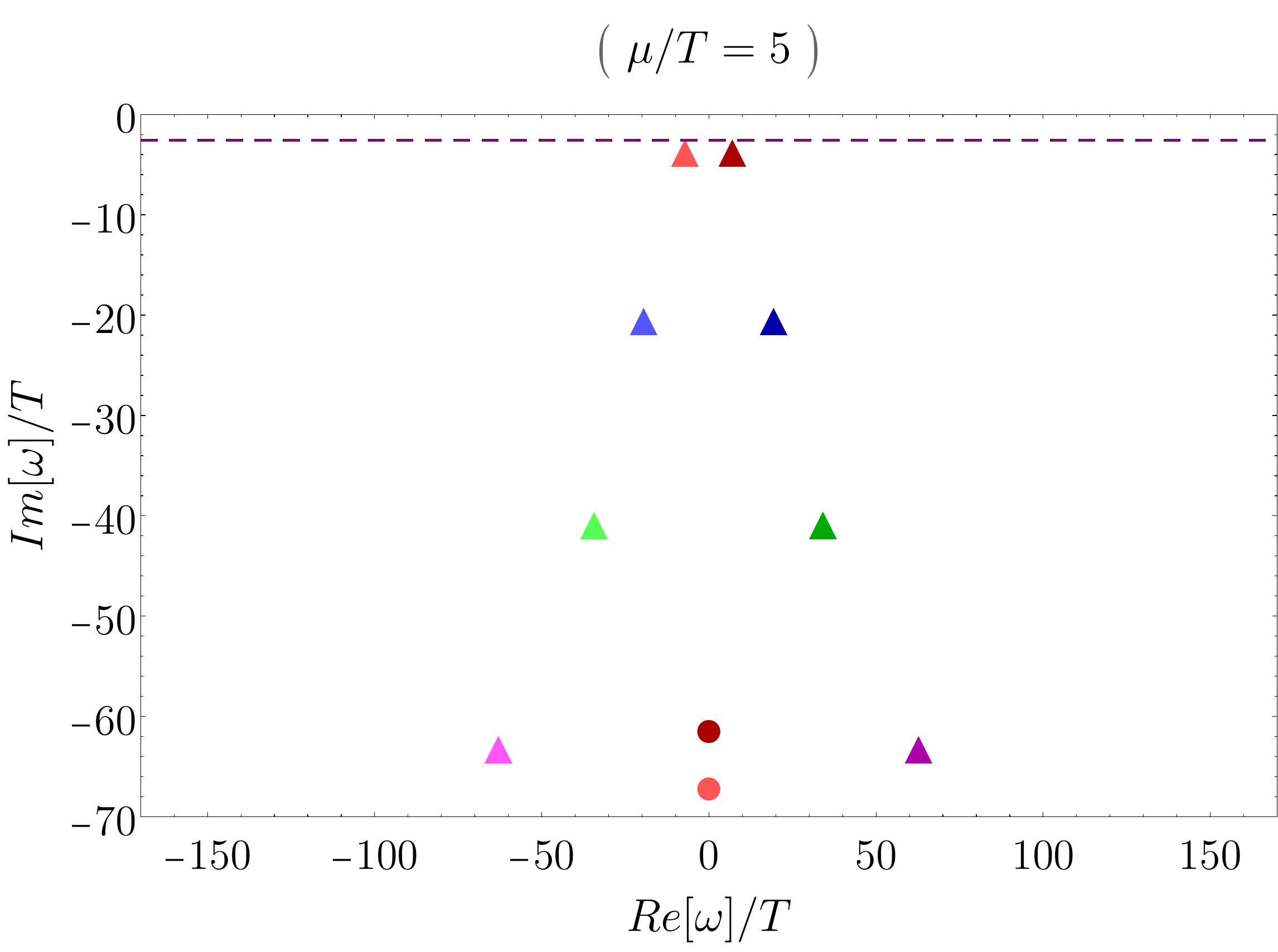}}
\subfigure[$\mu/T=10$]{\includegraphics[width=0.45\linewidth]{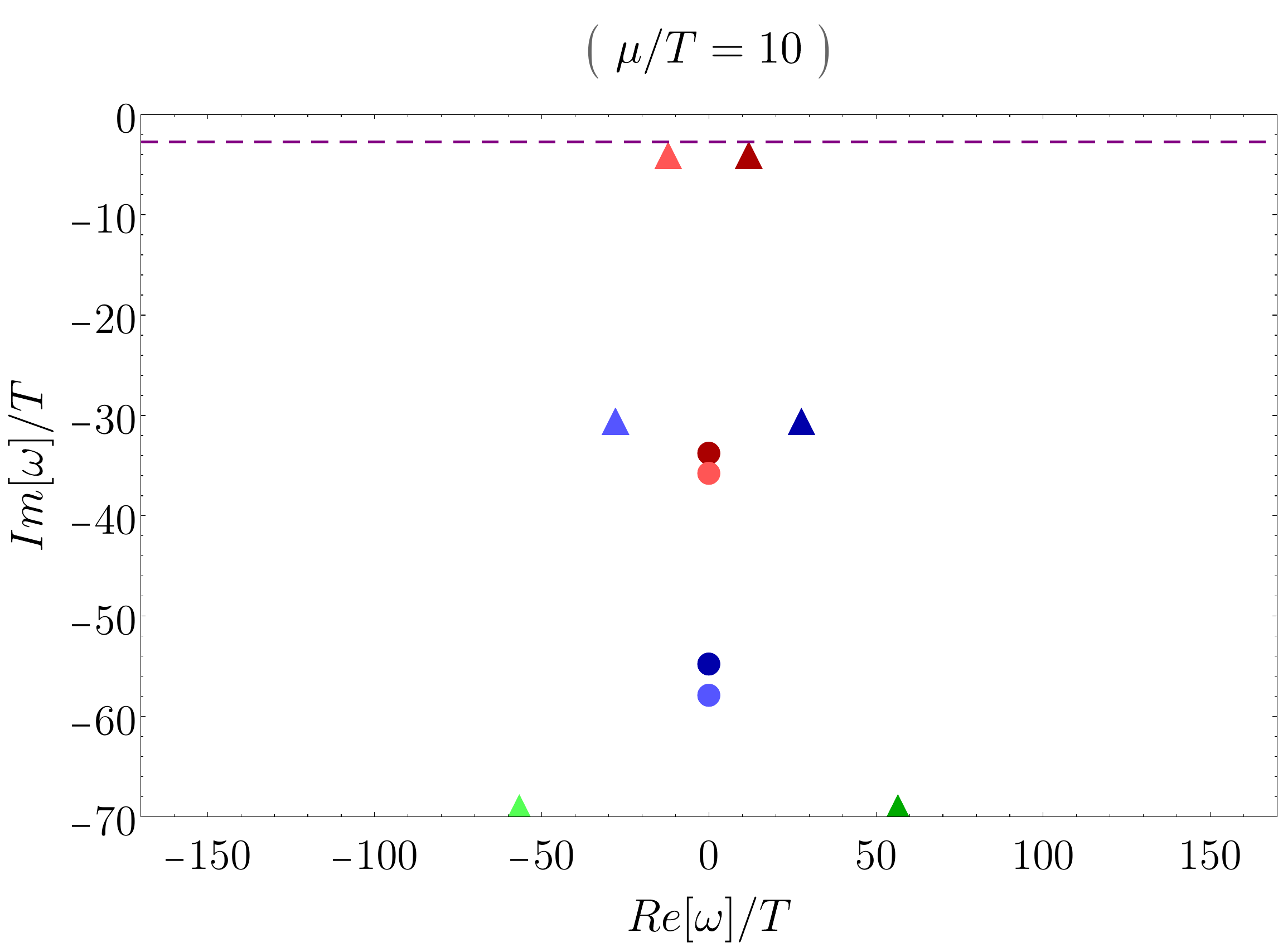}}
\subfigure[$\mu/T=30$]{\includegraphics[width=0.45\linewidth]{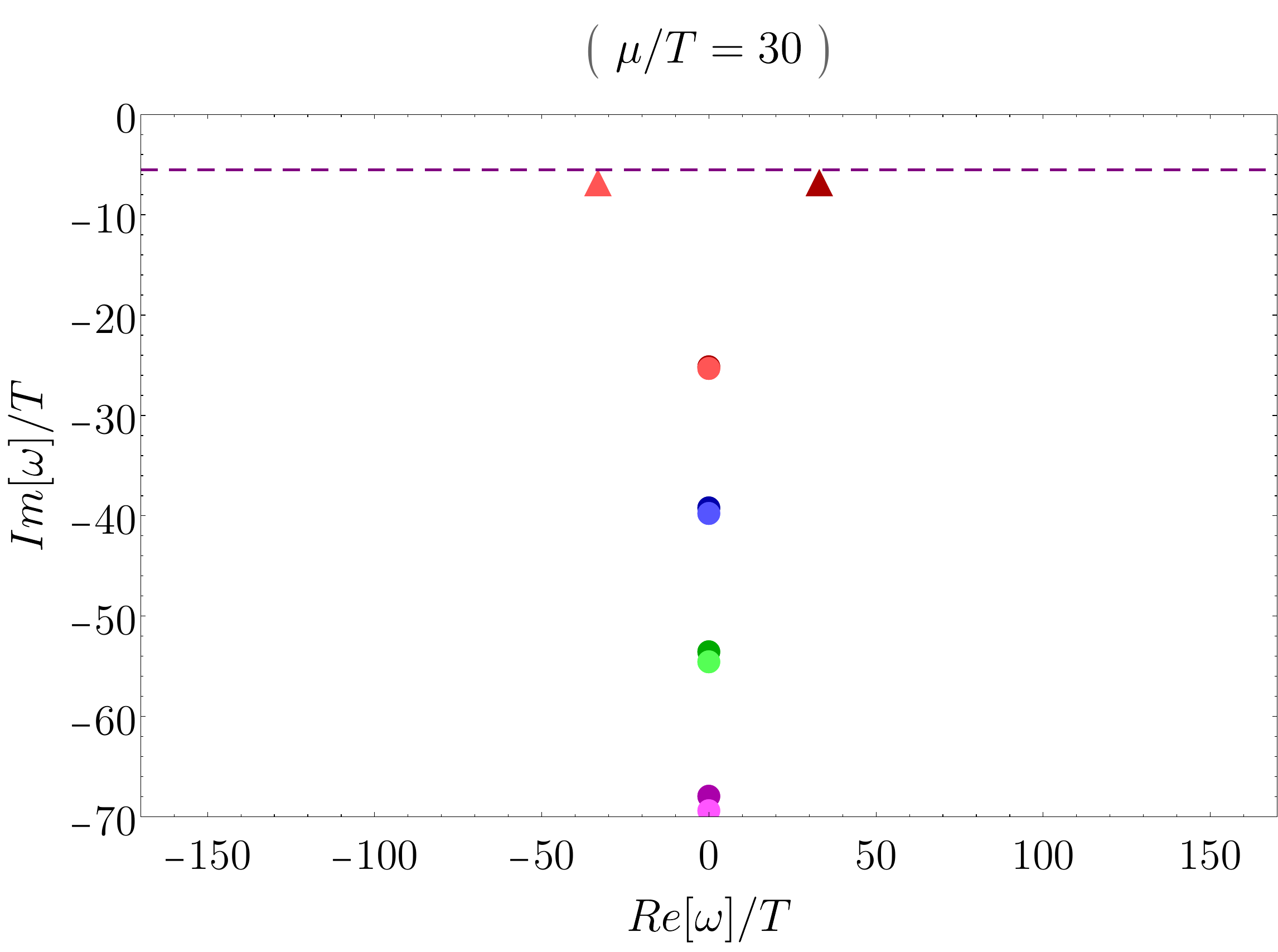}}
\subfigure[$\mu/T=133.68$]{\includegraphics[width=0.45\linewidth]{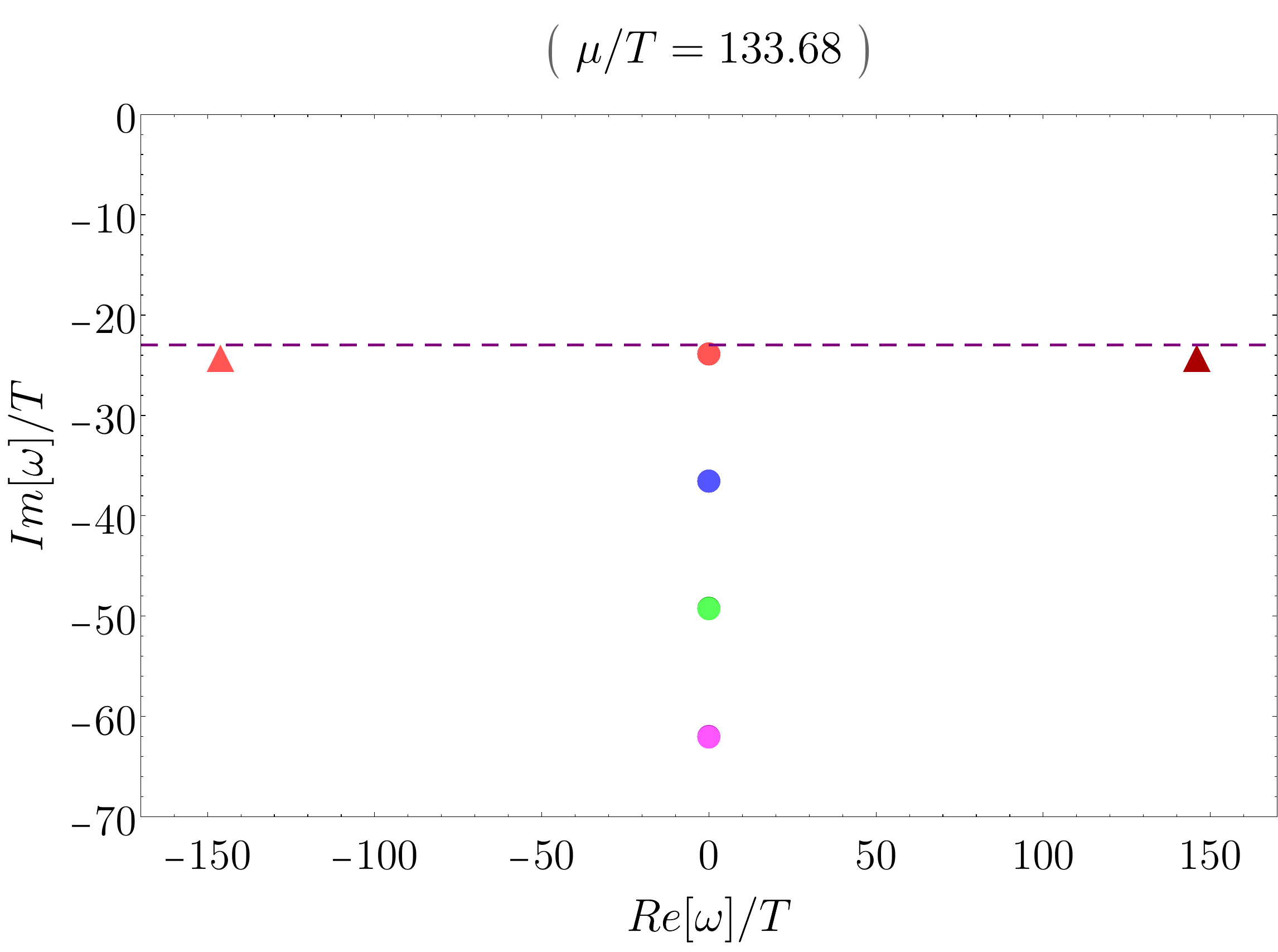}}
\subfigure[$\mu/T=150$]{\includegraphics[width=0.45\linewidth]{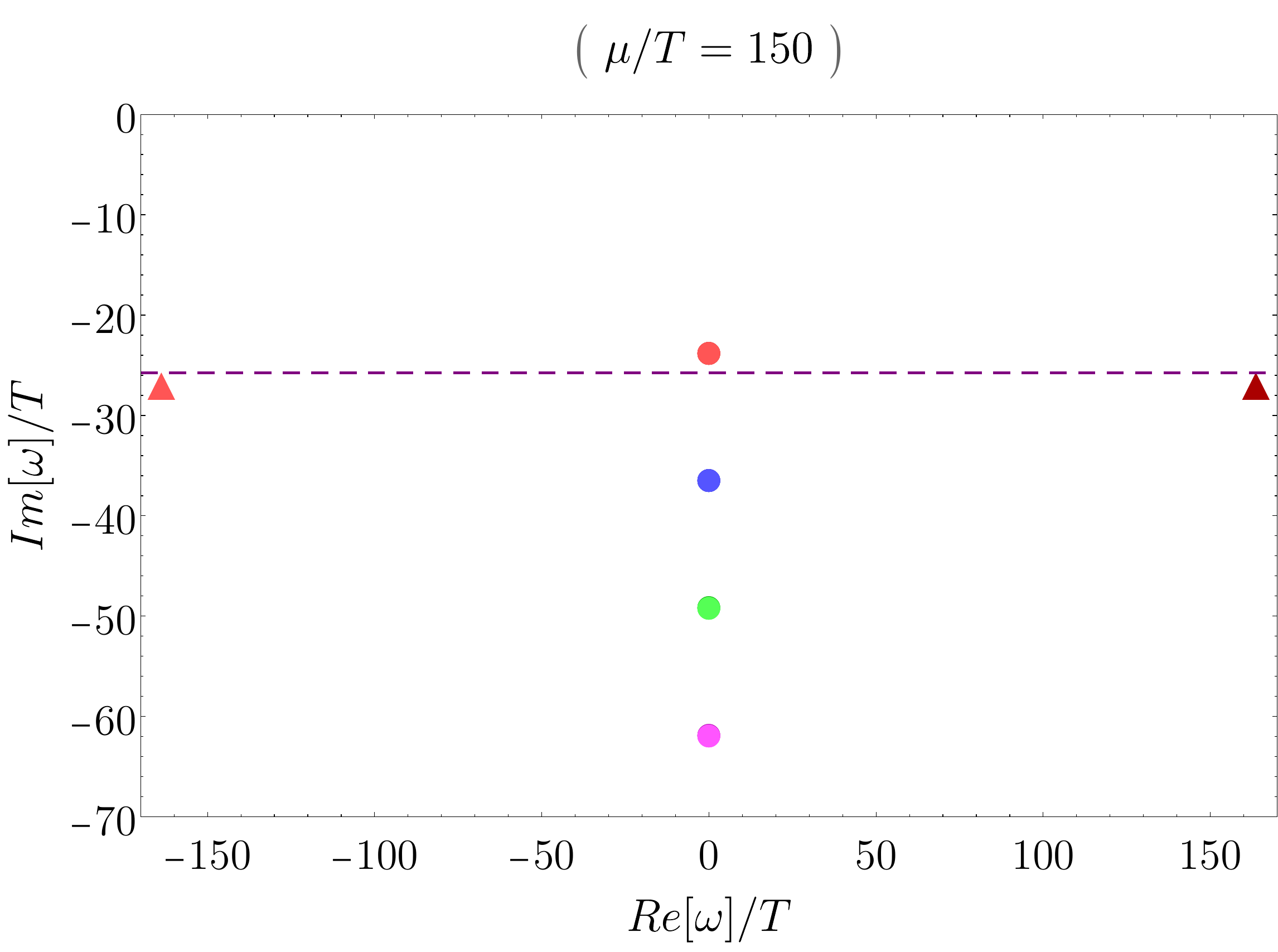}}
\caption{QNMs snapshots for 6 different values of $\mu/T$ in the $SO(3)$ singlet channel of the 2RCBH model. In the figure, colored triangles represent OQNMs (ordinary QNMs with nonzero real part), colored circles depict PIQNMs (purely imaginary QNMs) while the dashed horizontal line measures the imaginary part of the lowest OQNMs (i.e., the OQNMs with lowest imaginary part in modulus).}
\label{fig:StrQNMsSin}
\end{figure}
By substituting the background solutions~\eqref{eq:AnsatzAll} into Eq.~\eqref{eq:EOMSingletEE}, one obtains for the 1RCBH model the following expression,
\begin{align}
    \nonumber &S_1''+\left(\frac{Q^2 u^2 \left(u^2+1\right)+r_H^2 \left(u^4+3\right)}{u \left(u^2-1\right) \left(u^2 \left(Q^2+r_H^2\right)+r_H^2\right)}-\frac{2 i r_H \omega_1 \sqrt{\frac{Q^2 u^2}{r_H^2}+1}}{\left(u^2-1\right) \left(u^2 \left(Q^2+r_H^2\right)+r_H^2\right)}\right)S_1'\\
    &+ \left(\frac{i \omega_1 \left(2 Q^2 u^2+3 r_H^2\right)}{r_H u \left(u^2-1\right) \sqrt{\frac{Q^2 u^2}{r_H^2}+1} \left(u^2 \left(Q^2+r_H^2\right)+r_H^2\right)}-\frac{12 r_H^2 \left(-4 Q^4 u^4 \left(u^2-1\right)+2 Q^2 r_H^2 u^2 \left(3-2 u^4\right)+3 r_H^4\right)}{\left(u^2-1\right) \left(u^2 \left(Q^2+r_H^2\right)+r_H^2\right) \left(2 Q^2 u^3+3 r_H^2 u\right)^2}\right)S_1=0,
\end{align}
while for the 2RCBH model, one has,
\begin{align}
    \nonumber &S_2''+ \left(\frac{2 Q^2 u^2 \left(u^2+1\right)+r_H^2 \left(u^4+3\right)}{u \left(u^2-1\right) \left(u^2 \left(2 Q^2+r_H^2\right)+r_H^2\right)}-\frac{2 i \omega_2 \left(Q^2 u^2+r_H^2\right)}{r_H \left(u^2-1\right) \left(u^2 \left(2 Q^2+r_H^2\right)+r_H^2\right)}\right)S_2'\\
    &+ \left(\frac{12 \left(Q^6 u^6+Q^4 r_H^2 u^4 \left(4 u^2-1\right)+Q^2 r_H^4 u^2 \left(2 u^4-3\right)-3 r_H^6\right)}{\left(u^2-1\right) \left(u^2 \left(2 Q^2+r_H^2\right)+r_H^2\right) \left(Q^2 u^3+3 r_H^2 u\right)^2}+\frac{i \omega_2 \left(Q^2 u^2+3 r_H^2\right)}{r_H u \left(u^2-1\right) \left(u^2 \left(2 Q^2+r_H^2\right)+r_H^2\right)}\right)S_2=0.
\end{align}

\begin{figure}
\centering  
\subfigure[Real part of the First 8 Ordinary QNMs]{\includegraphics[width=0.45\linewidth]{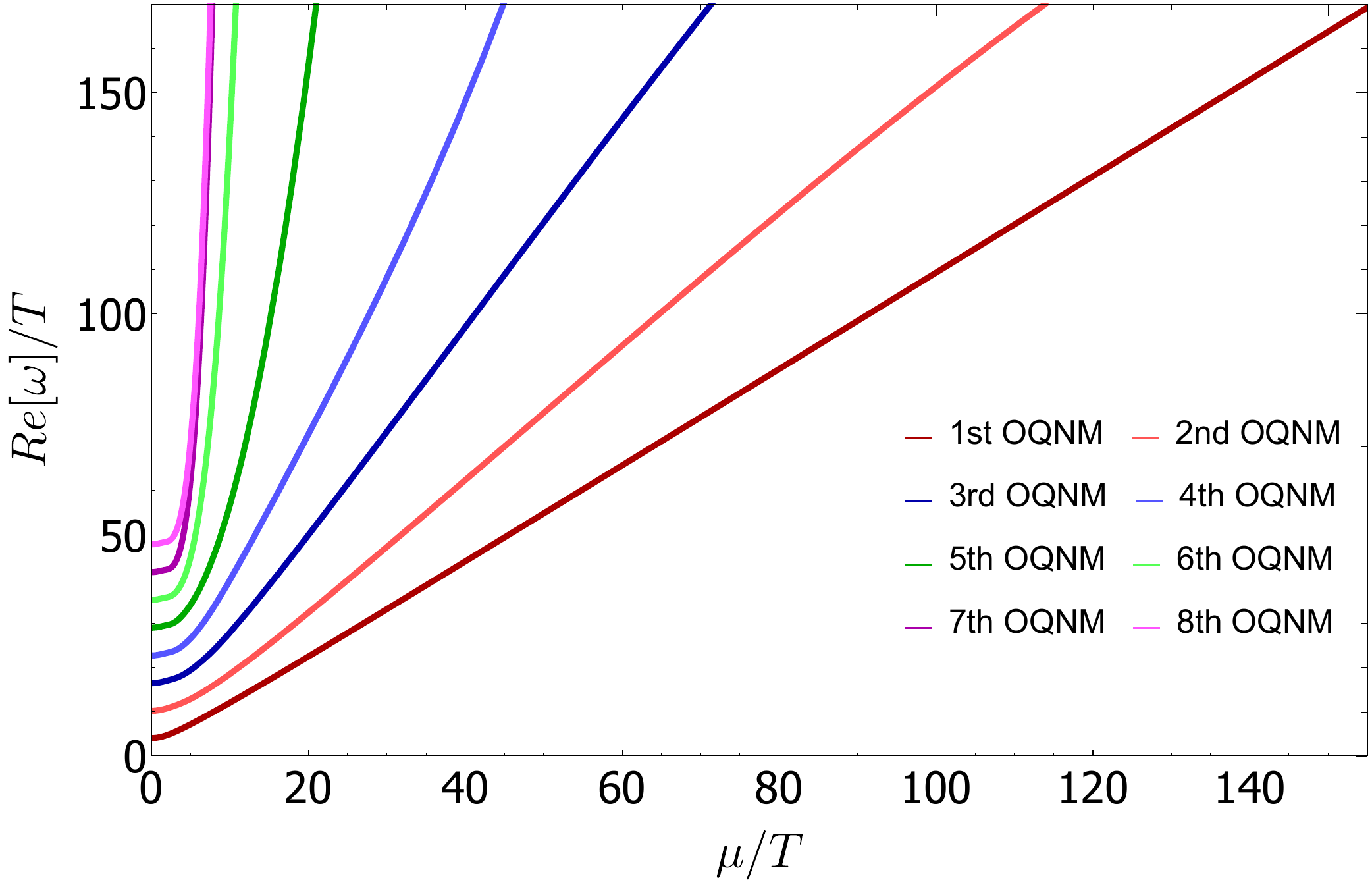}}
\subfigure[Imaginary part of the First 8 Ordinary QNMs]{\includegraphics[width=0.45\linewidth]{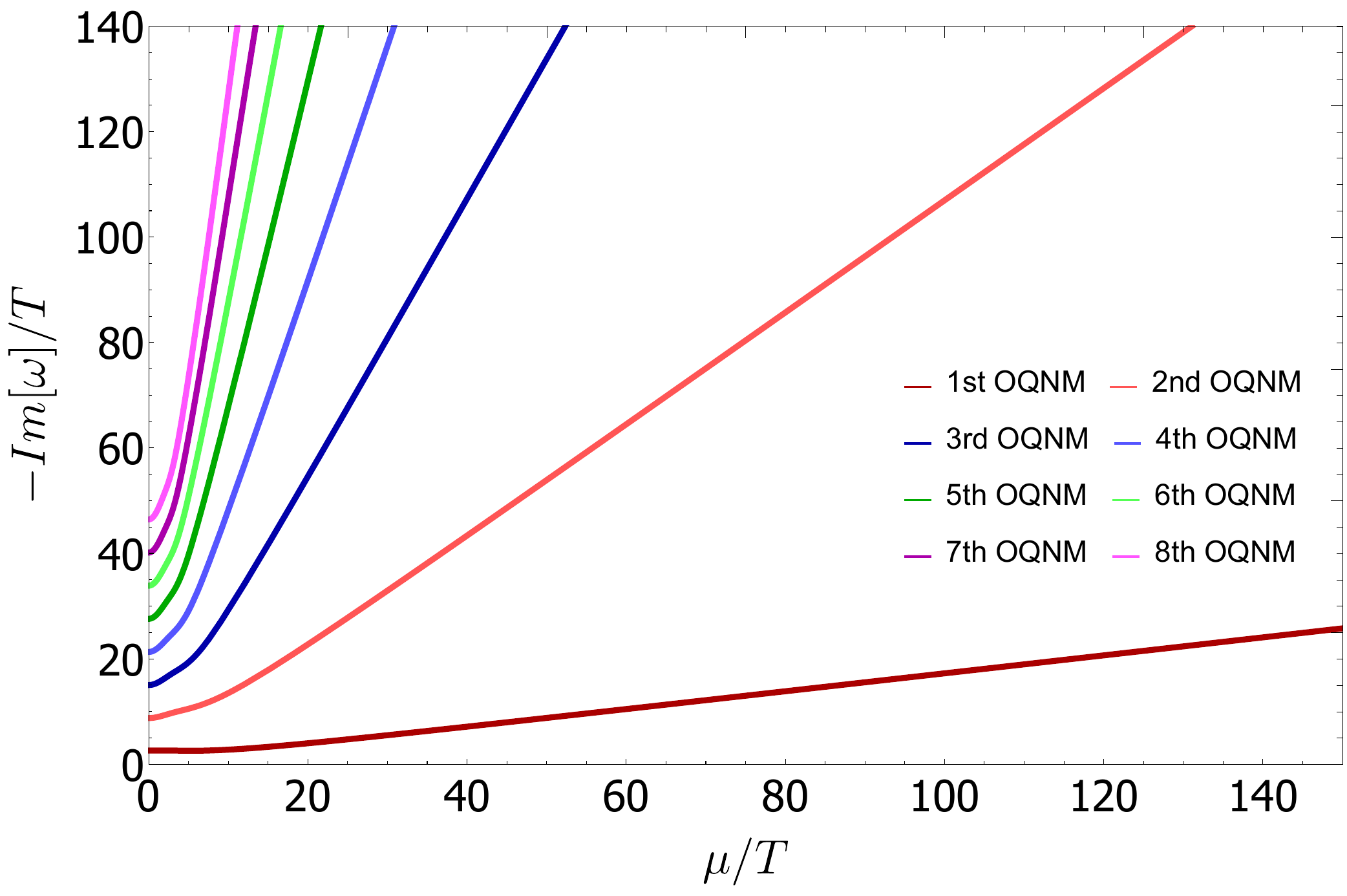}}
\subfigure[Imaginary part of the First 8 Purely Imaginary QNMs]
{\includegraphics[width=0.6\linewidth]{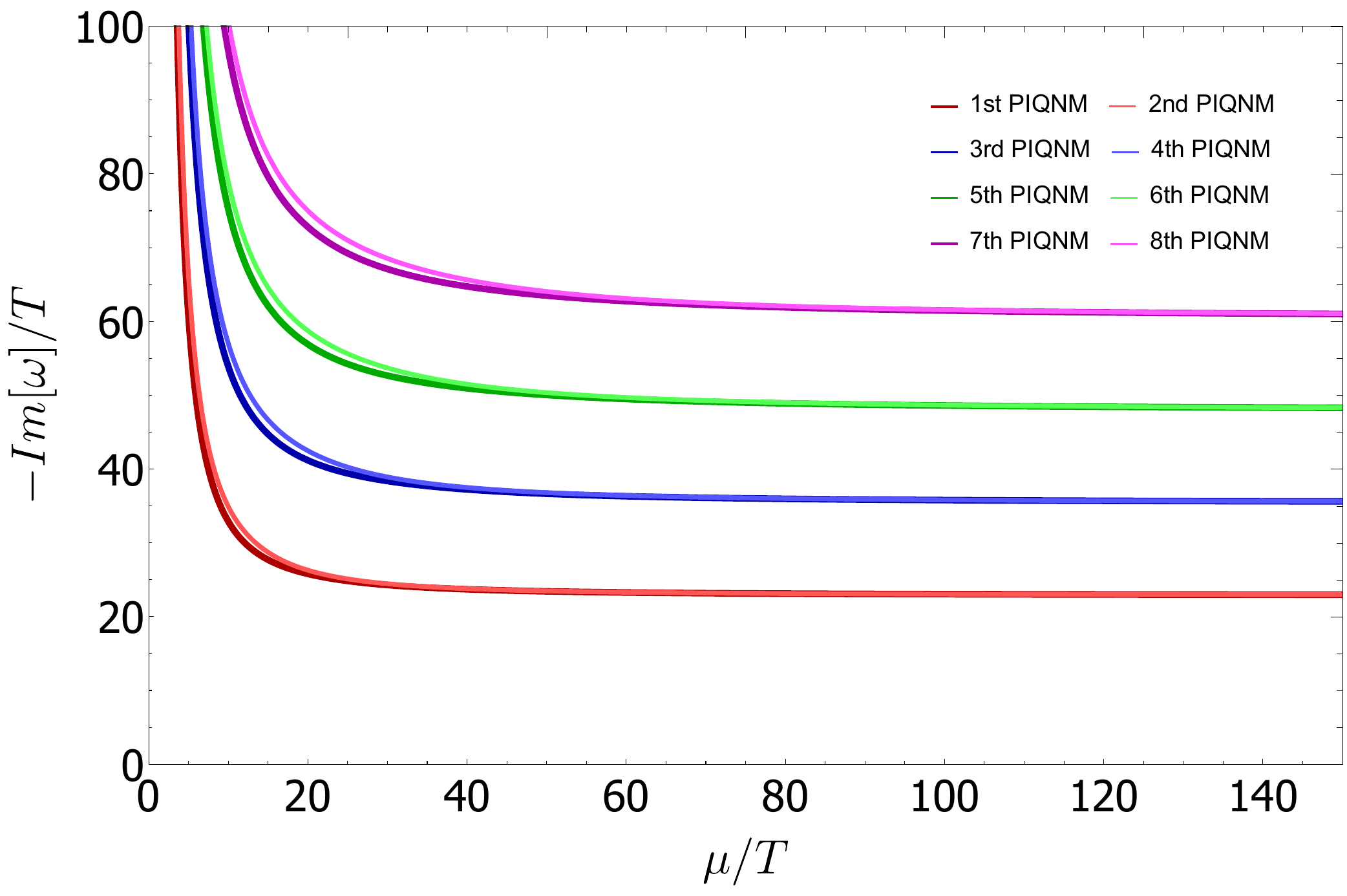}}
\caption{QNMs as functions of $\mu/T$ in the $SO(3)$ singlet channel of the 2RCBH model.}
\label{fig:PIQNMSin2RCBH}
\end{figure}

In the case of the scalar perturbation in the $SO(3)$ singlet channel, the normalizable mode at the boundary is obtained by setting $\mathcal{S}(u)=u^{\Delta_\phi} F(u)=u^2 F(u)$, with $F(0)\neq 0$ \cite{DeWolfe:2011ts,Critelli:2017euk}. Implementing the last prescription, one obtains for the 1RCBH model the following eigenfrequency problem \cite{Critelli:2017euk},\footnote{Notice that Eq. (B.3) in~\cite{Critelli:2017euk} is written in terms of the dimensionless frequency $\bar{\omega}\equiv\omega/T=2\pi\omega\sqrt{Q^2+r_H^2}\bigg/(Q^2+2r_H^2)$.}
\begin{align}
    \nonumber F_1''+ \Bigg[&\frac{1}{u}+2 u \left(\frac{Q^2+r_H^2}{u^2 \left(Q^2+r_H^2\right)+r_H^2}+\frac{1}{u^2-1}\right)-\frac{2 i r_H\omega_1\sqrt{\frac{Q^2 u^2}{r_H^2}+1}}{\left(u^2-1\right) \left(u^2 \left(Q^2+r_H^2\right)+r_H^2\right)}\Bigg]F_1'\\
    + \Bigg[&\frac{4 \left(4 Q^6 u^6+4 Q^4 r_H^2 u^2 \left(u^4+6 u^2-2\right)+3 Q^2 r_H^4 \left(8 u^4+3 u^2-2\right)+9 r_H^6 u^2\right)}{\left(u^2-1\right) \left(2 Q^2 u^2+3 r_H^2\right)^2 \left(u^2 \left(Q^2+r_H^2\right)+r_H^2\right)}\nonumber\\
    &-\frac{i\omega_1\left(2 Q^2 u^2+r_H^2\right)}{r_H u \left(u^2-1\right) \sqrt{\frac{Q^2 u^2}{r_H^2}+1} \left(u^2 \left(Q^2+r_H^2\right)+r_H^2\right)}\Bigg]F_1=0,
\label{eq:SingF1}
\end{align}
while in the case of the 2RCBH model, one obtains,
\begin{align}
    \nonumber F_2''+\Bigg[&\frac{1}{u}+2 u \left(\frac{2 Q^2+r_H^2}{u^2 \left(2 Q^2+r_H^2\right)+r_H^2}+\frac{1}{u^2-1}\right)-\frac{2 i\omega_2\left(Q^2 u^2+r_H^2\right)}{r_H \left(u^2-1\right) \left(u^2 \left(2 Q^2+r_H^2\right)+r_H^2\right)}\Bigg]F_2' \\
    +\Bigg[&\frac{4 \left(Q^6 u^4 \left(2 u^2+3\right)+Q^4 r_H^2 u^2 \left(u^4+24 u^2-2\right)+3 Q^2 r_H^4 \left(4 u^4+6 u^2-1\right)+9 r_H^6 u^2\right)}{\left(u^2-1\right) \left(Q^2 u^2+3 r_H^2\right)^2 \left(u^2 \left(2 Q^2+r_H^2\right)+r_H^2\right)}\nonumber\\
    &-\frac{i\omega_2\left(3 Q^2 u^2+r_H^2\right)}{r_H u \left(u^2-1\right) \left(u^2 \left(2 Q^2+r_H^2\right)+r_H^2\right)}\Bigg]F_2 =0.
\label{eq:SingF2}
\end{align}

\subsection{QNM Spectra}

Figs. \ref{fig:QNMSinRe} and \ref{fig:QNMSinIm} provide a direct comparison for, respectively, the real and imaginary parts of the first four OQNMs in the $SO(3)$ singlet channel of the 1RCBH and 2RCBH models, plotted as functions of $\mu/T$.

The behavior of both real and imaginary parts of the 2RCBH OQNMs for a much broader range of values of $\mu/T$ can be seen in Figs.~\ref{fig:PIQNMSin2RCBH}(a) and \ref{fig:PIQNMSin2RCBH}(b), where, as before, we plot the first few OQNMs with positive real part. Again, the initial growths for the real and imaginary parts of the OQNMs turn into linearly increasing asymptotic behaviors for large values of $\mu/T$.

Likewise the quintuplet and triplet channels, from Fig.~\ref{fig:PIQNMTri2RCBH}(c) we also observe the presence of a structure of pairs of PIQNMs in the singlet channel of the 2RCBH model. Differently from the quintuplet channel and similarly to the triplet channel, the moduli of the PIQNMS start at infinity and continuously decrease as $\mu/T$ increases until stabilizing at some constant values, without ever receding. Each purely imaginary QNM within a given pair of modes seems to converge to the same asymptotic value, while the different pairs are themselves evenly spaced apart. In fact, we obtained the numerical estimate $\delta_\textrm{1et}\approx 12.5665$ for the separation between the first two pairs of merged PIQNMs in the singlet channel at $\mu/T=1500$, calculated with $2500$ collocation points. As in the quintuplet and in the triplet channels, the separation value found in the singlet channel is notably close to $4\pi$.

\begin{figure}[h]
\centering  
\subfigure[Equilibration time]{\includegraphics[width=0.45\linewidth]{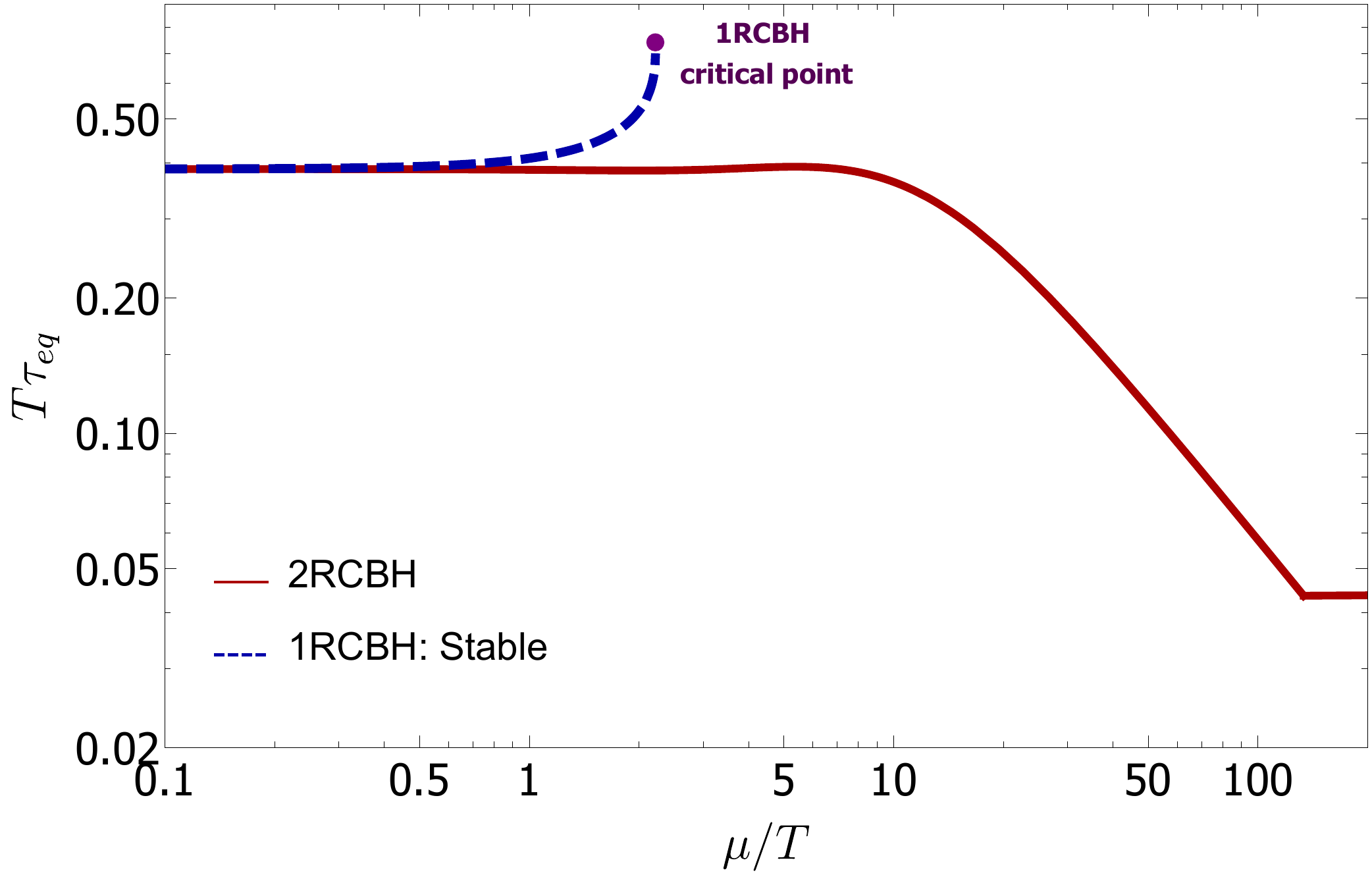}}
\subfigure[First $(\mu/T)$-derivative of the equilibration time]{\includegraphics[width=0.45\linewidth]{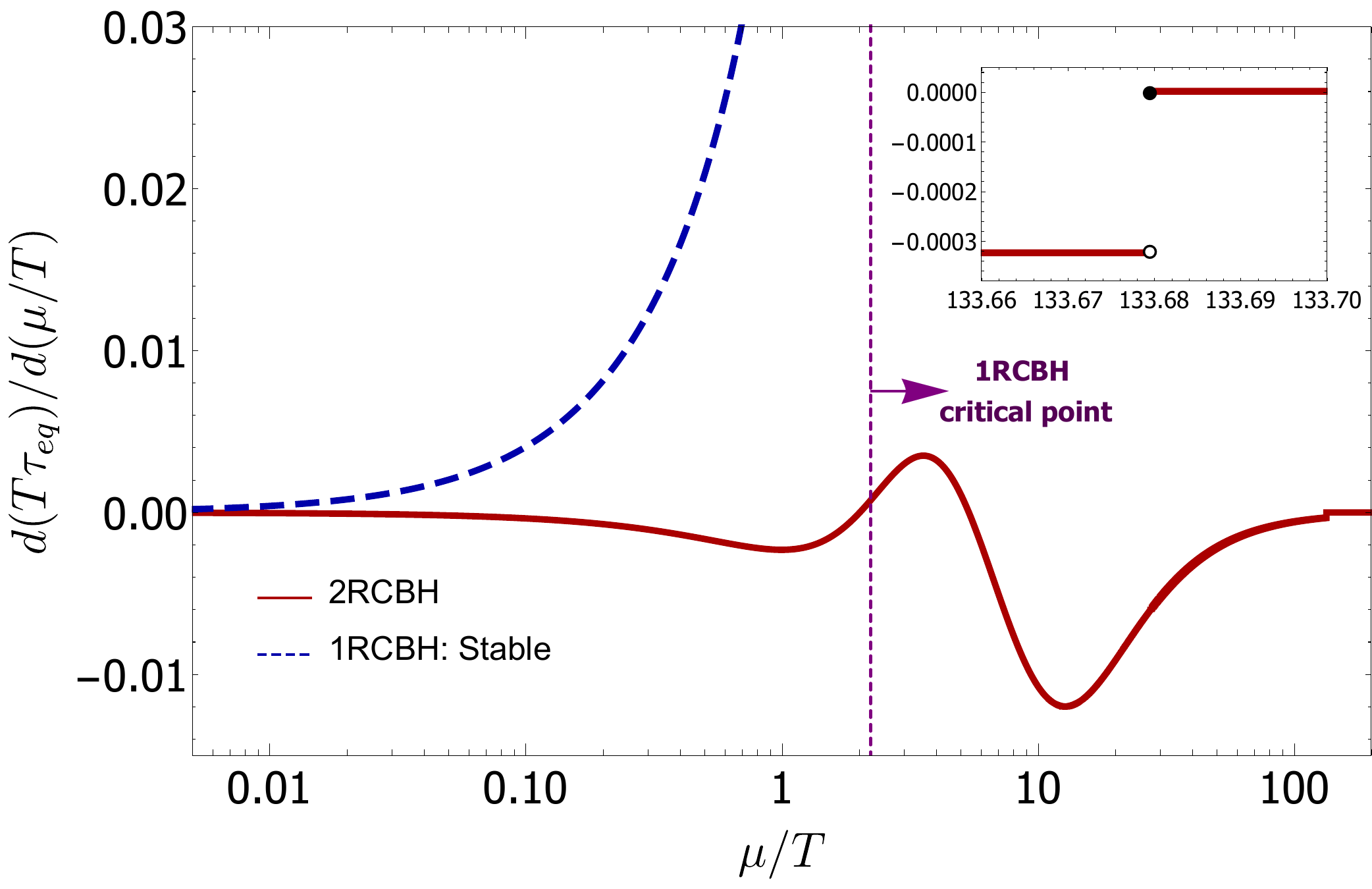}}
\caption{Characteristic equilibration time and its first $(\mu/T)$-derivative for the $SO(3)$ singlet channel of the 1RCBH and 2RCBH models. The inset plot on the right shows the discontinuity in the first $(\mu/T)$-derivative of the equilibration time of the 2RCBH model around $\mu/T\approx 133.68$, where the first PIQNM becomes the lowest QNM of the singlet channel.}
\label{fig:eqtime&derSin}
\end{figure}

The overall structure of QNMs in the $SO(3)$ singlet channel of the 2RCBH model is organized as displayed in Fig.~\ref{fig:StrQNMsSin}, where we plot six different key values of $\mu/T$. The first PIQNM, which rapidly came from $-i\infty $, takes the place as the lowest QNM of the singlet channel at $\mu/T\approx 133.68$, which is a much higher value than $\mu/T\approx 5.065$ and $\mu/T\approx 19.58$, for which the first PIQNM of the quintuplet and triplet channels become, respectively, the fundamental QNMs of those channels.

In what regards the plots displayed in the present work for the $SO(3)$ singlet channel of the 2RCBH model, the basic configuration employed in our numerical pseudospectral routine is presented in Table~\ref{TabSing}.
\begin{table}[]
\begin{tabular}{|c|c|c|c}
\cline{1-3}
$\mu/T$ & \# of collocation points & \# of data points &  \\ \cline{1-3}
{[}0, 5{]} & 150 & 200 &  \\
(5, 20{]} & 200 & 50 &  \\
{(}20,50] & 300 & 25 & \multicolumn{1}{l}{} \\
{(}50, 100{]} & 450 & 20 & \multicolumn{1}{l}{} \\
{(}100, 150{]} & 650 & 15 &  \\ \cline{1-3}
\end{tabular}
\caption{Some numerical details used to calculate the QNMs of the $SO(3)$ singlet channel.}
\label{TabSing}
\end{table}

Fig.~\ref{fig:eqtime&derSin} displays a comparison between the 1RCBH and 2RCBH models regarding the characteristic equilibration time of the $SO(3)$ singlet channel and its first derivative with respect to $\mu/T$. As in the quintuplet and triplet channels, the curves for both models in the singlet channel agree at small values of $\mu/T$. On the other hand, for the 1RCBH model the equilibration time acquires an infinite slope at the critical point of the model \cite{Critelli:2017euk}, with a corresponding divergence in the associated first $(\mu/T)$-derivative, while for the 2RCBH model there is an abrupt change in the equilibration time at $\mu/T \approx 133.68$, when the first PIQNM becomes the lowest QNM of the singlet channel, with a corresponding discontinuity in the associated first $(\mu/T)$-derivative.

\section{Conclusions and Perspectives}
\label{sec:conc}

In the present work, we investigated the homogeneous QNM spectra of the top-down holographic 2RCBH model, corresponding to a strongly-coupled quantum fluid defined at finite temperature and R-charge chemical potential, and also compared them with previous results for the 1RCBH model. The main discovery we report here is a new structure of pairs of purely imaginary quasinormal modes that emerge in the 2RCBH model at nonzero chemical potential, in all the $SO(3)$ channels analyzed, corresponding to the quintuplet (spin 2), the triplet (spin 1), and the singlet (spin 0) perturbations of the bulk EMD black hole backgrounds. Remarkably, the two poles within each pair of purely damped quasinormal modes converge to the same constant value at asymptotically large values of the dimensionless ratio of chemical potential over temperature, $\mu/T$. Another very interesting fact is that the asymptotic values of the different pairs of fused purely imaginary poles are themselves evenly space apart within each $SO(3)$ channel by the same value of $4\pi$. This phenomenon of \textit{asymptotic pole fusion} of purely imaginary pairs of QNMs is, as far as we know, an entirely new discovery which adds to the literature of quasinormal modes in holographic systems, deserving further investigation in future works. In particular, the physical mechanism inducing the pairing of purely imaginary quasinormal modes as the value of $\mu/T$ is increased, and the associated asymptotic pole fusion observed with a constant spacing of $4\pi$ between the different pairs of merged poles in all $SO(3)$ channels, is something still to be understood. Interestingly, the fact that the purely imaginary QNMs saturate at constant values at asymptotically large values of the chemical potential indicate that there are limited upper bounds in the characteristic equilibration times of the different channels of the system, which cannot be surpassed by further doping the system with increasing values of the chemical potential.

Another remarkable feature observed in the QNM spectra of the 2RCBH model is the fact that the first purely imaginary pole of each $SO(3)$ channel becomes the lowest/fundamental QNM of the system in the corresponding channel at some value of $\mu/T$, causing abrupt changes in the characteristic equilibration times of the quintuplet, the triplet, and the singlet channels of the 2RCBH model. This feature will certainly have a relevant impact on the late time evolution of different physical observables of the 2RCBH model undergoing homogeneous isotropization dynamics, when viewed as a functions of $\mu/T$, which we shall investigate in an upcoming work. In fact, as explicitly checked in~\cite{Critelli:2017euk,Rougemont:2024hpf} for the 1RCBH model, the lowest homogeneous non-hydrodynamic QNM of the $SO(3)$ quintuplet channel dominates the late time equilibration of the pressure anisotropy of the strongly-coupled plasma (corresponding to the isotropization of the medium), while the lowest QNM of the singlet channel dominates the late time equilibration of the one-point function (condensate) of the scalar field operator dual to the bulk dilaton field (corresponding to the thermalization of the system, since this is the slowest observable to equilibrate). In the case of the 1RCBH model, since there are no purely imaginary QNMs in those channels,\footnote{As previously discussed, the $SO(3)$ triplet channel of the 1RCBH model has a single purely imaginary QNM which becomes the fundamental QNM of that channel close to the critical point of the 1RCBH model \cite{Finazzo:2016psx}. However, the physical observable associated with the triplet channel, which is the R-charge density, is a conserved constant set by the initial conditions in the homogeneous isotropization dynamics \cite{Critelli:2017euk}.} the respective fundamental QNMs have, besides an imaginary part, a real part such that the aforementioned observables oscillate around their asymptotic equilibrium values at late times. In the case of the 2RCBH model disclosed here, the fundamental QNMs of the quintuplet and singlet channels are ordinary QNMs with imaginary and real parts until certain values of $\mu/T$, but above those threshold values the lowest QNMs are described by purely imaginary quasinormal modes, corresponding to purely damped modes with no oscillatory contribution. Therefore, one may speculate in principle that perhaps the pressure anisotropy and the scalar condensate of the medium will tend to their respective equilibrium values without presenting oscillations when the value of $\mu/T$ is above the respective thresholds for having the first purely imaginary quasinormal modes as the lowest QNMs of the quintuplet and singlet channels. However, since the QNM analysis is based on linearized field equations, the precise manifestation of such a qualitative change in the late time equilibration of the pressure anisotropy and the scalar condensate of the 2RCBH model undergoing homogeneous isotropization dynamics requires the numerical simulation of the full nonlinear dynamics of initially far-from-equilibrium anisotropic states, which is something that we expect to report soon in an upcoming work. Besides that, we shall investigate whether in the 2RCBH model one also observes the subtle near-equilibrium stairway structure disclosed in~\cite{Rougemont:2024hpf} for the out-of-equilibrium entropy in the late time equilibration of the homogeneous isotropization dynamics of the 1RCBH model. In the same vein of the aforementioned speculation, perhaps the stairway structure for the near-equilibrium entropy will be present at lower values of $\mu/T$, when the lowest QNM of the system has a nonzero real part, while it may not be present at higher values of $\mu/T$, when the lowest QNM is purely imaginary, since in~\cite{Rougemont:2024hpf} the period of plateau formation in the stairway to equilibrium entropy was found to be half the period of oscillation of the lowest QNM of the medium, which is associated to a nonzero real part. That will be investigated in detail.

\acknowledgments
G.O. acknowledges financial support by Coordination of Superior Level Staff Improvement (CAPES). R.R. acknowledges financial support by National Council for Scientific and Technological Development (CNPq) under grant number 407162/2023-2.


\bibliographystyle{apsrev4-2}
\bibliography{bibliography,extrabiblio} 

\end{document}